\documentclass[aps,notitlepage,nofootinbib,longbibliography,twocolumn,superscriptaddress,10pt]{revtex4-2}

\usepackage{bbold}
\usepackage{simplewick}

\usepackage[usenames,dvipsnames]{color}
\usepackage[colorlinks=true,citecolor=PineGreen,linkcolor=blue]{hyperref}

\usepackage{amsmath}
\usepackage{amsthm, amssymb}
\usepackage{physics}
\usepackage{enumitem}
\usepackage{slashed}
\usepackage{graphicx}
\usepackage{xcolor}
\usepackage{tikz}
\usetikzlibrary{calc,fadings,decorations.pathreplacing,shapes,shapes.multipart,arrows,shapes.misc,intersections,positioning,patterns}
\usepackage{bm}
\usepackage{bbm}
\usepackage[colorlinks=true,citecolor=PineGreen,linkcolor=blue]{hyperref}
\usepackage{dsfont}
\usepackage{changepage}
\usepackage{array}
\usepackage{soul}
\usepackage[capitalise]{cleveref}
\crefname{section}{Sec.}{Secs.}
\Crefname{section}{Sec.}{Secs.}
\usepackage{xifthen}
\usepackage{xargs}
\usepackage{subcaption}
\usepackage{braket}
\newcommand{\rB}{\boldsymbol{r}}
\newcommand{\sB}{\boldsymbol{s}}

\newcommand{\kB}{\boldsymbol{k}}

\begin{document}
\title{Monitored Quantum Dynamics and the Kitaev Spin Liquid}
\author{Ali Lavasani}
\affiliation{Joint Quantum Institute, NIST/University of Maryland, College Park, MD 20742, USA}
\affiliation{Condensed Matter Theory Center, University of Maryland, College Park, MD 20742, USA}
\author{Zhu-Xi Luo}
\affiliation{Kavli Institute for Theoretical Physics, University of California, Santa Barbara, CA 93106, USA}
\author{Sagar Vijay}
\affiliation{Department of Physics, University of California, Santa Barbara, CA 93106, USA}

\begin{abstract}

Quantum circuit dynamics with local projective measurements can realize a rich spectrum of entangled states of quantum matter.  Motivated by the physics of the Kitaev quantum spin liquid \cite{kitaev2006anyons}, we study quantum circuit dynamics in (2+1)-dimensions involving local projective measurements, in which the monitored trajectories realize (i) a phase with topological quantum order or (ii) a ``critical" phase with a logarithmic violation of area-law-scaling of the entanglement entropy along with long range tripartite entanglement.  A Majorana parton description of these dynamics, which provides an out-of-equilibrium generalization of the parton description of the Kitaev honeycomb model, permits an analytic understanding of the universal properties of these two phases, including the entanglement properties of the steady-state, the dynamics of the system on the approach to equilibrium, and the phase transition between these states. In the topologically-ordered phase, two logical qubits can be encoded in an initial state and protected for a time which scales exponentially in the linear dimension of the system, while no robust encoding of quantum information persists in the critical phase. Extensive numerical simulations of these monitored dynamics confirm our analytic predictions. 


\end{abstract}
\maketitle
\tableofcontents
\section{Introduction}
\label{sec:intro}


Quantum spin liquids \cite{Anderson1,Anderson2,Savary_2016} and topological orders characterized by long-range entanglement \cite{WenBook} have advanced our understanding of the possible phases of quantum matter. In equilibrium, quantum spin liquids can arise from frustration; examples include  geometric frustration in non-bipartite lattices such as the nearest-neighbor Heisenberg model on the Kagom\'{e} lattice \cite{PhysRevLett.62.2405,PhysRevB.45.12377,PhysRevB.56.2521,PhysRevB.76.180407,YanHuseWhite,PhysRevLett.109.067201,PhysRevB.83.224413,PhysRevB.84.020407,PhysRevB.84.020404,Jiang_2012}, competition between longer-ranged and shorter-ranged interactions such as the nearest and second-nearest neighbor Heisenberg model on the triangular lattice \cite{PhysRevB.92.041105,PhysRevB.92.140403,Kaneko,PhysRevB.93.144411}, and anisotropic interactions such as the celebrated Kitaev honeycomb model \cite{kitaev2006anyons}. 

The search for material realizations of quantum spin liquids remains challenging due to the requirement of strong correlations, while definitive signatures of spin liquids can be difficult to experimentally probe. On the other hand, digital quantum simulators \cite{altman2021quantum} have recently emerged as a platform for using controllable operations to simulate interesting phases of quantum condensed matter.  Therefore, it is natural to consider constructive way to design long-range entangled states in this setting using local unitary operations and projective measurements. 

In this work, we exploit non-commutative, projective measurements as a new source of frustration to generate out-of-equilibrium analogues of spin liquid states. Specifically, we consider projective measurements of the competing anisotropic interactions in the Kitaev honeycomb model \cite{kitaev2006anyons}.  When these measurements are performed in a spatially random fashion, we find that the monitored trajectories of the quantum many-body system produce analogues of the two phases in the original Kitaev model, as the relative rate of the different kinds of measurements are tuned: ($i$) a topologically-ordered phase with long-range entanglement, and area-law scaling of the entanglement entropy and ($ii$) a ``critical" phase with a logarithmic violation of area-law entanglement and long-range tripartite entanglement.  The latter phase appears to resemble a state in which the Majorana partons of the Kitaev spin liquid  have formed a Fermi surface, which cannot occur in the ground-state of the Kitaev spin liquid on the honeycomb lattice while preserving time-reversal and translation symmetry \cite{kitaev2006anyons}.  
Through analytical arguments and numerical studies, we argue that these two phases are robust in the presence of other perturbations, such a small rate of additional projective measurements.

 An analytic understanding of these  phases is obtained by invoking a Majorana-fermion parton representation, which has been previously used to re-cast the honeycomb model as a model of Majorana fermions coupled to a static $\mathbb{Z}_{2}$ gauge field  \cite{kitaev2006anyons}.  The dynamics that we consider can be formulated as the dynamics of the Majorana partons evolving under the action of local measurements, followed by a projection back into the physical spin Hilbert space.  A limit of these measurement-only dynamics admits a particularly simple representation in terms of the Majorana partons, allowing for an understanding of the universal properties of the topologically-ordered and critical phases, as well as the phase transition between them. This  representation is also used to argue in general terms that certain high-symmetry regimes of these measurement-only dynamics must give rise to either a topologically-ordered phase for the monitored pure-states or a phase in which these states have super-area-law scaling of the entanglement entropy. 
We note that the far-from-equilibrium parton construction invoked in this work could potentially be useful to study other long-range-entangled phases of quantum matter that can emerge in monitored quantum dynamics.

We note that quantum circuits with both projective measurements and local unitary gates (``hybrid" circuits) have been extensively studied to uncover a rich behavior of the monitored pure-state trajectories. In (1+1)-dimensions, these dynamics can give rise to a new phase of volume-law-entangled quantum matter \cite{li2021entanglement} with close connections to the theory of quantum error-correcting codes  \cite{fan2021self,li2021statistical,choi2020quantum,fidkowski2021dynamical,yoshida2021decoding}. Dynamics in which the measurements and unitary gates commute preserve a global symmetry can give rise to monitored trajectories that exhibit symmetry-protected topological order \cite{Ali} and spin-glass order \cite{PhysRevResearch.3.023200}. In (2+1)-dimensions, hybrid quantum circuits can yield monitored pure-states with truly long-range entanglement \cite{PhysRevLett.127.235701}. 

The remainder of the paper is organized as follows. After a detailed summary of the key results in Sec. \ref{sec:summary_of_results}, the dynamics considered in this work are introduced in Section \ref{sec:circuitModel}. Sections \ref{sec:prop_of_std_state} and \ref{sec:steady_state_ee} present the Majorana parton representation of the dynamics of the monitored pure-states, as well as analytical results regarding the entanglement properties of these trajectories. In Section \ref{sec:purification_dyn} we discuss the purification dynamics of the circuit model in each phase. The numerical results are presented in Section \ref{sec:numerics}. In Section \ref{sec_perturbation} we study small perturbations to the monitored dynamics studied in Sec. \ref{sec:circuitModel}, and show through a combination of numerical and analytic arguments that the critical and topologically-ordered phases remain stable.  We conclude with a short discussion and outlook in Section \ref{sec:discussion}.

\subsection{Summary of Results}
\label{sec:summary_of_results}

\begin{figure}
         \includegraphics[width=.8\columnwidth]{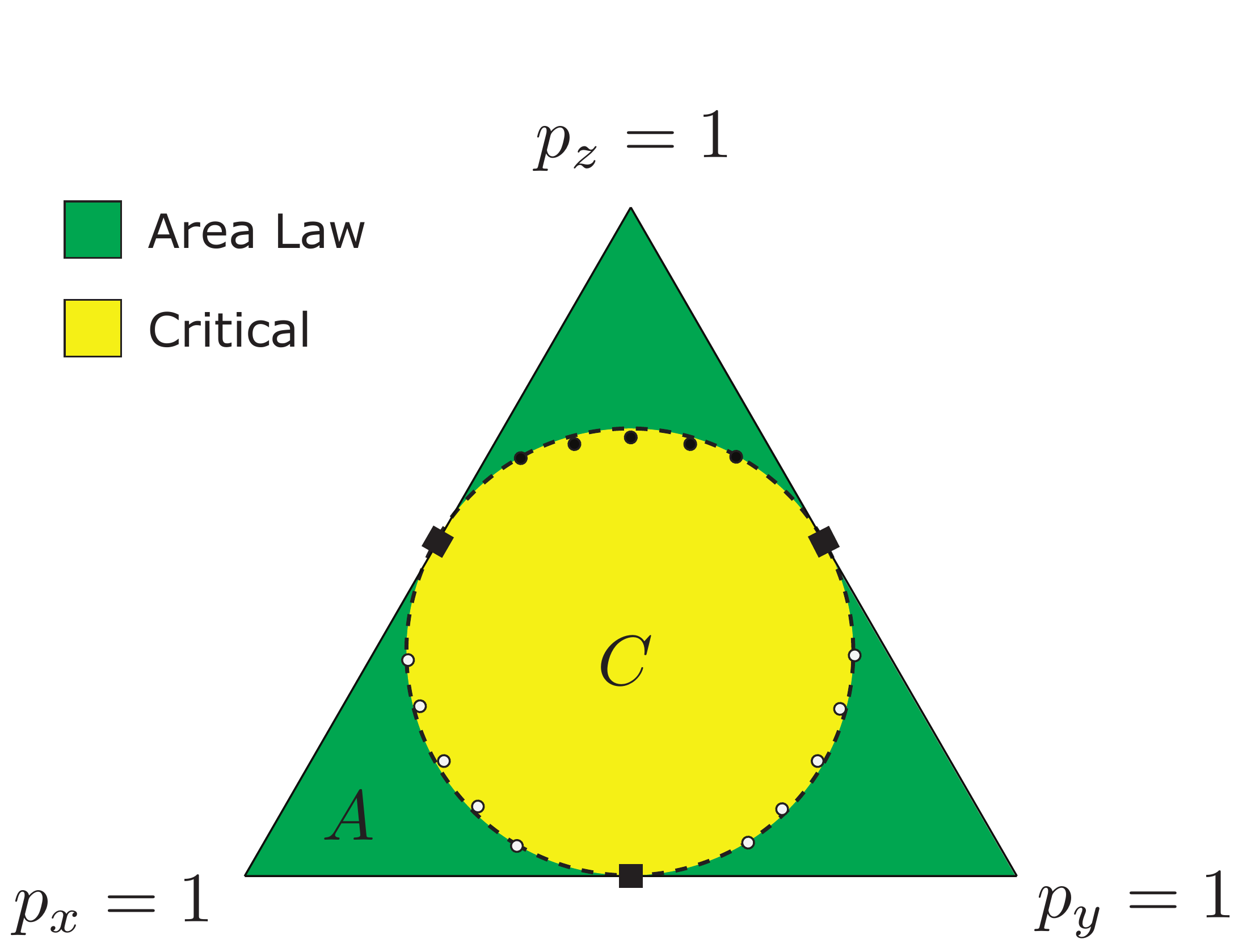}
        \caption{{\bf Phase diagram:}. As the probabilities of measuring different bond operators ($p_{x}$, $p_{y}$, $p_{z}$ with $p_{x}+p_{y}+p_{z}=1$) are tuned, the monitored pure-state trajectories realize ($i$) a topologically-ordered phase with an area-law scaling of the entanglement entropy (green regions) or ($ii$) a ``critical" phase with a  logarithmic violation of area-law entanglement scaling along with long-range tripartite entanglement (yellow regions). The black dots show the numerically-extracted phase transition from  simulations of these dynamics, while the remaining dots are obtained by $\pi/3$ rotations. The topologically-ordered and critical phases are robust when the dynamics are perturbed, e.g by adding a small rate of single-qubit projective measurements. }\label{fig_phasediagram}
\end{figure}

The monitored random circuit studied here is inspired by the Kitaev honeycomb model \cite{kitaev2006anyons}. We consider the same arrangement of qubits on an $L\times L$ honeycomb lattice with periodic boundary conditions. We first study dynamics involving projective measurements of the nearest-neighbor interaction terms in the Kitaev honeycomb model \cite{kitaev2006anyons}, which depend on the orientation $j\in\{x,y,z\}$ of the bond connecting the two qubits: at each timestep, a random two-qubit operator along a bond of type $j$ is measured with  probability $p_{j}$ (with $p_x+p_y+p_z=1$); these operators and the arrangement of qubits are discussed in greater detail in Sec. \ref{sec:circuitModel}.  As these measurement rates are tuned, the monitored pure-state trajectories of the system support two distinct phases, as summarized in Fig. \ref{fig_phasediagram}:



\begin{enumerate}[label=(\roman*)]
    \item \textit{Topologically-Ordered Phase}: For highly biased measurement probabilities (e.g. $p_z \gg p_x,p_y$) corresponding to the green regions in Fig. \ref{fig_phasediagram}, the monitored pure-states exhibit topological order.  In addition to exhibiting area-law scaling of the von Neumann entanglement entropy\footnote{We use the following definition of the von Neumann entropy of a density matrix $\rho$ throughout this work as $S(\rho) = -\mathrm{Tr}\left[\rho\log_{2}\rho\right]$}, the steady-state of the monitored dynamics exhibits long-range entanglement, as evidenced by a quantized topological entanglement entropy \cite{kitaev2006topological,levin2006detecting} $S_{\mathrm{topo}} = 1$ as well as a bipartite mutual information between two non-contractible regions on the torus and separated by $O(L)$ distance, which is equal to $1$. 
    
    The evolution of a maximally-mixed initial state in this dynamical regime further reveals the persistent, long-ranged entanglement in this phase. Projective measurements have the effect of rapidly disentangling the maximally-mixed initial state, resulting in an exponential decay of the entanglement entropy until it plateaus at $2$. This plateau persists until a time $t_{\mathrm{purif}} \sim O(\exp(L))$, when the system completely purifies. This protected information which survives until exponentially long times in the system size corresponds to the entanglement between the environment and two operators which loop around the topologically non-trivial cycles of the torus, and suggests the emergence of a dynamically generated quantum code with two logical qubits. 
    
    In contrast, we note that the Kitaev honeycomb model does not encode any logical qubits as a subsystem code \cite{suchara2011constructions}, while the recently-discovered honeycomb code \cite{Hastings2021dynamically} uses a time-periodic schedule of measurements to protect two logical qubits. In contrast, the topologically-ordered phase that we find demonstrates that randomness in measurement order can also lead to the emergence of a long-lived logical subspace.  
    
    \item \textit{Critical Phase}: A ``critical" phase is found in region $C$ in Fig. \ref{fig_phasediagram}. In this phase, the entanglement entropy of a large subsystem $A$ scales with its linear dimension $L_{A}$ as 
    \begin{align}
        S_{A} = c_{0} L_{A}\log L_{A} + \cdots
    \end{align}
    where the ellipsis denotes sub-leading corrections, and $c_{0}$ is a non-universal constant. Furthermore, on the torus, the bipartite mutual information between two non-contractible regions of width $O(L)$ separated by a similar distance, denoted $I_{2}(L)$, scales linearly with $L$. 
    The critical phase also supports multi-partite long range entanglement as quantified by the negative tripartite mutual information between non-contractible  regions, in contrast to the area-law phase. Finally, the magnitude of the expectation value of open Wilson lines with endpoints separated by a distance $r$ decays as a power law $1/r^\Delta$ with $\Delta=3$.
    
    The purification dynamics in the critical phase is quite different from the area law phase. Starting from a totally mixed state, there is an initial exponential drop, followed by $1/t$ decay of entanglement entropy of the system, which can be understood as a ``L\'{e}vy flight" annihilation process of unpaired Majorana partons. The system completely purifies at a time $t_{\mathrm{purif}}\sim O(L^2)$. No logical information is protected in this phase.
\end{enumerate}

We also consider small perturbations to these dynamics, e.g. single-qubit or other multi-site measurements, which have the important effect of removing the extensive number of conserved quantities which are present in dynamics involving only bond measurements. Through a combination of numerical and analytical arguments, we show that the critical and topologically-ordered phases persist in the presence of these perturbations.

We may compare the entanglement properties of these monitored trajectories with the phases of the Kitaev spin liquid.  In the presence of time-reversal and translational symmetry, the Kitaev spin liquid on the honeycomb lattice  \cite{kitaev2006anyons} has two phases: a gapless phase where the Majorana partons have a semi-metallic dispersion, and three gapped phases which each exhibit $\mathbb{Z}_{2}$ topological order, and are related to each other by three-fold ($\theta = 2\pi/3$) spatial rotations. The area-law-entangled phase that arises in the monitored dynamics that we study is quantitatively similar to the $\mathbb{Z}_{2}$ topological order that arises in the ground-state of the Kitaev honeycomb model, due to the quantized topological entanglement entropy, and the ability to encode two logical qubits on the torus.  As discussed in Sec. \ref{sec:prop_of_std_state}, the quantized bipartite mutual information between a pair of non-contractible regions on the torus is also shared by a particular ground-state sector of the two-dimensional $\mathbb{Z}_{2}$ toric code on the torus, {into which we argue that the steady-state of the measurement-only dynamics purifies when the measurement rates are highly anisotropic}. 

In contrast to these area-law-entangled phases \cite{yao2010entanglement}, the critical phase quantitatively resembles a state in which the Majorana partons in the Kitaev spin liquid have formed a Fermi surface, due to both ($i$) the scaling of the entanglement entropy with subsystem size and ($ii$) the extensive scaling of the bipartite mutual information $I_{2}(L)\sim L$.  Such a state cannot emerge in the ground-state of the Kitaev spin liquid in the presence of time-reversal and translation symmetry.  Since the critical phase of the monitored dynamics arises when measurements are applied randomly in space, translation symmetry is not present in any typical realization of the dynamics, though it is preserved by the statistical ensemble of monitored pure-states. As a result, a more appropriate comparison may be to the Kitaev honeycomb model with broken translation symmetry, e.g. by introducing randomness in the spin exchange interaction \cite{knolle2019bond}, which can produce an effective description of the Majorana partons with random hopping matrix elements on a bipartite lattice, which could lead to a ``metallic" phase for the partons  \cite{motrunich2002particle} in which states at zero-energy are delocalized. The entanglement properties of such a phase are not known to us, however, and we are unable to draw any further quantitative comparisons between this phase and the critical phase of the monitored dynamics.

Finally, perturbations which break time-reversal symmetry, such as an external magnetic field, can drive the gapless phase of the Kitaev honeycomb model into a gapped, chiral spin liquid supporting non-Abelian anyons \cite{kitaev2006anyons}. Furthermore, in the presence of both strong disorder and broken time-reversal symmetry, the Majorana partons can form a metallic state \cite{Andreas} in which the fermionic parton wavefunctions $\psi_n(\bm{r})$ are spatially-extended  \cite{chalker2001thermal,Thermal_MC}, as quantified by a non-trivial scaling of the inverse participation ratios (IPR) with system size $L$ as $I_q=\int d^2r |\psi_n(\bm{r})|^{2q}\sim L^{-(q-1)D_q}$, with $D_{q}$ a $q$-dependent constant. Analogous perturbations to the measurement-only dynamics that we consider, such as single-qubit projective measurements, do not appear to qualitatively alter the entanglement properties of the critical phase. Furthermore, the stabilizer nature of the dynamics that we study, as clarified in Sec. \ref{sec:circuitModel},  necessarily prevents the parton wavefunctions from being spatially extended in a manner which is characteristic of this disordered, metallic phase. 


We also study phase transitions between the critical and topologically-ordered phases in this monitored dynamics. We argue that the generic nature of the phase transition  is related to a geometrical phase transition in a three-dimensional classical loop model, which has arisen in a different context in the study of measurement-only dynamics of free fermions \cite{nahum2020entanglement}. The tripartite mutual information, which serves as an order parameter for the phase transition between the critical and topologically-ordered phases, is used to numerically extract the correlation length critical exponent $\nu\approx 0.9$ consistently at distinct phase transition points in the bulk of the phase diagram. This value is close to the numerically-obtained value of this exponent from previous studies of this phase transition  \cite{ortuno2009random,serna20213d}. We note that the  transitions at the boundary of the phase diagram, e.g. $p_z=0$, can be understood by mapping the growth of local operators which stabilize the evolving wavefunction to the problem of bond percolation in two spatial dimensions, which is in a different universality class from the phase transition between the critical and topologically-ordered phases.

\section{Monitored Dynamics}\label{sec:circuitModel}

\begin{figure}
$\begin{array}{cc}    \includegraphics[width=0.45\columnwidth]{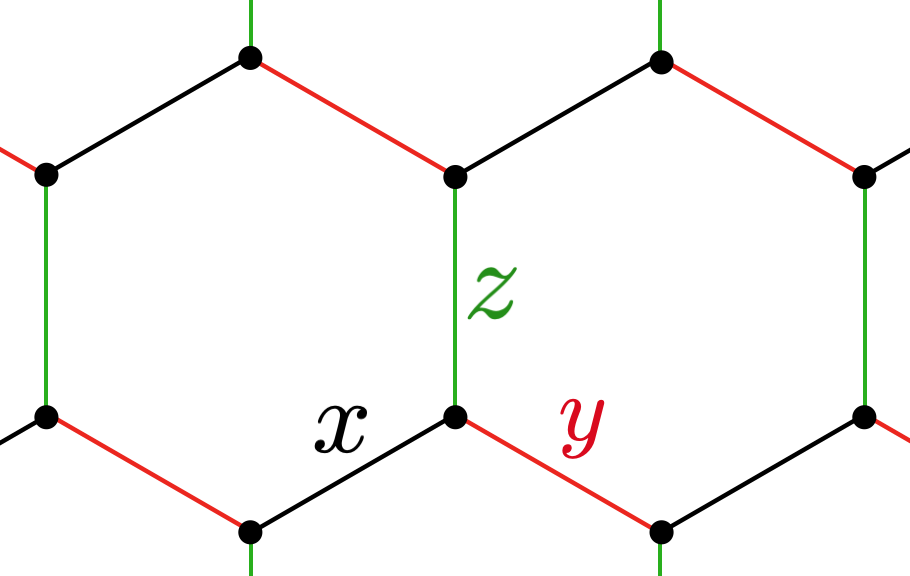} & 
\includegraphics[width=0.28\columnwidth]{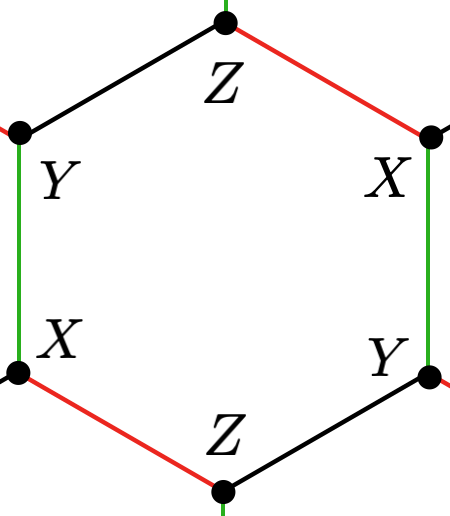}\\
\text{(a)} & \text{(b)}
\end{array}$
    \caption{{\bf Dynamics:} Qubits degrees of freedom reside on the sites of the honeycomb lattice.  Bonds of the lattice are of type $j\in\{x,y,z\}$ shown in (a) and a bond operator is defined as the product of the two Pauli operators of type $j$ along that bond.  A plaquette operator $W_{p}$ is defined as the product of bond operators over an elementary plaquette, as shown in (b).}
    \label{fig:linkops}
\end{figure}

We now describe the monitored dynamics in detail. Consider a system of $N=2L^2$ qubits arranged on vertices of a $L\times L$ honeycomb lattice with periodic boundary condition. On each edge of the honeycomb lattice connecting sites $\rB$ and $\rB'$, we define a \textit{bond operator} depending on the orientation of the edge and which acts on the two qubits as $X_{\rB}X_{\rB'}$, $Y_{\rB}Y_{\rB'}$ or $Z_{\rB}Z_{\rB'}$ (see Fig.\ref{fig:linkops}).   Starting from a maximally-mixed initial state $\rho=\mathbbm{1}/2^{N}$, we choose a bond of type $x$, $y$, or $z$ randomly with probability $p_x$, $p_y$ or $p_z$ respectively ($p_x+p_y+p_z=1$), and we measure the corresponding two-qubit bond operator, with measurement outcomes obtained according to Born's rule.  A time step of these dynamics corresponds to $N$ consecutive measurements.  

We note that product of bond operators around an elementary plaquette $p$ of the honeycomb lattice, as shown in Fig. \ref{fig:linkops}b, commutes with all of the bond operators; we refer to this as the \textit{plaquette operator} $W_{p}$. As a result, a monitored dynamics involving measurements of bond operators contains extensively many conserved quantities. In Sec. \ref{sec_perturbation}, we will consider perturbing away from these dynamics in such a way that this conservation law is no longer microscopically maintained.

\subsection{Parton Description of the Steady-State}\label{sec:prop_of_std_state}
We now explore the steady-state entanglement properties of the monitored pure-state of the system as a function of the probabilities $p_{x}$, $p_{y}$, $p_{z}$.  To make analytic progress, we replace each qubit with four Majorana fermions with fixed fermion parity so that the  Pauli spin operators at lattice site $r$ are represented as $X_{\rB} = ib^{x}_{\rB}c_{\rB}$, $Y_{\rB} = ib^{y}_{\rB}c_{\rB}$, and $Z_{\rB} = ib^{z}_{\rB}c_{\rB}$ \cite{kitaev2006anyons}.  It is convenient to arrange the fermions within each lattice site so that each $b^{j}$ Majorana fermion lies at the end of a bond of type $j$. By requiring that the fermion parity is constrained at each site as  $b^{x}_{\rB}b^{y}_{\rB}b^{z}_{\rB}c = -iX_{\rB}Y_{\rB}Z_{\rB} = 1$, we faithfully recover the spin Hilbert space.

\begin{figure}
$\begin{array}{c}
 \includegraphics[width=.47\textwidth]{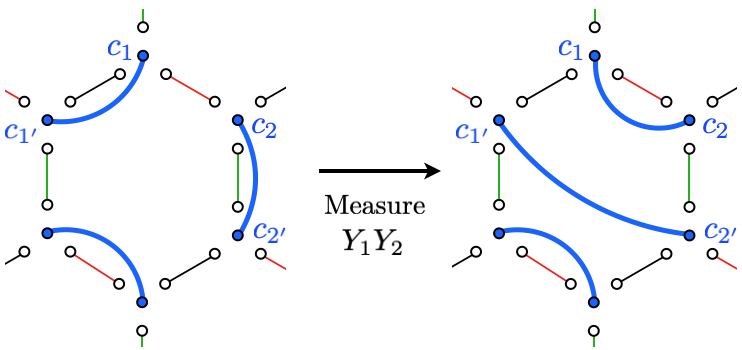}\\
 \text{(a)}\\
 \includegraphics[width=.43\textwidth]{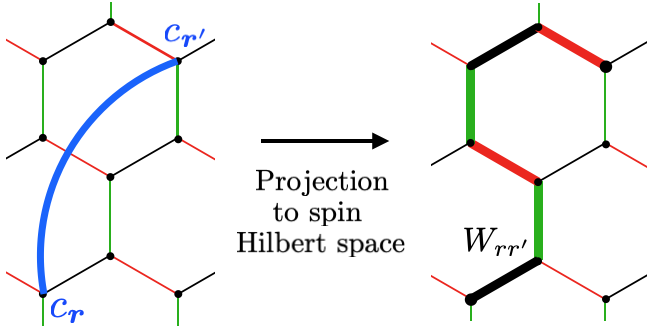}\\
 \text{(b)}
 \end{array}$
         \caption{{\bf Parton Dynamics:} A projective measurement of a bond operator gives rise to a dynamics of the Majorana partons as shown in $(a)$ and described in Sec. \ref{sec:prop_of_std_state}. Each pair of dimerized Majorana fermions $ic_{\rB}c_{\rB'} = \pm 1$ in the parton wavefunction gives rise to a stabilizer in the spin Hilbert space, which has support along a string connecting $\rB$, $\rB'$ as shown in $(b)$. }
         \label{fig:dimer_dynamics}
 \end{figure}
 
The monitored pure state of the spins evolves through the application of local projectors, which correspond to a sequence of local measurements and outcomes, which are drawn according to Born's rule.  Within a particular monitored trajectory, the dynamics of the spins may be re-cast as a dynamics for the Majorana partons, followed by a projection 
$P \equiv \prod_{\rB}({1 + c_{\rB}b^{x}_{\rB}b^{y}_{\rB}b^{z}_{\rB}})/{2}$ back into the Hilbert space of the spins, due to the fact that the projector $P$ commutes with any operator acting on the spin degrees of freedom.  As a concrete example, let $\ket{\psi_{f}}$ be a state of the Majorana partons, while $\ket{\Psi} \sim P\ket{\psi_{f}}$ be the corresponding wavefunction for the spins.  A measurement of a single-qubit operator $X_{\rB}$ which yields the outcome $+1$ modifies the state of the spins, up to normalization, as $[(1+X_{\rB})/2]\ket{\Psi} \sim P [(1 + ib^{x}_{\rB}c_{\rB})/2]\ket{\psi_{f}}$.  The latter expression may be interpreted as a dynamics for the Majorana parton wavefunction $\ket{\psi_{f}}$ followed by a projection into the spin Hilbert space. 

While a monitored trajectory of the spins corresponds to a monitored evolution of the Majorana partons, we note an important difference in the probability distribution of monitored trajectories between these two kinds of dynamics. If $\rho_{f}$ is the density matrix of the Majorana partons, so that $\rho = P\rho_{f}P/\Tr(P\rho_{f})$ is the corresponding state of the spins, then a measurement of a Pauli operator $\mathcal{O}$ yields the two possible outcomes with probability $p_{\pm} = \Tr(P\Pi_{\pm}\rho_{f})/\Tr(P\rho_{f})$, respectively, where $\Pi_{\pm} = (1\pm\mathcal{O})/{2}$.  This is \emph{a priori} different from the  probability distribution $\Tr(\Pi_{\pm}\rho_{f})$ of obtaining these outcomes when measuring the corresponding operator in the state of the Majorana partons\footnote{A simple example of such an observable is the operator $b^{x}_{\rB}b^{y}_{\rB}b^{z}_{\rB}c_{\rB}$; this operator is mapped to the identity within the spin Hilbert space, and thus has only one possibe outcome when ``measured". In contrast, the operator $b^{x}_{\rB}b^{y}_{\rB}b^{z}_{\rB}c_{\rB}$ can take on various values depending on the precise state of the Majorana partons.}.  Below, however, we will focus on understanding the monitored pure-state trajectories when $\rho$ describes a stabilizer state \cite{gottesman1997stabilizer}, for which the entanglement properties of the evolving pure-state are sensitive to the Pauli operators which are measured, but insensitive to their outcomes.

Starting with a maximally-mixed initial state, we perform measurements of the bond operators. The plaquette operators, given by the product of bond operators around an elementary plaquette, are all measured after a short time $t_{*}\sim O(\log L)$, as we argue in Sec. \ref{sec:purification_dyn}. For times $t> t_{*}$, it is convenient to take as an {ansatz} that the density matrix for the Majorana fermions $\rho_{f}(t)$, before projecting back into the spin Hilbert space, is given by
\begin{align}\label{eq:dynamics_rho}
\rho_{f}(t) = \rho_{c}(t) \otimes \rho_{b}
\end{align}
Here, $\rho_{b} = \ket{\Psi_{b}}\bra{\Psi_{b}}$ is a {pure state} in which each $b$ Majorana fermion is ``dimerized" with its nearest neighbor, i.e. so that $ib^{j}_{\rB}b^{j}_{\rB'}\ket{\Psi_{b}} = \ket{\Psi_{b}}$ where $r$, $r'$ are lattice sites which are connected by a $j$-type bond with $j\in\{x,y,z\}$.  This {ansatz} for the density matrix is particularly convenient, since the resulting density matrix for the spin degrees of freedom $\rho(t) = P\rho_{f}(t)P/\Tr(P\rho_{f})$ is stabilized by all of the plaquette operators so that $\mathrm{Tr}(W_{p}\rho(t)) = 1$.\footnote{The density matrix in Eq. (\ref{eq:dynamics_rho}) on the torus is also stabilized by the product of the bond operators around any non-contractible cycle of the torus.  We will return to this point at the end of this section. }  Any bond operator may be represented as the product of the four Majorana fermions along that bond.  A measurement of this operator manifestly commutes with the density matrix $\rho_{b}$.  As a result, these measurements give rise to dynamics of the $c$ Majorana partons, while leaving the state of the $b$ fermions invariant. 

A partial understanding of the steady-state entanglement properties of the monitored pure-states is obtained by taking the $c$ Majorana partons to be in a pure state $\ket{\Psi_{c}}$ in which all of the Majoranas are ``dimerized", i.e. for each site $r$ there is another site $r'$ such that $ic_{\rB}c_{\rB'}\ket{\Psi_{c}} = \pm\ket{\Psi_{c}}$; the sign of the fermion parity will be unimportant for determining the entanglement properties of the steady-state.  These operators can be visualized as forming a unique pairing of distinct sites on the honeycomb lattice which evolves under the monitored dynamics generated by the two-spin measurements. We may see this by considering a concrete example, which is summarized in Fig. \ref{fig:dimer_dynamics}a: let the pure state $\ket{\Psi_{c}}$ be given such that $ic_{1}c_{1'}\ket{\Psi_{c}}=ic_{2}c_{2'}\ket{\Psi_{c}}=1$ where the sites $1$ and $2$ are nearest-neighbors connected by an $y$-type bond.  The two-spin measurement $Y_{1}Y_{2} = c_{1}b^{y}_{1}b^{y}_{2}c_{2}$ is now performed, yielding the outcome that $Y_{1}Y_{2} = +1$.  Applying the projection operator $\Pi \equiv (1 + c_{1}b^{y}_{1}b^{y}_{2}c_{2})/2$ 
\begin{align}
&\Pi\frac{1 + ic_{1}c_{1'}}{2}\frac{1 + ic_{2}c_{2'}}{2}\frac{1 + ib^{y}_{1}b^{y}_{2}}{2}\Pi\nonumber\\
&= \frac{1 - ic_{1}c_{2}}{2}\frac{1 + ic_{1'}c_{2'}}{2}\frac{1 + ib^{y}_{1}b^{y}_{2}}{2}\label{eq:free_fermion}
\end{align}
yields the result that $-ic_{1}c_{2}$ and $ic_{1'}c_{2'}$ are the new stabilizers for the evolving pure state of the $c$ fermions.  In terms of the original configuration, the endpoints of the two dimerized Majoranas have now been paired together as a result of the measurement.  

The evolving density matrix for the original spins $\rho(t) \sim P\rho_{f}(t)P$ is stabilized by all elements of the stabilizer group of $\rho_{f}(t)$ which commute with $P$. Each Majorana dimer $ic_{\rB}c_{\rB'}$ stabilizing $\rho_{f}(t)$ gives rise to a stabilizer $W_{\rB\rB'}$ for $\rho(t)$, which is given, up to an overall sign, by the product of bond operators along any path connecting the two points, as shown in Fig. \ref{fig:dimer_dynamics}b.  Physically, the $c$ Majorana partons carry gauge charge, and must be connected by a Wilson line in the physical  Hilbert space.

 \subsection{Entanglement Properties of the Monitored Trajectories}\label{sec:steady_state_ee}
For any qubit stabilizer wavefunction, the von Neumann entanglement entropy of a subsystem $A$ is given by $S_{A} = -\Tr\left[\rho_{A}\log_{2}\rho_{A}\right] = I_{A} - |A|$ where $|A|$ is the number of qubits in the $A$ subsystem and $I_{A}$ counts the number of stabilizers which are linearly-independent, after restricting their support to the region $A$.  Using this expression, we show in Appendix \ref{app:entanglement_parton} that if the $A$ subsystem is semi-infinite with a perfectly ``flat" boundary that cuts across the $z$-type bonds of the honeycomb lattice, that the entanglement entropy of this region 
\begin{align}\label{eq:S_A_scaling}
    S_{A} = \frac{n_{p} + n_{\ell}}{2} - 1
\end{align}
where $n_{p}$ is the number of plaquettes that cross the entanglement bipartition, and $n_{\ell}$ counts the number of string stabilizers $W_{\rB\rB'}$ which contain only one of their endpoints in the $A$ subsystem (equivalently, the number of dimerized pairs of the $c$ Majorana partons which cross the entanglement cut).  Though the above expression is not exact when the boundary of the $A$ subsystem is not perfectly flat, we conjecture that for any sufficiently large subsystem, that $S_{A}$ and $n_{\ell}$ scale in the same way with subsystem size, up to area-law corrections.  

The above relation may be used to determine the entanglement properties of the monitored pure-states as the relative measurement rates $p_{x}$, $p_{y}$, $p_{z}$ are tuned.  First, using (\ref{eq:S_A_scaling}), we argue on general grounds that when $p_{x} = p_{y} = p_{z} = 1/3$, the steady-state must exhibit topological order or super-area-law scaling of the entanglement entropy.  Consider a particular monitored pure-state, which is specified by the collection of plaquette stabilizers $\{W_{p}\}$ and the string stabilizers $\{W_{\rB\rB'}\}$ between distinct pairs of lattice sites.  In the steady-state, let $Q(r)$ be the probability distribution over the ensemble of monitored pure-states for a string stabilizer to connect two sites a Cartesian distance $r$ apart from each other; we assume that when $r$ is much larger than the lattice spacing, this distribution is only a function of this Cartesian distance $r$ due to the statistical three-fold rotational symmetry enjoyed by the ensemble of monitored pure states when $p_{x} = p_{y} = p_{z}$.  If the string stabilizers are \textit{short-ranged} so that the typical Cartesian distance between their endpoints $\int dr \,r\,Q(r) \sim \mathrm{const.}$ in the thermodynamic limit, then  $Q(r)$ must decay faster than $r^{-2}$ at long distances.  In this case, we may define an \emph{orientation} for each string stabilizer $W_{\rB\rB'}$, by connecting the endpoints $\rB$, $\rB'$ by a product of bond operators along a short path. With this orientation, it is meaningful to define the number of string stabilizers that pass through a cut that wraps around a cycle of the torus.  As we demonstrate in the Appendix \ref{app:dimer_covering}, the \textit{number parity} of strings crossing such a cut is ($i$)  preserved by the dynamical rules evolving a dimer configuration and ($ii$) the same for all other vertical cuts on the lattice.  This number parity can be detected in the steady-state by an operator whose support extends along the cut and is localized in the direction transverse to the cut.  Therefore, a short-ranged ensemble of string stabilizers in the steady-state implies that the dynamics at the isotropic point can encode non-local information about the initial state of the system, similar to a topological quantum order.  On the other hand, if $Q(r) \sim r^{-\alpha}$ with $\alpha \le 2$ at long distances, the typical string stabilizers are long-ranged, and the dimer number parity across a vertical cut is ill-defined, even in a thermodynamically large system.  However in this case, the system will exhibit super-area-law scaling of the entanglement entropy; for example, given a disk of diameter $\ell$, the contribution to the von Neumann entropy from string stabilizers which extend a distance larger than $\ell$ is $S_{A} \sim \ell^{2}\int_{r\ge \ell}Q(r)dr \sim \ell^{3-\alpha}$.

\begin{figure}
 \includegraphics[width=.34\textwidth]{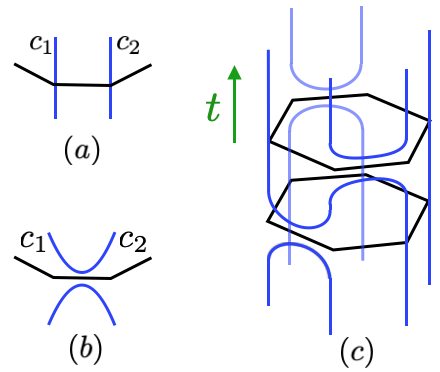}\\
     \caption{{\bf Loop Representation of the Stabilizer Evolution:} The spacetime evolution of the string stabilizers $W_{\rB\rB'}$ can be understood by keeping track of their endpoints. A projective measurement of a bond operator leads to a dynamical rule that the loops incident upon the vertices of bond are connected, and a new pair of loops is sourced at the bond, as shown in $(b)$. These endpoints evolve as in $(a)$ when no measurement is performed. The collective evolution of these endpoints can be interpreted as a loop model on a bipartite, three-dimensional lattice, as shown in $(c)$. }
     \label{fig:loop_model}
 \end{figure}
 
A universal understanding of the entanglement properties of the steady-state may be obtained by studying the dynamics of the Majorana partons as the relative measurement rates $p_{x}$, $p_{y}$, $p_{z}$ are tuned.  Given the initial state (\ref{eq:dynamics_rho}), the bond measurements give rise to a free-fermion evolution of the $c$ Majorana partons, as derived in Eq. (\ref{eq:free_fermion}). After projecting back into the spin Hilbert space, each pair of dimerized partons can be viewed as forming the endpoints of a string which has zero line tension, due to the fact that all of the plaquette operators $\{W_{p}\}$ belong to the stabilizer group (the $\mathbb{Z}_{2}$ fluxes are pinned and static within the steady-state of each realization of the monitored evolution involving bond measurements).  A spacetime representation of the evolving string operators $\{W_{\rB\rB'}\}$ may now be invoked, by keeping track of the evolving endpoints of each string stabilizer.  The spacetime evolution of these endpoints trace out loops in three dimensions, as shown in Fig. \ref{fig:loop_model}.  A projective measurement of a bond operator leads to the dynamical rule that the loops incident upon the vertices of the bond are connected and a new pair of loops are sourced at this bond, as shown in Fig. \ref{fig:loop_model}a and b.  We note that this representation has been previously used to understand the free-fermion evolution of Majorana fermions in two spatial dimensions under frequent measurements \cite{nahum2020entanglement}, where it was shown that the spacetime evolution of the endpoints of the paired Majorana fermions leads to a representation of the ensemble of evolving pure-states of the Majorana fermions as a fully-packed loop model on a bipartite lattice in three dimensions. 

A description of this loop model as a non-linear sigma model with target space CP$^{n-1}$ and in the replica limit $n\rightarrow 1$ was studied in Ref.  \cite{nahum20113d,nahum2013loop} and it is known that this model supports two phases: one in which the loops behave as a Brownian walk, so that the probability distribution of the Cartesian distance between endpoints of a loop, denoted $Q(r)$, scales as $Q(r)\sim r^{-2}$ \cite{nahum2013length}, and another phase in which the loops are ``short" and $Q(r)\sim \exp(-r/\xi)$.  Much is also known about the continuous phase transition separating these two phases \cite{ortuno2009random}. In conjunction with Eq. (\ref{eq:S_A_scaling}), we then conjecture that the monitored pure-states exhibit either ($i$) logarithmic violation of area-law scaling of the entanglement $S_{A}\sim L_{A}\log L_{A}$ or ($ii$) area-law scaling $S_{A} \sim L_{A}$, respectively.  We note that the scaling ($i$) is generally not possible in the presence of time-reversal and translation symmetry in the ground-state of the Kitaev honeycomb model\footnote{In the Kitaev honeycomb model, time-reversal symmetry can be chosen to act on the Majorana partons as $c_{\rB}\rightarrow \pm c_{\rB}$ where the sign depends on whether $\rB$ belongs to the $A$ or $B$ sublattice of the honeycomb lattice, respectively \cite{kitaev2006anyons}, which requires that the Majorana partons can only hop between different sublattices.  In the presence of translational symmetry, the Hamiltonian describing the Majorana partons in momentum space then takes the form $h(\kB) = f_{x}(\kB)\tau^{x} + f_{y}(\kB)\tau^{y}$ where the $\tau$ Pauli matrices act in sublattice space, so that the dispersion vanishes when both $f_{x}(\kB) = 0$ and $f_{y}(\kB) = 0$.  The solution to these equations generically gives a set of points in momentum space, as opposed to a gapless manifold of states, which would be required to obtain a logarithmic violation of area-law scaling of the entanglement. }. Finally, we note that when one of the probabilities is zero, e.g. $p_z=0$, the two dimensional system is effectively comprised of $L$ decoupled one-dimensional systems. The entanglement dynamics of each one-dimensional system can be mapped to a two-dimensional classical loop model (See Refs.\cite{nahum2020entanglement,PhysRevResearch.3.023200, Ali} and Appendix \ref{apx_perc}).

The area-law-entangled phase exhibits topological quantum order. 
The easiest way to see this is to observe that the probability of having a stabilizer which is given by the product of bond operators around a non-contractible cycle of the torus in the steady-state is exponentially small in the linear dimension of the system in the area-law-entangled phase. Starting with a maximally-mixed initial state and evolving the system in this phase, the time for this Wilson loop operator to be measured should then scale as $t_{\mathrm{purif}} \sim O(\exp(L))$.  The dynamics can then lead to a non-local encoding of information about the initial state which is exponentially long-lived.  

Other universal entanglement properties of the critical and topologically-ordered phases, which follow from the two phases of the loop-model representation of the stabilizer evolution, (e.g. measures of long-ranged  tripartite entanglement, the expectation values of Wilson loop operators, topological entanglement entropy) will be discussed in Sec. \ref{sec:numerics} alongside numerical simulations of these dynamics.


\section{Purification Dynamics}\label{sec:purification_dyn}

We now study the purification dynamics of the system, starting from a maximally-mixed initial state. As we shall see, different degrees of freedom in the system get disentangled from the environment with different rates. Thus we start by identifying the degrees of freedom which are relevant to the purification dynamics. Consider the following set of commuting operators: 1) Plaquette operators, 2) the set of bond operators of a specific type, (e.g. $z$-type bond operators) and 3) the two long-cycle stabilizers, which are given by the product of bond operators along the non-trivial cycles of the torus (one horizontal and one vertical). It is easy to see that there are $L^2-1$ independent plaquette operators and $L^2-1$ independent $z$-bond operators in this set, and thus by including the two long cycle stabilizers, we have a complete set of $N=2L^2$ commuting Pauli operators. Therefore, we may view the Hilbert space of the system as the tensor product of the Hilbert spaces of $N$ virtual qubits consisted of $L^2-1$ plaquette qubits, $L^2-1$ bond qubits, and $2$ long cycle qubits~\cite{zanardi2004quantum}.  Accordingly, the entanglement between the environment and the system  can be viewed to be consisted of three different parts: 1) entanglement with the plaquette qubits, 2) entanglement with the bond qubits and 3) entanglement with the long cycle qubits. 
Bellow we discuss how each part of the system gets disentangled from the reference system on the approach to the steady-state.

\textit{1. Plaquette stabilizers:} Due to their local nature, the plaquette stabilizers are  measured with constant relative rate for any point inside the phase diagram with $0<p_x,p_y,p_z$. Hence, the entanglement entropy of the system will drop exponentially until $O(\log(L))$ times when all plaquettes are measured. This is true in both the topologically-ordered  and the critical phase. 

\begin{figure}
    \includegraphics[width=.4\textwidth]{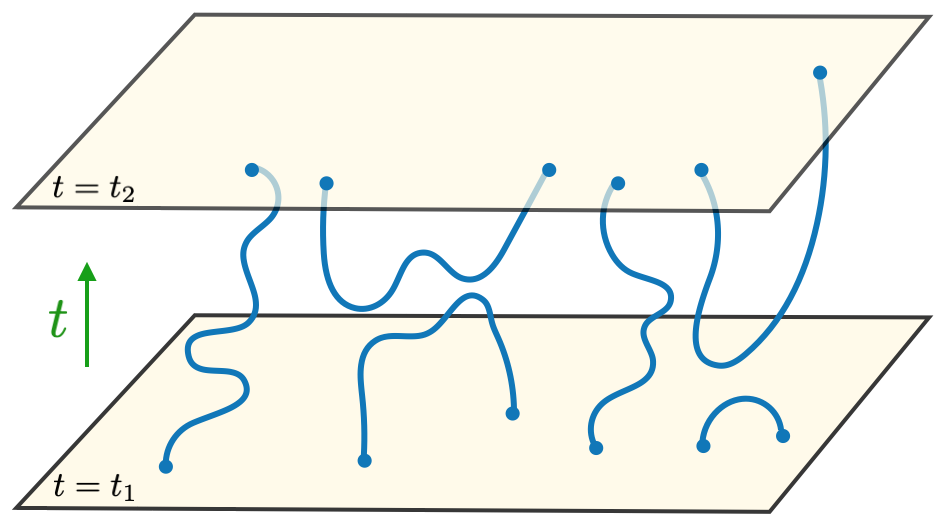}
    \caption{{\bf Spanning Number:} The purification dynamics of the bond operators may be understood by in the fully-packed loop model, by considering the number of strands of loops which connect an initial time $t_{1}$ to a final time $t_{2}$. This \emph{spanning number} decreases polynomially (exponentially) in $t_{2}-t_{1}$ in the critical (topologically-ordered) phases.}
    \label{fig:spanning_number}
\end{figure}

\textit{2. Bond Operators:} The purification dynamics of the bond qubits may be understood by starting with a density matrix for the Majorana partons
\begin{align}\label{eq:mixed_parton_state}
\rho_{f} \sim \ket{\Psi_{b}}\bra{\Psi_{b}}\otimes\mathds{1}
\end{align}
 where $\ket{\Psi_{b}}$ describes a dimerized pure-state of the $b$ Majorana fermions as described below Eq. (\ref{eq:dynamics_rho}).  The corresponding density matrix for the spin degrees of freedom is volume-law-entangled, and is stabilized by all of the plaquette operators.

\begin{figure}
    \includegraphics[width=.4\textwidth]{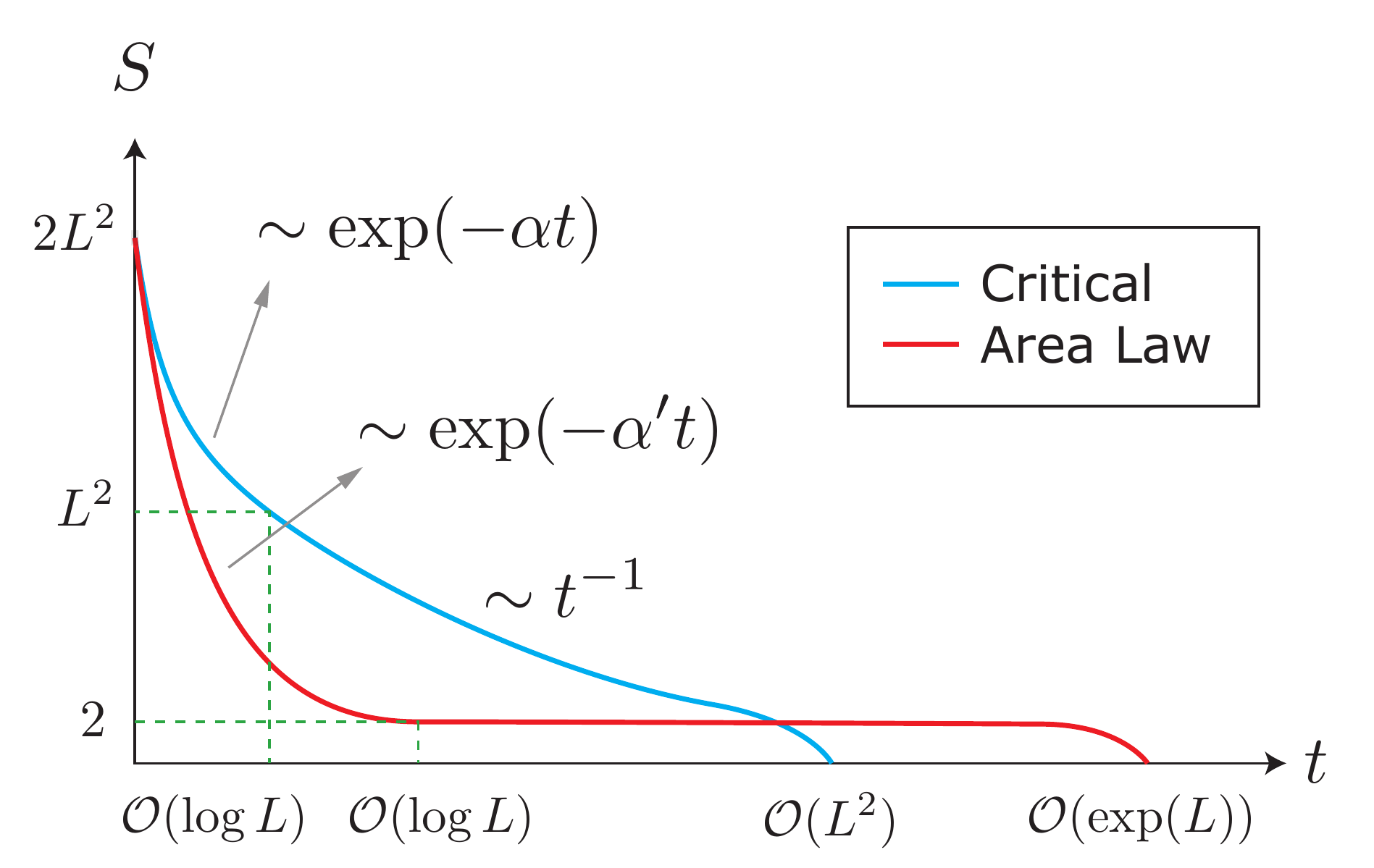}
       \caption{{\bf Purification Dynamics:} A summary of the evolution of the entropy of the system, starting from a maximally-mixed initial state in the critical phase (blue) and topologically-ordered phase (red).}
        \label{fig_schematic}
    \end{figure}
    
A subsequent measurement of a bond operator such as $X_{\rB}X_{\rB'}$ -- where $\rB$ and $\rB'$ are adjacent sites separated by an $x$-type bond -- is equivalent to adding $\pm i c_{\rB}c_{\rB'}$ as a stabilizer for the evolving state of the Majorana partons.  We refer to $c_{\rB}$ and $c_{\rB'}$ as ``paired" Majorana partons since they are now dimerized.  The ``unpaired" $c$ Majorana partons each provide an $O(1)$ contribution to the entropy of the evolving state; once a bond operator connecting two such unpaired partons is measured, these degrees of freedom become ``paired" and the entropy of the state decreases.  In Appendix \ref{app:Levy_Flight}, we show that the dynamics of these unpaired Majorana partons can be understood as a classical diffusion/annihilation reaction, in which the partons can take random steps whose lengths  are power-law distributed in the critical phase.  We show that this leads to a power-law decay of the entropy in the critical phase, as $S(t)\sim L^2/t$. As a result, it takes a polynomial time of $O(L^2)$ to disentangle the bond operators. On the other hand, the bond qubits will be disentangled exponentially fast in the topologically-ordered phase. 
 
 We note that an alternate understanding of the purification dynamics in the critical and topologically-ordered phases comes from the fully-packed loop model, which clarifies the universal nature of these dynamics in both phases.  We again start from the state (\ref{eq:mixed_parton_state}) of the parton degrees of freedom. This state may be viewed as the reduced density matrix of a pure-state in which each $c$ Majorana parton has been dimerized with a reference Majorana degree of freedom. As a result, we may view each unpaired $c$ Majorana parton as the endpoint of a loop which is attached to a reference Majorana.  As the measurements proceed, these loops evolve according to the rules presented in Sec. \ref{sec:steady_state_ee}, and as shown schematically in Fig. \ref{fig:spanning_number}.  In a time-interval $t$, we may follow the spacetime trajectory of a loop; if the loop does not return to the initial time-slice within this interval, then the corresponding unpaired Majorana parton remains unpaired after time $t$, and contributes to the entropy of the system.  Since each loop resembles a Brownian path in the critical phase, the motion of the loop in the time direction resembles a one-dimensional random walk.  The probability that a given loop does not return to the initial interface is then given by the probability that a random walk on the interval $(0,t)$ initialized near the origin reaches the point $t$ first, before reaching the origin.  This probability decays as $t^{-1}$ at long times \cite{krapivsky2010kinetic}.  From this reasoning, the entropy of the purifying state of the system may be identified with the \emph{spanning number} of the loop model in a system with dimensions $L\times L\times t$, which counts the number of strands of loops which connect the bottom and top layers which are separated by a distance $t$.  Since there are $O(L^{2})$ such strands (one for each unpaired Majorana parton) in the initial time, the entropy should decay as $L^{2}/t$ in the critical phase. In the topologically-ordered phase on the other hand, the spanning number decays exponentially in time, due to the exponentially small probability of having long loops \cite{nahum2014critical}.

\textit{3. Long cycle stabilizers:} In the topologically-ordered phase, the long cycle stabilizers will remain entangled with reference qubits until times which are exponentially large in system size, as discussed in Sec. \ref{sec:prop_of_std_state}, due to the fact that in the topologically-ordered phase, the probability of having a long loop is exponentially small. In the critical phase, this entanglement survives only up to polynomial times.

\begin{figure}
     \centering
     \begin{subfigure}[b]{0.49\columnwidth}
         \centering
         \includegraphics[width=\textwidth]{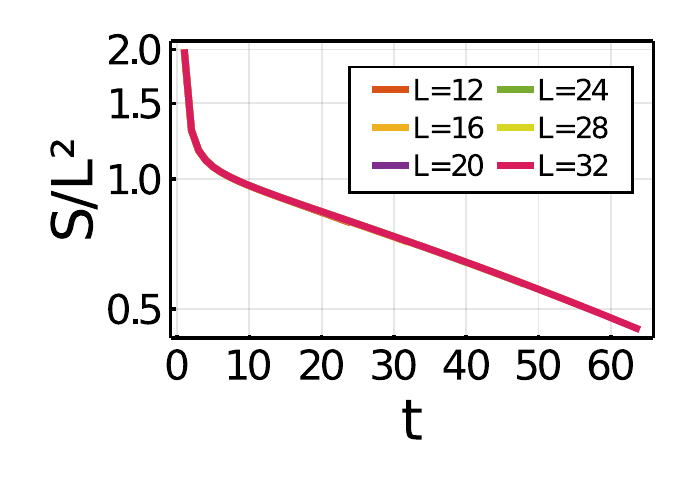}
         \caption{}
         \label{fig_St_totmix_iso}
     \end{subfigure}
     \hfill
     \begin{subfigure}[b]{0.49\columnwidth}
         \centering
         \includegraphics[width=\textwidth]{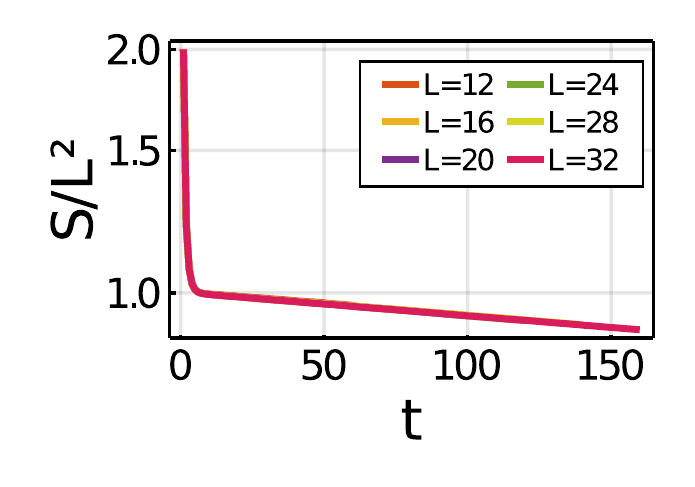}
         \caption{}
         \label{fig_St_totmix_px01}
     \end{subfigure}
        \caption{{\bf Purification of the Plaquette Stabilizers:} Entropy density as a function of time, starting from a totally mixed density matrix at (a) isotropic point $p_x=p_y=p_z=1/3$ and (b) highly biased measurements with $p_x=p_y=0.1$ and $p_z=0.8$. The exponential decays in both plots are related to plaquette stabilizers being measured at constant rate. }
\end{figure}

\begin{figure}
     \centering
     \begin{subfigure}[b]{\columnwidth}
         \centering
         \includegraphics[width=.8\textwidth]{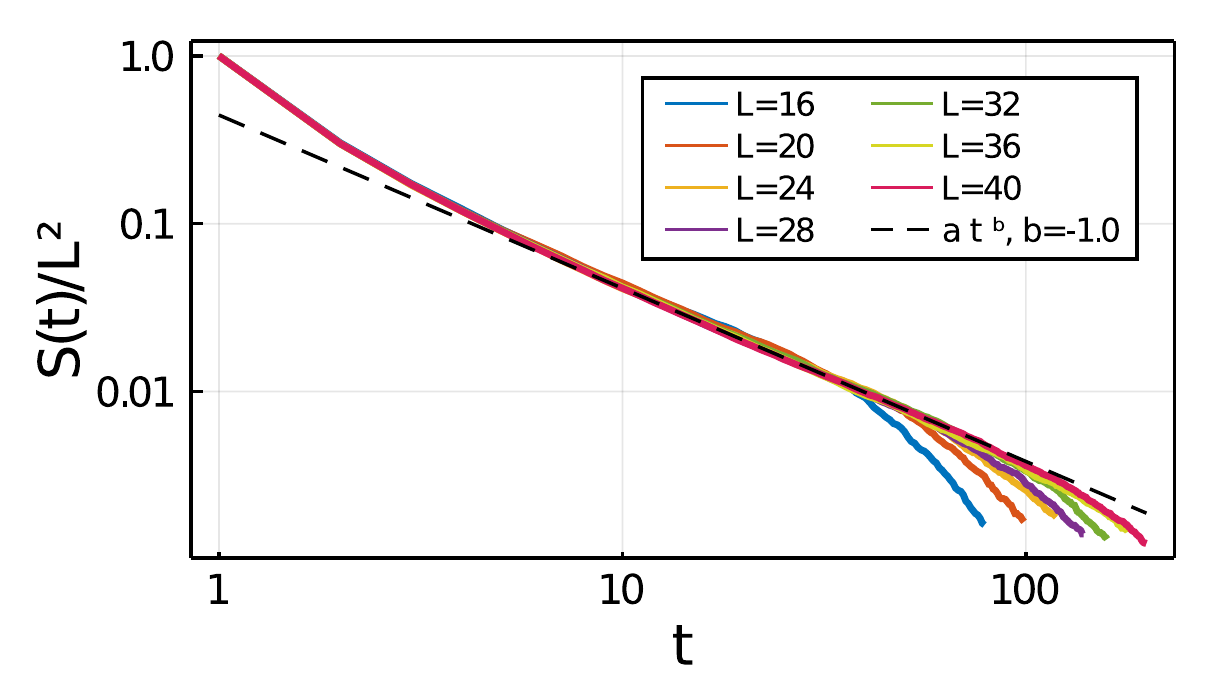}
          \caption{}
         \label{fig_St_prtmix_iso}
         
     \end{subfigure}

     \begin{subfigure}[b]{\columnwidth}
         \centering
         \includegraphics[width=.8\textwidth]{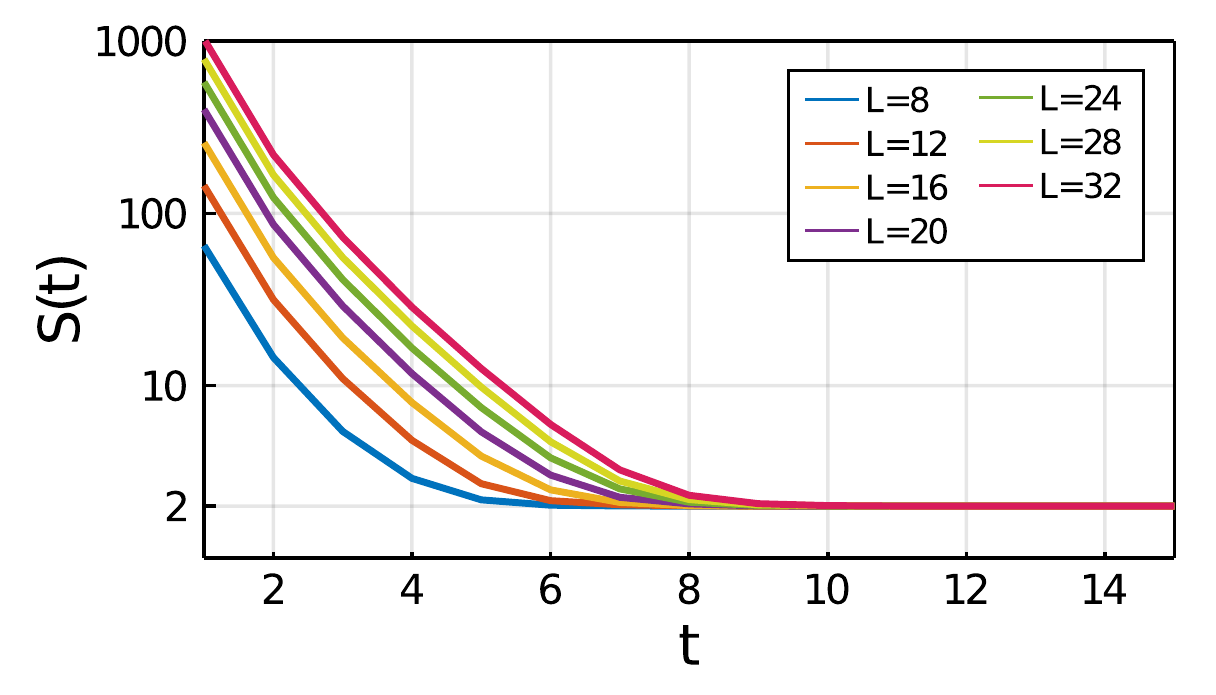}
         \caption{}
         \label{fig_St_parmix_px01}
     \end{subfigure}
        \caption{ {\bf Purification of the Bond Operators} a) Entanglement entropy density versus time at the isotropic point $p_x=p_y=p_z$ and (b) Entanglement entropy versus time at $p_x=p_y=0.1$ and $p_z=0.8$. The initial state is the projection onto the subspace where all plaquette operators have definite values, say $+1$. }

\end{figure}

The purification dynamics in both phases are summarized schematically in Fig.\ref{fig_schematic}.


\begin{figure}
     \centering
     \begin{subfigure}[b]{\columnwidth}
         \centering
         \includegraphics[width=0.84\textwidth]{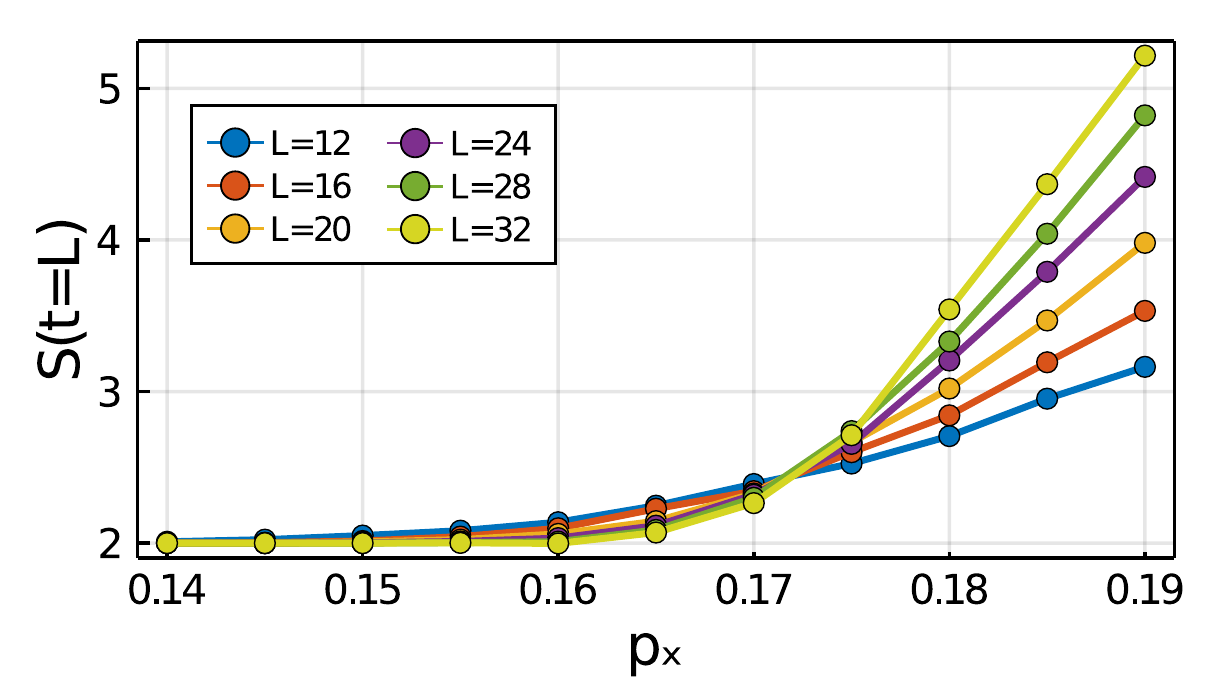}
         \caption{}
         \label{fig_SL-px}
     \end{subfigure}

     \begin{subfigure}[b]{\columnwidth}
         \centering
         \includegraphics[width=0.84\textwidth]{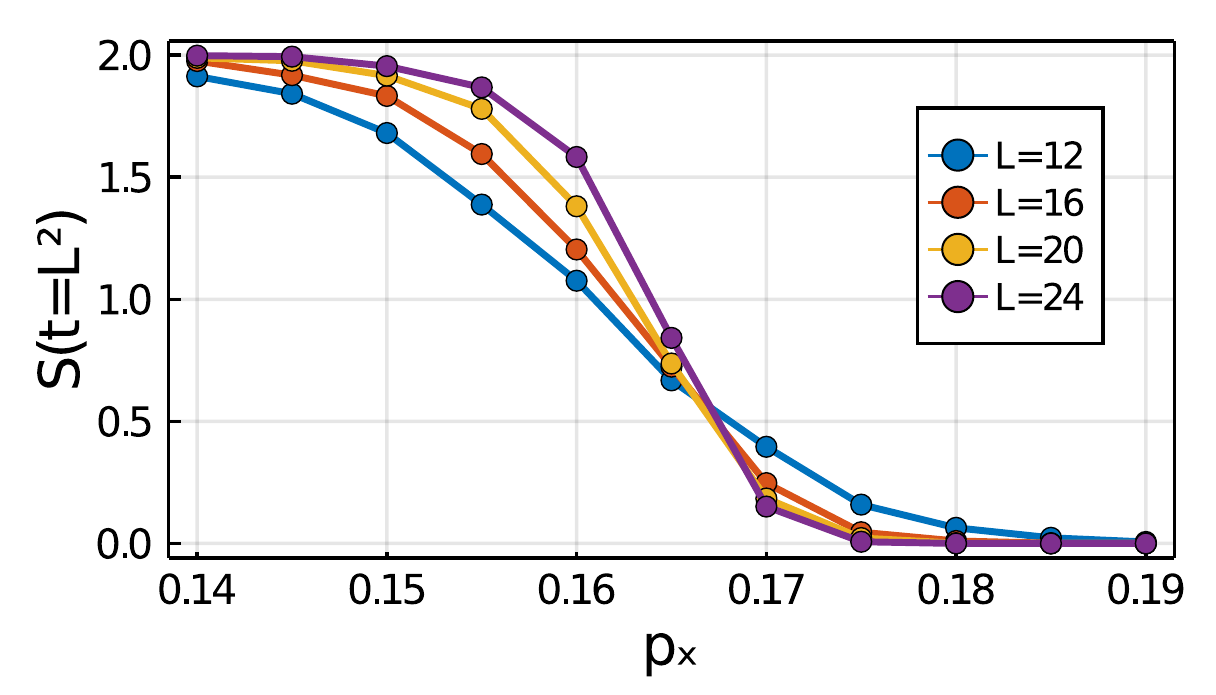}
         \caption{}
         \label{fig_SL2-px}
     \end{subfigure}
        \caption{Entropy of the purifying state of the system $S(t)$ for different values of $p_x$ on the  line $p_x=p_y$ at time (a) $t=L$ and (b) $t=L^2$. }
\end{figure}

\section{Numerical results}\label{sec:numerics}
In this section, we numerically study the measurement-only dynamics considered in the previous section
The measurement-only dynamics here are studied in $L\times L$ systems with periodic boundary conditions, with $L \lesssim 40$ sites.  Large-scale simulations of the monitored pure-states are made possible by the Clifford nature \cite{gottesman1997stabilizer} of these dynamics, which permit an efficient storage of the evolving state of the spin degrees of freedom.  

\subsection{Purification Dynamics}
We first numerically study the purification dynamics of the spins, starting from a maximally-mixed initial state.  We observe that the plaquette stabilizers are added in a time $O(\log L)$ in both the topologically-ordered and critical phases, as shown in Fig. \ref{fig_St_totmix_px01} and \ref{fig_St_totmix_iso}, respectively.  We note that starting with a maximally-mixed initial state, the rate at which plaquette stabilizers are {being measured is} very small; a plaquette stabilizer is added to the stabilizer group after the bond measurements are performed in a particular sequence around a given plaquette. As a result, the pre-factor appearing in the logarithmically-large timescale is quite large. 
For this reason, to study the purification dynamics of the bond operators in the topologically-ordered and critical phases, we initialize the system in a state in which all of the plaquette stabilizers belong to the stabilizer group.  The system still contains a finite entropy density.  The reduction of the entropy of the system follows a power-law in time (Fig. \ref{fig_St_prtmix_iso}) in the critical phase. In contrast, the entropy of the system decreases exponentially in time in the topologically-ordered phase before saturating at $2$ (Fig. \ref{fig_St_parmix_px01}) for an exponentially long time before the system completely purifies.  We can use the purifying dynamics as an order parameter to distinguish the topologically-ordered phase from the critical phase \cite{gullans2020scalable,gullans2020dynamical}. To this end, we may look at $S(t)$ for $t=O(L)$, which is $ 2$ in the topologically-ordered phase while it is $O(L)$ in the critical phase(see Fig.\ref{fig_SL-px}). Otherwise, we can look at $S(t)$ at $T=O(L^2)$ which is $0$ in the critical phase and $ 2$ in the topologically-ordered phase (see Fig.\ref{fig_SL2-px}).

\subsection{Scaling of subsystem entanglement entropy}
We now study properties of the steady-state of the monitored dynamics.  To study the properties of the steady-state, we measure all stabilizers (plaquettes and long cycle stabilizers) and all $z$-type bond operators. Then we run the circuit for $O(L)$ time step. We check the time dependence of the averaged quantities of interest to make sure they are saturated by this time.

\begin{figure}
     \centering
     \includegraphics[width=\columnwidth]{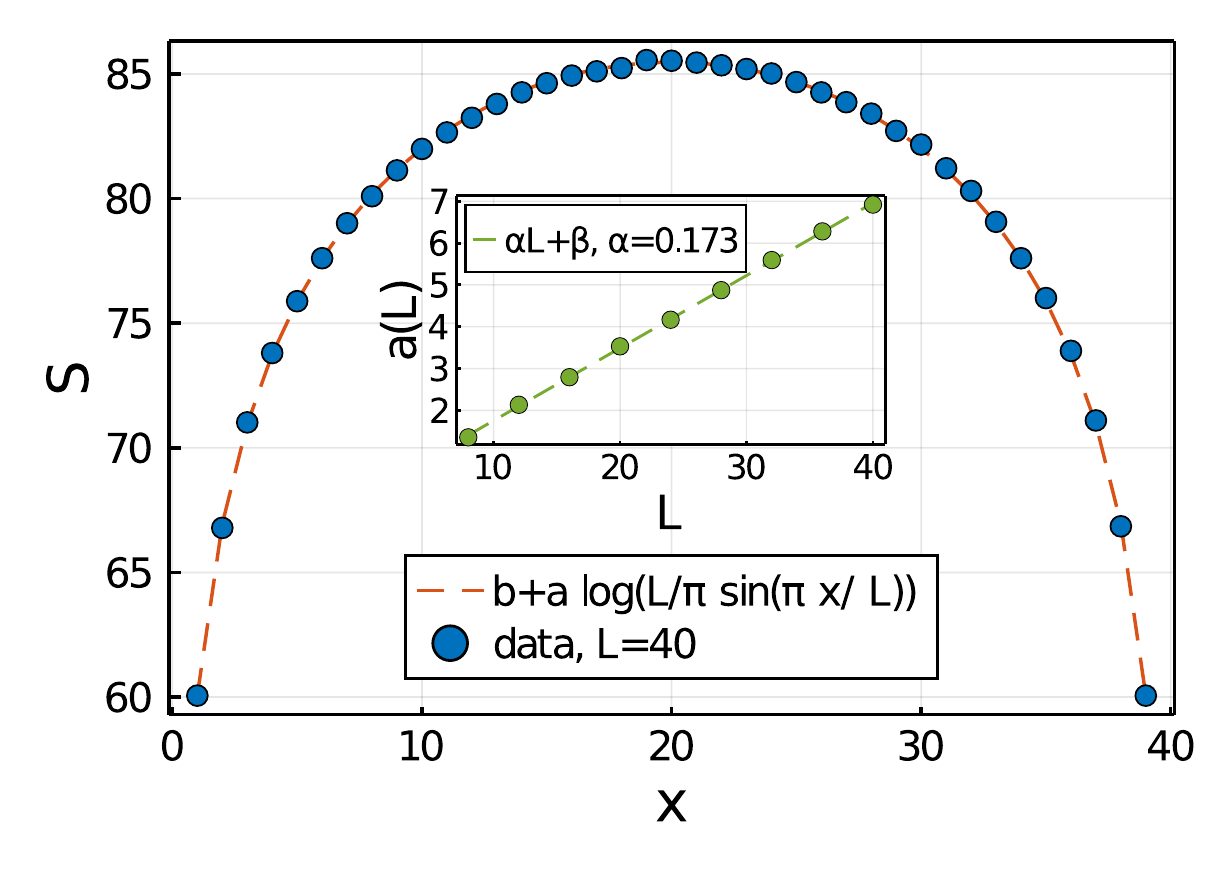}
     \caption{{\bf Entanglement in the Critical Phase.} Entanglement entropy of a cylindrical region of size $x\times L$ as a function of $x$ for the steady state of the circuit at the point $p_{x}=p_{y}=p_{z}=1/3$. The dashed line shows the best fit to the functional form in Eq.  (\ref{eq:ee_log_violation}). The inset shows the best-fit value of the parameter $a(L)$ for different system sizes, which scales  linearly in $L$.} 
     \label{fig_Sr_iso}
\end{figure}

At the critical phase, we expect the string operators to have the length distribution of $~1/l^2$ based on the 3D loop model. This in turns implies that the entanglement entropy of a subregion of linear size $R$ should diverge as $R\log R$, for $R \ll L$. Fig. \ref{fig_Sr_iso} shows the entanglement entropy of a cylindrical region as a function of its length $x$, at a fixed system size $L=40$, at the isotropic point $p_x=p_y=p_z=1/3$ after $t=L$ time steps. Fig. \ref{fig_S_half_t} in Appendix \ref{apx_sup_figs} shows the entanglement entropy of the cylinder of size $L/2$ as a function of time, to make sure the value we are reading is already saturated to its steady state value. As is clear from the figure, the entanglement entropy fits very well to the following,
\begin{equation}\label{eq:ee_log_violation}
S(x)=b(L)+a(L) \log \left[\frac{L}{\pi}\sin\left(\frac{\pi x}{L}\right)\right],
\end{equation}
where $a(L)$ and $b(L)$ have been used as fitting parameters. Moreover, by changing the system size we find that the best fit parameter $a(L)$ scales linearly with system size as is shown in the inset of Fig.\ref{fig_Sr_iso}. We find the factor of proportionality at the isotropic point to be $a(L) \approx 0.173 L$ . Note that if we are at one of the percolation points,  
e.g. $(p_x,p_y,p_z)=(1/2,1/2,0)$, the system decouples into $L$ 1d critical chains, resulting in the same $L\log L$ violation but with a prefactor $\alpha=\frac{\ln(2)\sqrt{3}}{2\pi}\approx 0.191$.

In general $\alpha$ seems to change inside the critical region. Moreover, it seems that it depends on the direction of the cylinderical region. Fig. \ref{fig_SRZ_px02} and Fig. \ref{fig_SRX_px02} in Appendix \ref{apx_sup_figs} show the entanglement entropy of a cylindrical regions at $(p_x,p_y,p_z)=(0.2,0.2,0.6)$ which is still in the critical phase. The data in Fig. \ref{fig_SRZ_px02} is for a cylindrical region where the boundary cuts through $z$ bonds while  the data in Fig. \ref{fig_SRX_px02} is for a cylindrical region where the boundary cuts through $x$ bonds. As one can see, not only the value of $\alpha $ is different from that of the isotropic point, but its value also depends significantly on the configuration of the region, which shows that the steady state does not have rotational symmetry. 

In the topologically-ordered phase, the entanglement entropy exhibits area-law-scaling as is shown in Fig. \ref{fig_SRZ_px01} in Appendix \ref{apx_sup_figs} for  $(p_x,p_y,p_z)=(0.1,0.1,0.8)$.

\subsection{Mutual Information}\label{sec_mutual_info}
\begin{figure}
     \centering
     \includegraphics[width=0.6\columnwidth]{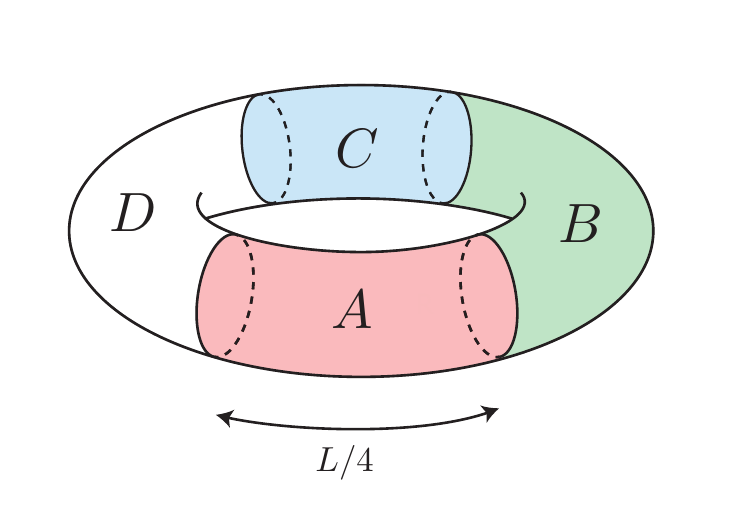}
     \caption{The non-contractible regions $A$, $B$ and $C$ which are used for the calcuation of the bipartite mutual information $I_2(A:C)$ and the tripartite mutual information $I_3(A:B:C)$ in Sec. \ref{sec_mutual_info}.  }
     \label{fig_ABC}
\end{figure}

To detect the phase transition between the critical phase and the area law phase, we may look at how the information is shared between distant parts of the system. To this end, consider slicing the torus into four cylinders with equal length of $L/4$ as is shown in Fig. \ref{fig_ABC}. A natural diagnostic for the phase transition between the topologically-ordered and critical phases is the mutual information between $A$ and $C$ \cite{li2019measurement}:
\begin{equation}
    I_2(A:C)=S_A+S_C-S_{AC},
\end{equation}
as well as the tripartite mutual information between $A$, $B$ and $C$ \cite{zabalo2020critical}, which is defined as:
\begin{align}
    &I_3(A:B:C)= I_2(A:B)+I_2(A:C)-I_2(A:BC)\label{eq_I3} \\
                &= S_A+S_B+S_C-S_{AB}-S_{BC}-S_{AC}+S_{ABC}.
\end{align}
$I_2(A:C)$ measures the  correlations between the information in $A$ and $C$, while $I_3(A:B:C)$, when negative, is indicative of information shared between three regions that can only be inferred by having access to all three regions. 

\begin{figure}
     \centering
         \centering
         \includegraphics[width=0.5\columnwidth]{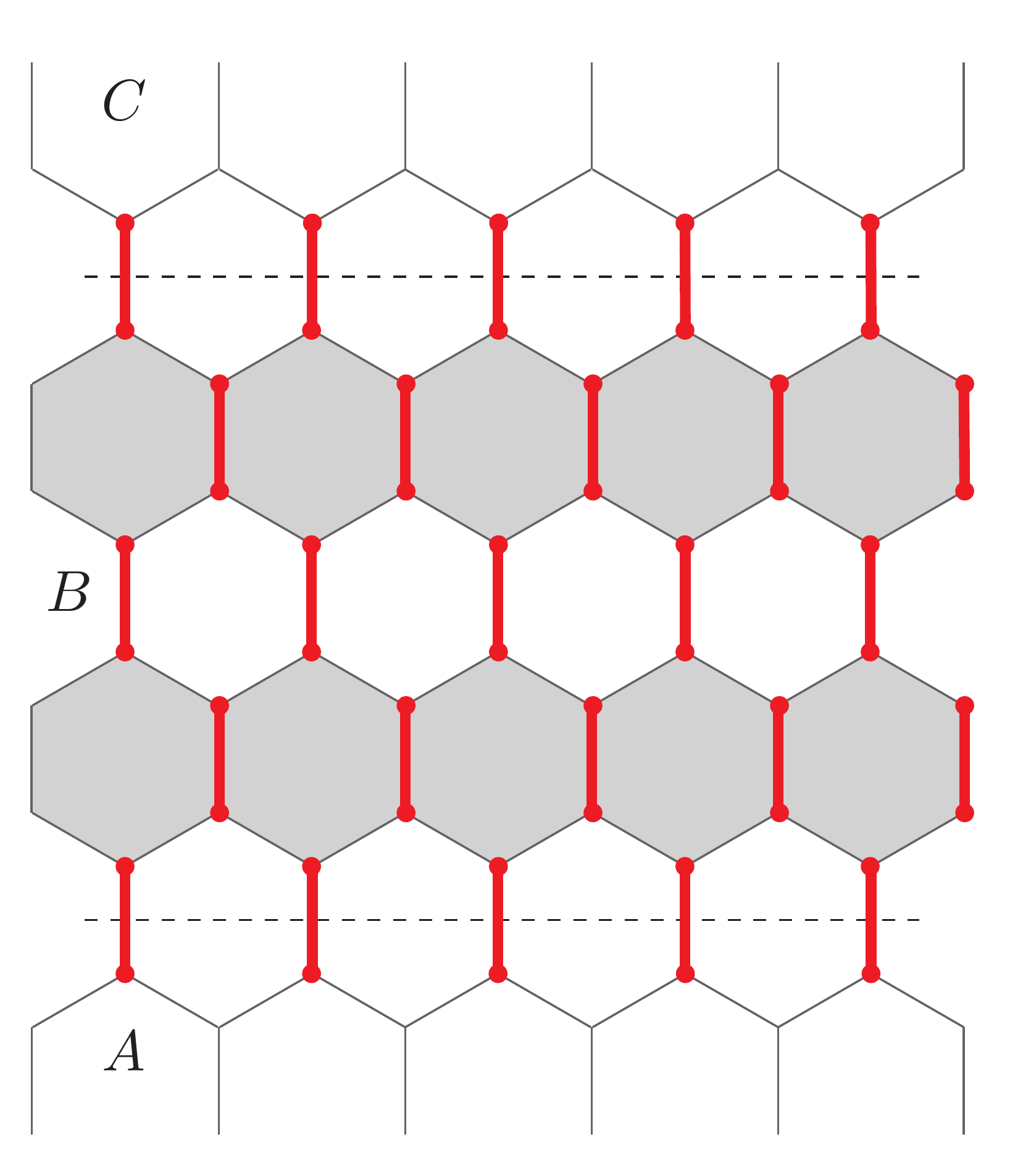}
        \caption{The product of the red bond-operators and the shaded plaquette operators act trivially in $B$ and only has support in $A$ and $C$. This is the operator which results in $I(A:C)=1$ in the limit that $p_x=p_y=0$.}\label{fig_MIga}
\end{figure}

\begin{figure}
     \centering
     \begin{subfigure}[b]{\columnwidth}
         \centering
         \includegraphics[width=.9\textwidth]{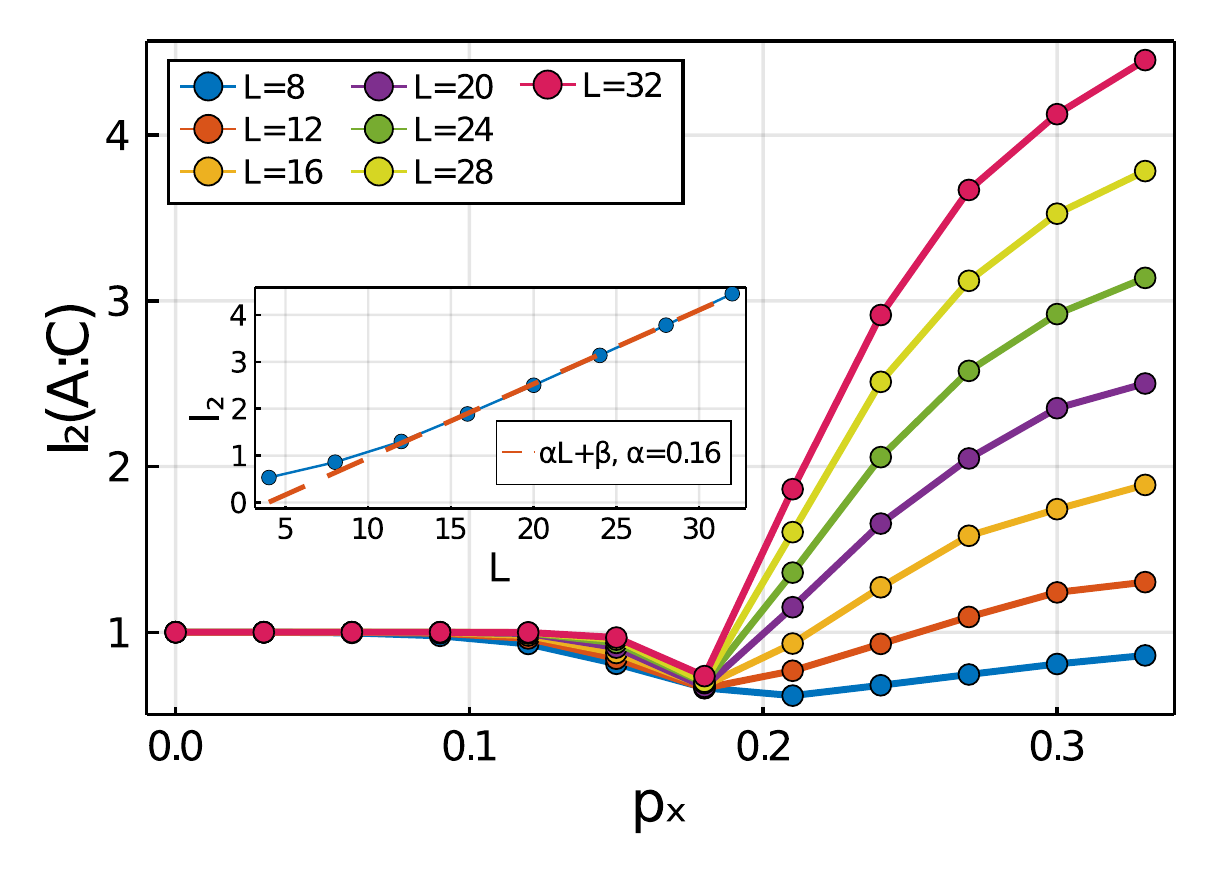}
         \caption{}
         \label{fig_I2}
     \end{subfigure}
     
    \begin{subfigure}[b]{\columnwidth}
         \centering
         \includegraphics[width=.9\textwidth]{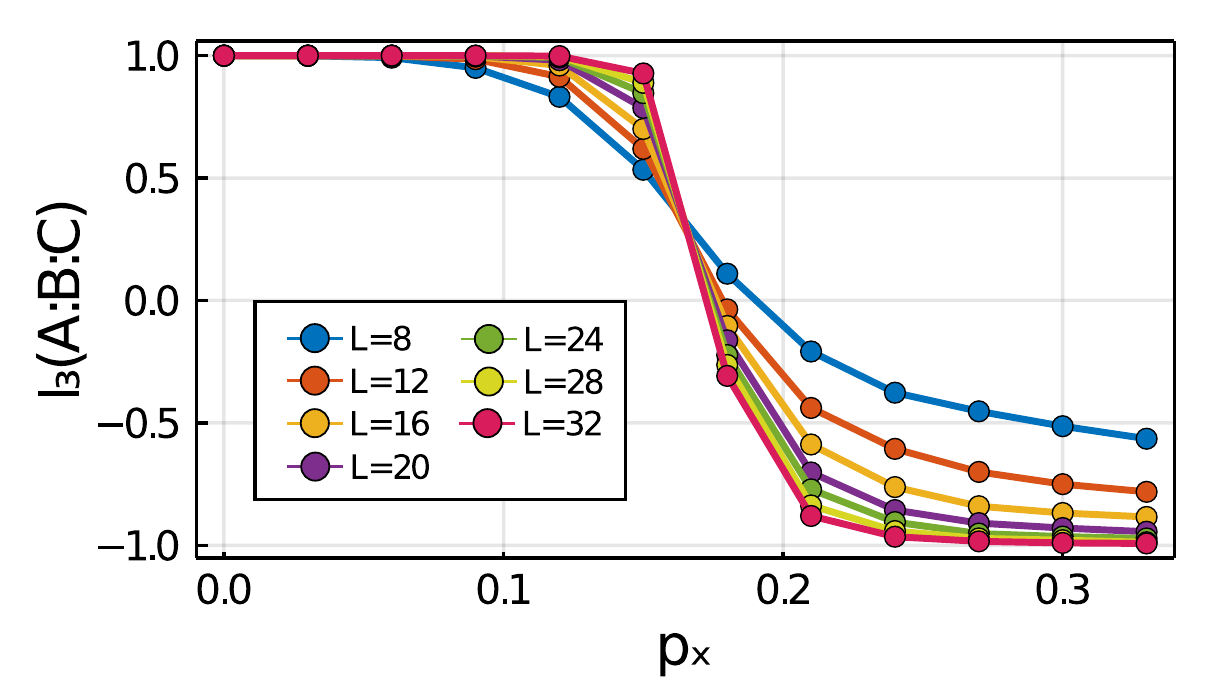}
         \caption{}
         \label{fig_I3}
     \end{subfigure}
        \caption{Steady state mutual information between cylindrical regions (shown in Fig.\ref{fig_ABC}) versus $p_x$ for points on the symmetric line $p_x=p_y$: (a) bipartite mutual information $I_2(A:C)$. The inset shows $I_2(A:C)$ at $p_x=p_y=1/3$ versus system size $L$ (b) Tripartite mutual information $I_3(A:B:C)$.}
\end{figure}

Fig. \ref{fig_I2} shows $I_2(A:C)$ as a function of $p_x$ on the symmetric line $p_x=p_y$.  As is clear form Fig. \ref{fig_I2}, $I_2(A:C)=1$ throughout the area law phase. It is straightforward to understand this result in the limit that $p_x=p_y= 0$. In this limit, $p_z=1$ hence all $z$ bond operators are in the stabilizer group of the state. Now consider the operator which is the product of all $z$ bond operators with a non-trivial support in $B$ (shown as thick red lines in Fig.\ref{fig_MIga}) multiplied by plaquette operators in every other row of $B$ (shaded  plaquettes in Fig.\ref{fig_MIga}). This operator acts trivially in $B$ and only has non-trivial support in $A$ and $C$, while it can not be expressed as a product of stabilizers which are localized in either $A$ or $C$. Hence it results in a unit mutual information between $A$ and $C$. In Appendix \ref{apx_mutual_info} we show that a similar operator exists even when $p_x$ and $p_y$ are finite.  

On the other hand, in the critical phase, long range string operators spanning between $A$ and $C$ cause $I_2(A:C)$ to grow with the system size. Note that product of two string operators that span between $A$ and $C$ can be deformed to a stabilizer that acts non-trivially only on $A$ and $C$ and hence contributes to $I_2(A:C)$. The inset of Fig. \ref{fig_I2} plots $I_2(A:C)$ as a function of system size $L$ at the isotropic point $p_x=p_y=p_z$. It is clear from the inset plot that $I_2(A:C)$ scales linearly with $L$ for large enough system sizes. This can be easily understood by computing the number of string operators going from $A$ to $C$, using their $1/r^3$ length distribution:
\begin{align}
   &I_2(A:C)\propto \nonumber \\
    &\int_{-L/2}^{L/2}\dd y'\int_{0}^{L/4}\dd x'\int_{-L/2}^{L/2}\dd y\int_{x'+L/4}^{x'+L/2} \frac{\dd x}{(x^2+y^2)^{3/2}}.
\end{align}
As can be seen from dimensional analysis, this integral is proportional to  $L$ which explains the linear scaling of the mutual information shown in the inset of Fig. \ref{fig_I2}.

Fig. \ref{fig_I3} shows the tripartite mutual information $I_3(A:B:C)$ as a function of $p_x$ on the symmetric line $p_x=p_y$. 
In the topologically-ordered phase, since the string operators are short ranged, the first and last terms in Eq. \eqref{eq_I3} cancel out and we find $I_3(A:B:C)=I_2(A:C)=+1$. 
However, deep in the critical phase, $I_2(A:BC)$ picks up a contribution from one string operator which spans the whole $ABC$ interval. The contribution from all the other strings cancels out with $I_2(A:C)$ and $I_2(A:B)$ and hence $I_3(A:B:C)$ ends up being $-1$.  Note that for two string operators that span the whole $ABC$, their product can be localized on $A$ and $C$ only, such that their support on $B$ cancel out. Therefore, only one string operator contributes to irreducible tripartite correlations while the correlations due to the rest of the strings can be expressed in terms of bipartite correlations. 

We contrast the behavior of $I_{3}(A:B:C)$ considered above with the topological entanglement entropy (TEE), which is equivalent to the tripartite mutual information between three contractible regions $A$, $B$, $C$ arranged in a particular geometry (see \cite{kitaev2006topological,levin2006detecting}), in which each region shares an edge with both of the remaining regions.  We may also numerically investigate the behavior of the TEE of the steady state.  We find that the TEE is always equal to $1$ regardless of which phase we are in (see Fig.\ref{fig_topoee_p} in Appendix \ref{apx_sup_figs}), due to the fact that any monitored trajectory contains a frozen configuration of the $\mathbb{Z}_{2}$ fluxes and is thus stabilized by each of the  plaquette operators.  
We note that the ground state of the Kitaev honeycomb model has also TEE equal to $1$ in both the gapped phase and the gapless phase.

\subsection{Wilson Line Correlators}\label{sec_wilson}

Let $W_{\boldsymbol{r}\boldsymbol{r}'}$ denote the string operator which is composed of product of bond operators along a path which starts at site $\boldsymbol{r}$ and ends at site $\boldsymbol{r}'$. We are interested in evaluating $\overline{\langle W_{\boldsymbol{r}\boldsymbol{r}'}\rangle^{2}}$ in the steady state, where the line indicates an average over  trajectories of the monitored dynamics using Born's rule.  We note that $\overline{\langle W_{\boldsymbol{r}\boldsymbol{r}'}\rangle}$ is always short-ranged due to the fact that each monitored pure-state hosts a random background of pinned $\mathbb{Z}_{2}$ fluxes. Since the monitored pure-state of the spins is always a stabilizer wavefunction, the quantity $\langle W_{\boldsymbol{r}\boldsymbol{r}'}\rangle^{2}$ is either $+1$ or $0$ within any monitored trajectory, and $\overline{\langle W_{\boldsymbol{r}\boldsymbol{r}'}\rangle^{2}}$ is proportional to the probability that the string operator connecting $\rB$ and $\rB'$ belongs to the stabilizer group of the steady state.

We study 
\begin{align}
    g(\rB - \rB')\equiv \overline{\langle W_{\rB\rB'}\rangle^{2}}.
\end{align}

\begin{figure}
         \includegraphics[width=.4\textwidth]{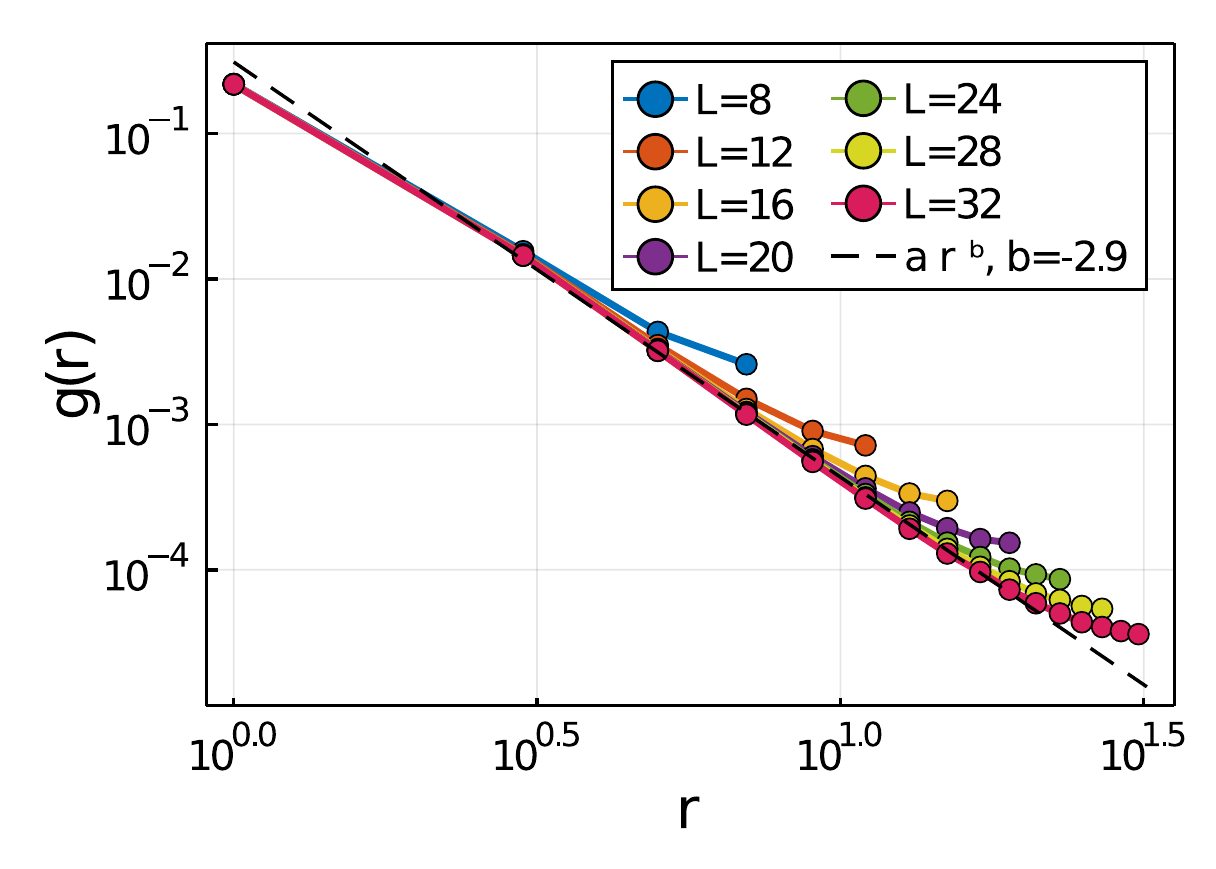}
    \caption{The expectation value $g(r)\equiv \overline{\expval{W_{\rB\rB'}}^{2}}$ as a function of $r \equiv |\rB - \rB'|$ at the isotropic point $p_x=p_y=p_z$. The dashed line is the best power law fit for system size $L=32$, which is consistent with analytic predictions.}\label{fig_gr}
\end{figure}

Due to the particular nature of the circuit model, $g(\rB - \rB')=0$ for any $|\rB - \rB'|=2k$ with $k\in \mathbb{Z}$ at any time. At the isotropic point $p_{x}=p_{y}=p_{z} = 1/3$, these correlations only depend on the graph distance $|\rB - \rB'|$. Fig. \ref{fig_gr} shows $g(r)$ for $r=2k+1$ at this point. We observe that this quantity falls as a power law $\sim r^{-\Delta}$, with an exponent $\Delta\approx 2.9$ until $r\sim L/2$ where it flattens. From the loop model picture, we expect it to be $\Delta=3$. 

\subsection{Phase Transition}
 If the point at $p_x \hat{x}+p_y \hat{y}+ p_z \hat{z}$ corresponds to the circuit parameters $(p_x,p_y,p_z)$, the phase diagram consists of the points inside a equilateral triangle whose three vertices lie at $\hat{x}$,$\hat{y}$ and $\hat{z}$ (see Fig.\ref{fig_phasediagram}). Based on the symmetries of the circuit model, the phase diagram should be symmetric under rotations around the center of the triangle by $\theta=2\pi/3$ as well as reflections about the perpendicular bisector of each sides. We may use this symmetries to simplify mapping out the phase diagram. 

To map out the phase diagram and determine the critical exponents associated with the corresponding phase transition, we make use of the tripartite mutual information $I_3(A:B:C)$ defined above, using the same partitioning of the torus as in Fig. \ref{fig_ABC}. 

\begin{figure}
     \centering
     \includegraphics[width=\columnwidth]{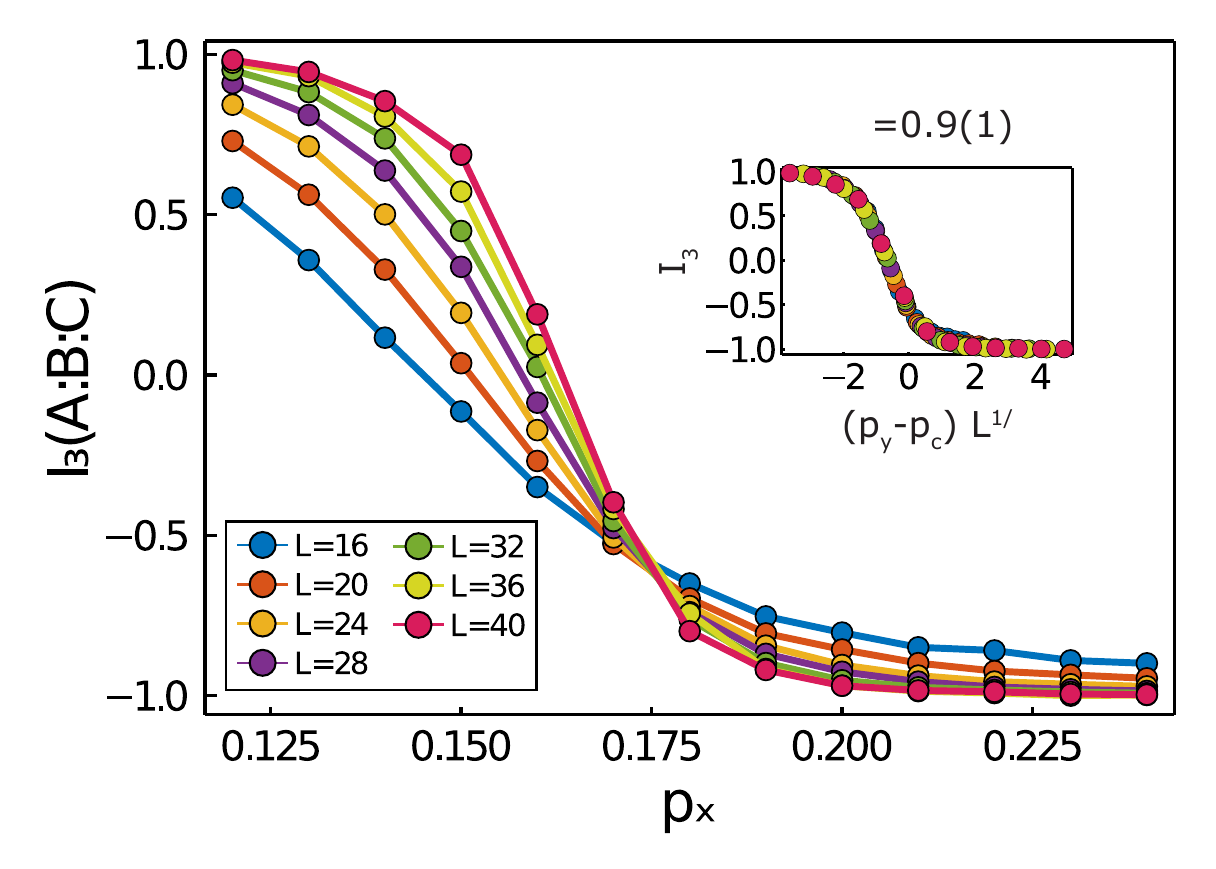}
     \caption{The tripartite mutual information of the steady state versus $p_x$ on the symmetric line $p_x=p_y$, near the phase transition at $p_c=0.172(5)$. The inset shows the data collapse of the same data using the scaling form of Eq.\eqref{eq_scalingform} with $\nu=0.9(1)$.}
     \label{fig_I3_collapse}
\end{figure}

Fig. \ref{fig_I3_collapse} shows $I_3(A:B:C)$ as a function of $p_y=p_x$ in the vicinity of the phase transition. 
We choose sections $A$, $B$ and $C$ such that their boundary cut through the $z$-bonds. On general grounds, we may assume that $I_3$ follows the scaling form 
\begin{equation}\label{eq_scalingform}
    I_3(p,L)=F((p-p_c) L^{1/\nu}),
\end{equation}
where $\nu$ is the correlation length critical exponent. By collapsing the data according to this scaling form, we find $p_c=0.172(5)$ and $\nu=0.9(1)$. 

By fixing $p_x$ and varying $p_y$ we can map out the whole phase diagram. Fig. \ref{fig_phasediagram} shows the result. The black dots are found via numerical simulation (see Fig.\ref{fig_I3_moreplots} in the Appendix for the corresponding plots) while the white dots are just the symmetric counter parts of the black dots. The black squares  on the sides correspond to 2D percolation fixed points. The three green regions are area law phases while the middle phase $C$ correspond to the critical phase with $L \log L$ violation of entanglement entropy. Based on the numerical results it seems that the phase boundary corresponds to the incircle of the triangular phase diagram although this needs further investigation. Along the phase boundary the numerical estimate for the value of $\nu$ changes between $0.7$ to $0.9$ but the variation is inside the margin of error. On the other hand, for the 2D percolation fixed points on the boundary, we have $\nu=4/3$. 


\section{Perturbations to the Critical and Topologically-Ordered Phases}\label{sec_perturbation}
In this section, we consider perturbing away from the limit in which only bond  operator measurements are performed in the monitored evolution, by adding in other kinds of measurements which have the effect of ($i$) removing the extensive number of conserved quantities in the dynamics considered previously and ($ii$) preventing a free-fermion description of the effective dynamics of the Majorana partons.  We find that the critical and topologically-ordered phases remain stable.

In particular, we consider two different types of perturbations to the circuit dynamics. First we study the effect of adding random single qubit measurements. In particular, we consider a circuit model where at each step a random qubit is measured in the $Z$ basis with probability $p_s$ or a bond operator is measured at random with probability $(1-p_s)$ such that $p_s+(1-p_{s})(p_x+p_y+p_z)=1$. As we will argue in the following, we expect both phases to be stable for small enough values of $p_s$. To see this, it is helpful to consider the effect of this perturbation on a slightly modified circuit model first. Assume that we measure all the plaquette operators after each measurement of either bond operators or of single qubit operators. Note that when $p_s=0$, measuring plaquette operators has no effect, since in the original model all of plaquette operators already belong to the stabilizer group of the steady state.

\begin{figure}
     \centering
     \includegraphics[width=0.5\columnwidth]{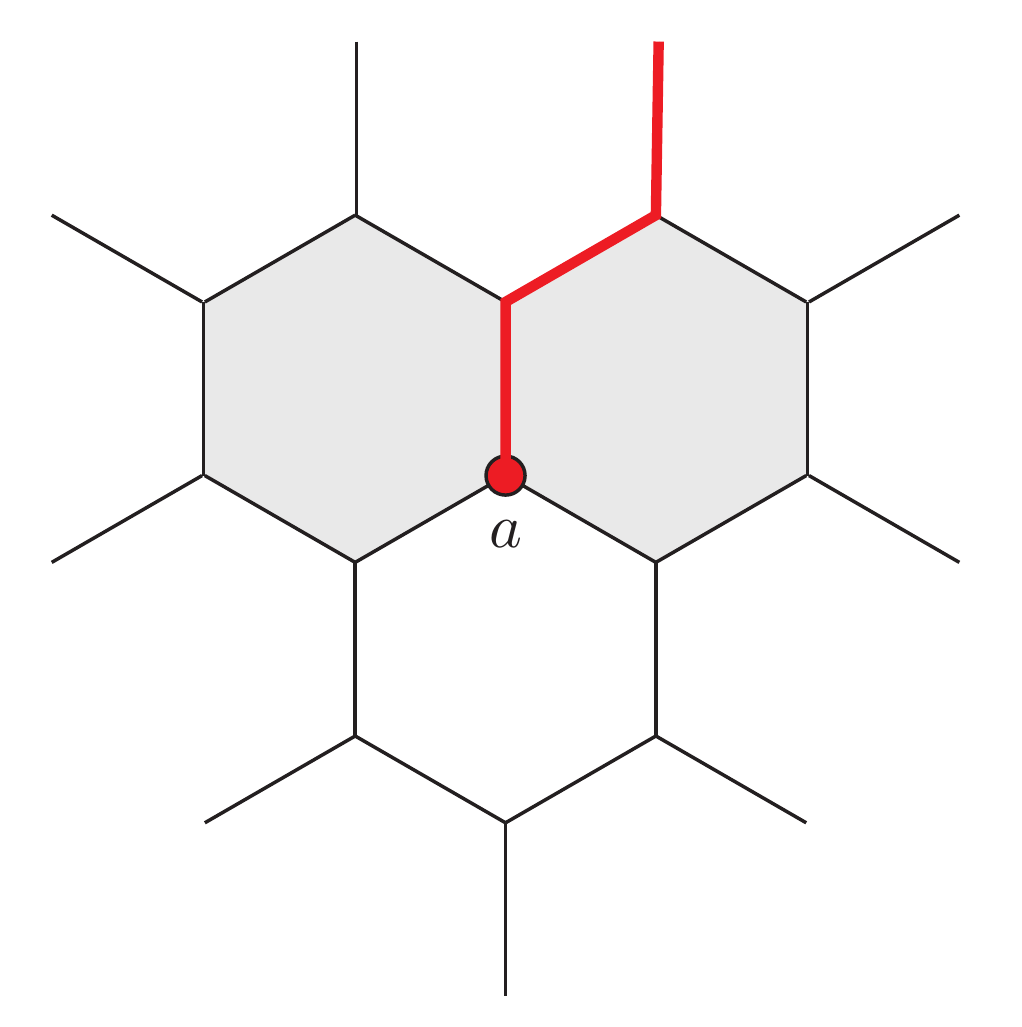}
     \caption{We may choose all string operators such that all commute with a given single qubit measurement. For example, in the shown example, if we choose the string operator that ends on site $a$ to come through the vertical $z$-bond, it will commute with the $Z_a$ measurement.}
     \label{fig_z_perturb_1}
\end{figure}

Now let us see how the stabilizer group changes when a single qubit measurement is performed. Imagine the qubit denoted by the red dot in Fig. \ref{fig_z_perturb_1} is measured in the $Z$ basis. First we note that we can choose all of the string stabilizers such that they all commute with the $Z$ measurement on the red  qubit. To this end, we may choose the string operator which ends on the red qubit to pass through the $z$-bond emanating from the red dot, and we may choose the other string operators such that none have support on the red  qubit. With this choice of generators for the stabilizer group, it is clear that after the single qubit measurement, the only change in the stabilizer group would be, ($i$) the single-qubit $Z$ operator is added to the stabilizer group, and ($ii$) the two shaded plaquette operators in Fig. \ref{fig_z_perturb_1} are replaced by their product. As such, single qubit measurements pin the endpoints of string operators ending at these sites in particular directions, which should not change the long-distance properties of the length distribution $Q(r)$ of stabilizer endpoints.  It is also clear that a subsequent measurement of all plaquette operators, will undo the effect of the $Z$ measurement. so for any $p_s$ we find the exact same phase diagram. Now let us relax the assumption of measuring all plaquette operators at each timestep of the circuit. Instead we assume that plaquette operators are measured randomly with probability $p_\text{plq}$, such that $p_s+p_\text{plq}+(1-p_s-p_{plq})(p_x+p_y+p_z)=1$. Based on the above arguments, one would expect to find the same phase diagram even in this model, as long as $p_s \ll p_\text{plq}$. This is indeed consistent with what we observe in numerical simulations. 

\begin{figure}
     \centering
          \begin{subfigure}[b]{\columnwidth}
         \centering
         \includegraphics[width=\textwidth]{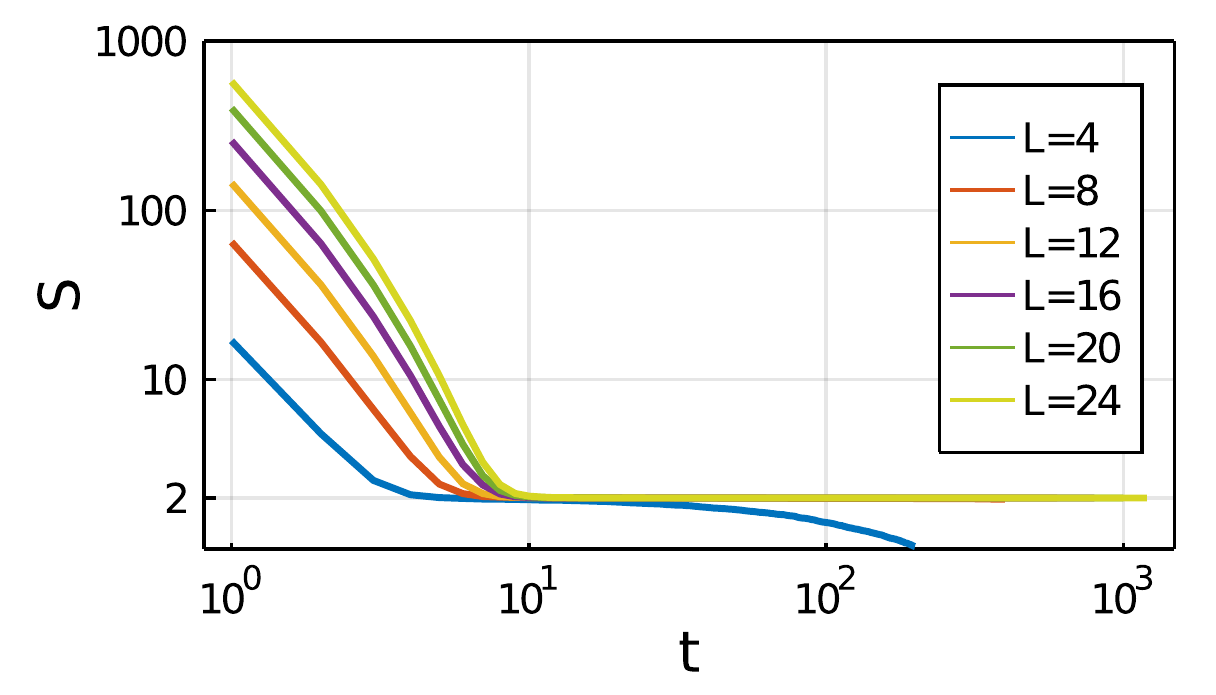}
         \caption{}\label{figs_perturb_arealaw_plq1}
     \end{subfigure}
     \begin{subfigure}[b]{\columnwidth}
         \centering
         \includegraphics[width=\textwidth]{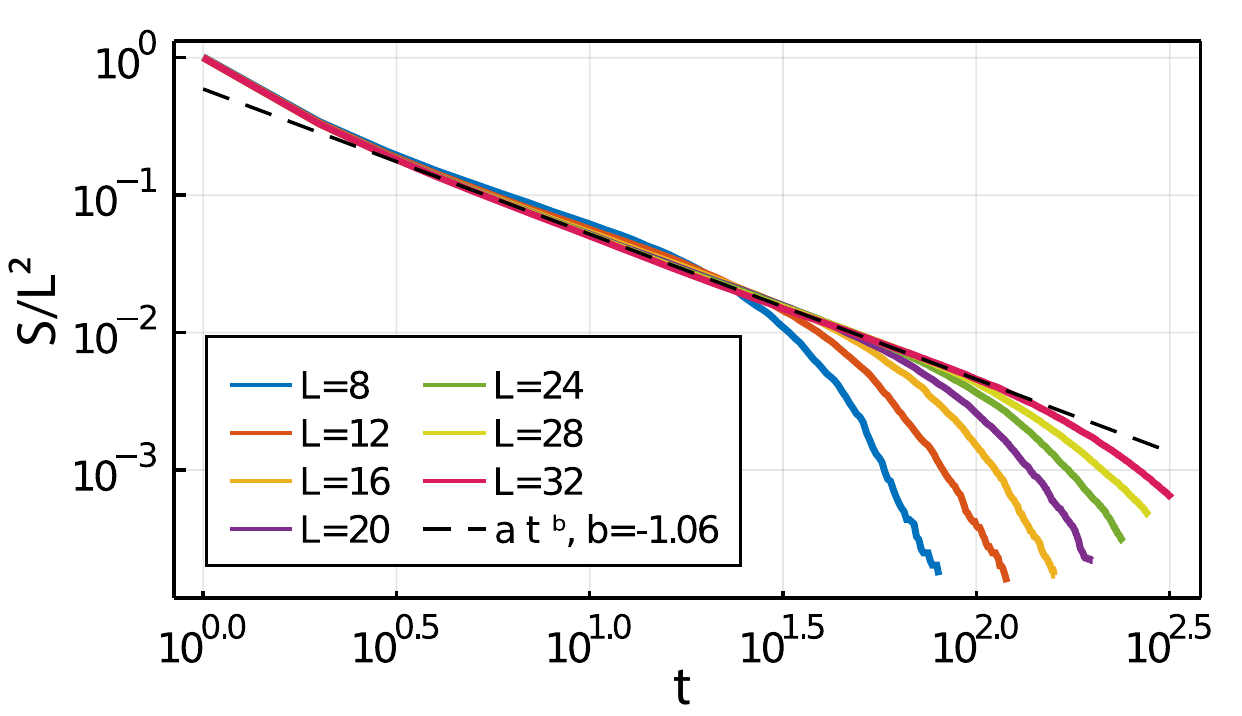}
         \caption{}
         \label{fig_perturb_iso_plq1}
     \end{subfigure}
    \caption{Purification dynamics in the presence of additional single-qubit measurements with $p_s=0.01$ and $p_\text{plq}=0.1$. Subplot (a) correspond to perturbing an area law point with $p_x=p_y=0.1$ and subplot (b) corresponds to perturbing the isotropic point $p_x=p_y=p_z$, as described in the text.}
\end{figure}

\begin{figure}
     \centering
          \begin{subfigure}[b]{0.48\columnwidth}
         \centering
         \includegraphics[width=\textwidth]{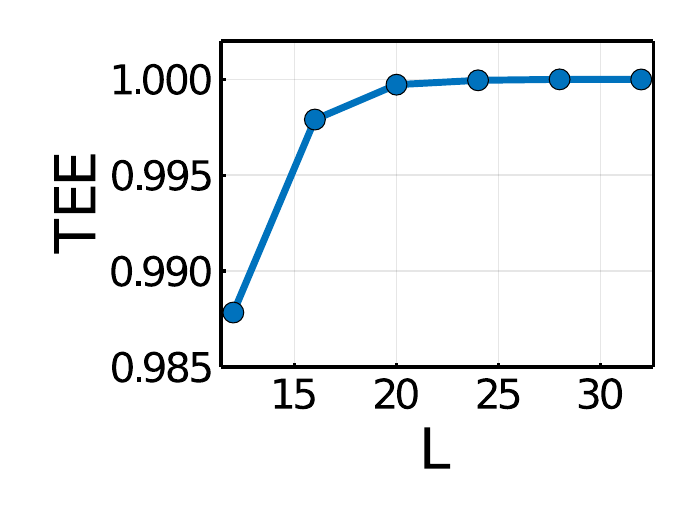}
         \caption{}\label{fig_perturb_plq1_arealaw_TEE}
     \end{subfigure}
     \hfill
     \begin{subfigure}[b]{0.48\columnwidth}
         \centering
         \includegraphics[width=\textwidth]{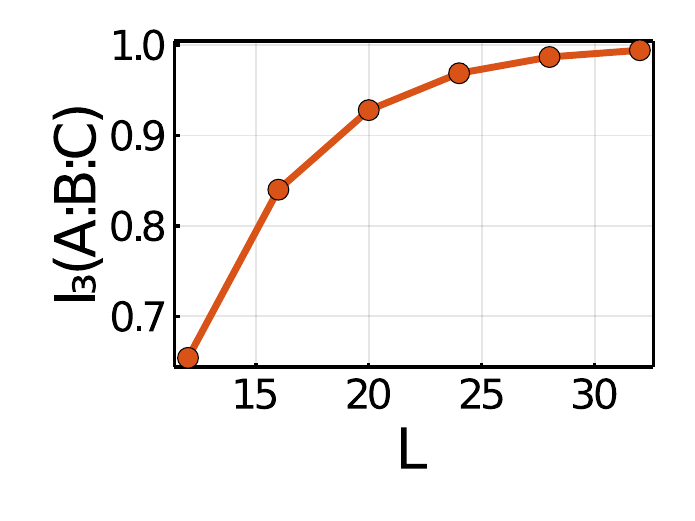}
         \caption{}
         \label{fig_perturb_plq1_arealaw_I3}
     \end{subfigure}
     
    \begin{subfigure}[b]{0.48\columnwidth}
         \centering
         \includegraphics[width=\textwidth]{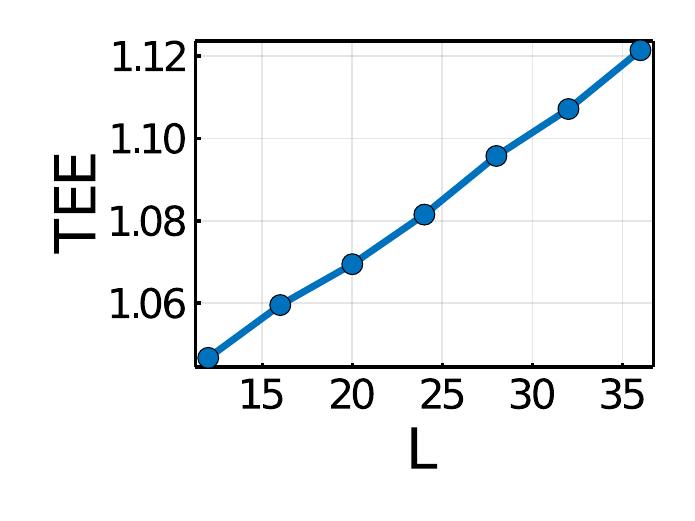}
         \caption{}\label{fig_perturb_plq1_iso_TEE}
     \end{subfigure}
     \hfill
     \begin{subfigure}[b]{0.48\columnwidth}
         \centering
         \includegraphics[width=\textwidth]{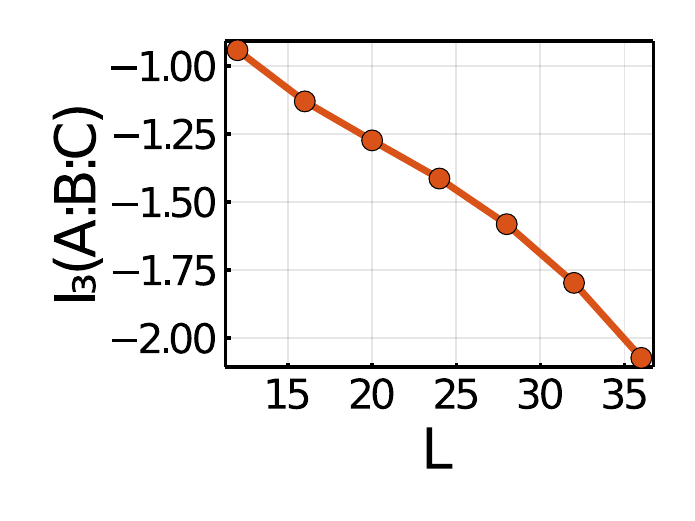}
         \caption{}
         \label{fig_perturb_plq1_iso_I3}
     \end{subfigure}
    \caption{(a) TEE and (b) tripartite mutual information versus system size in the steady state of an area law point with $p_x=p_y=0.1$ perturbed with $p_\text{plq}=0.1$ and $p_s=0.01$. (c) and (d) show analogous  quantities when the same perturbation is applied to the isotropic point $p_x=p_y=p_z$.}
\end{figure}

Fig. \ref{figs_perturb_arealaw_plq1} shows the purification dynamics of the system, starting from a maximally-mixed initial state, when $p_x=p_y=0.1$ with perturbation parameters $p_s=0.01$ and $p_\text{plq}=0.1$. As is clear from the figure, the long cycle stabilizers remain entangled to the environment until long times, while the remaining of degrees of freedom rapidly disentangle, as was observed to be the case when $p_{s}=0$. Moreover, TEE and $I_3(A:B:C)$ also remain equal to $+1$ for large system sizes as shown in Fig. \ref{fig_perturb_plq1_arealaw_TEE} and Fig. \ref{fig_perturb_plq1_arealaw_I3} respectively.

\begin{figure}
     \centering
     \includegraphics[width=\columnwidth]{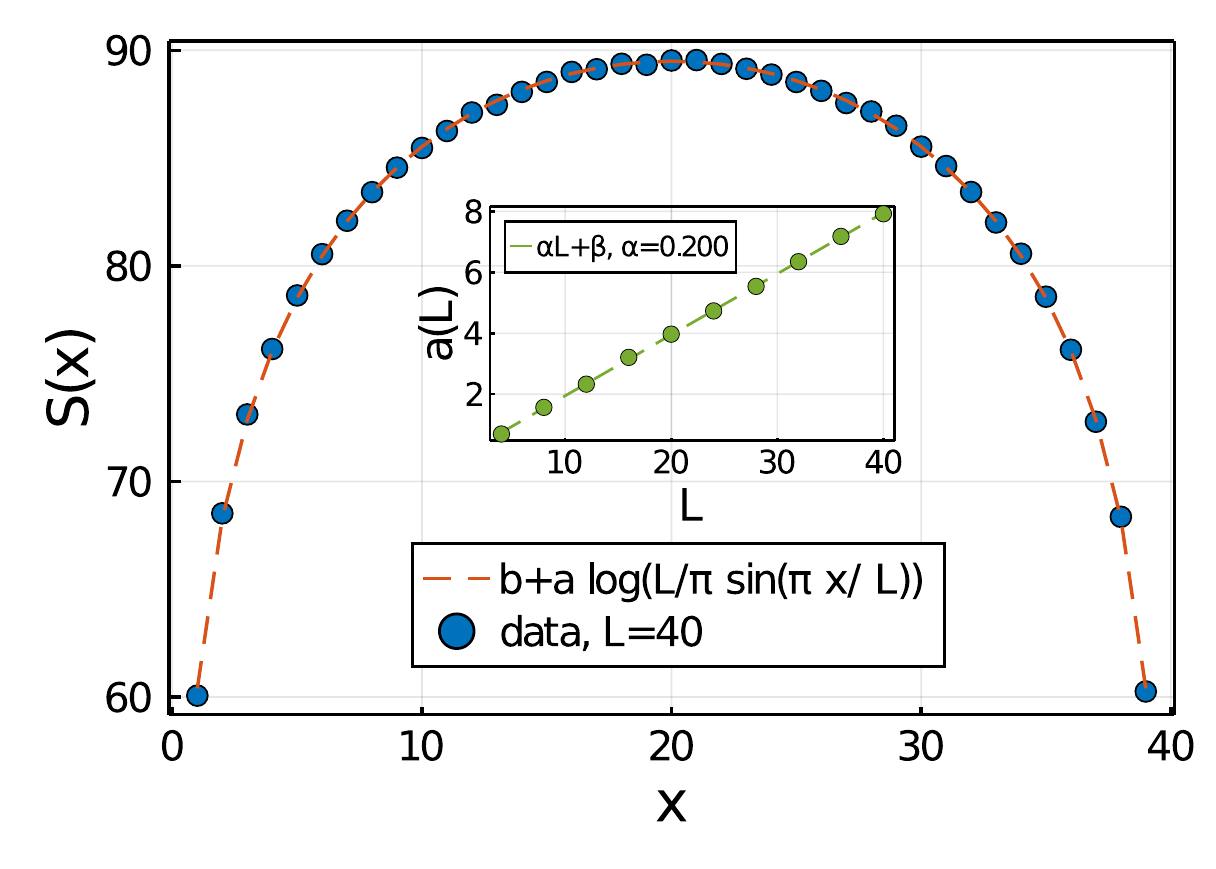}
     \caption{Entanglement entropy of a cylindrical region of size $x\times L$ as a function of $x$ for the steady state of the circuit at the isotropic point $p_x=p_y=p_z$ perturbed with $p_s=0.01$ and $p_\text{plq}=0.1$. }
     \label{fig_perturb_plq1_Sx}
\end{figure}

\begin{figure}
     \centering
     \includegraphics[width=\columnwidth]{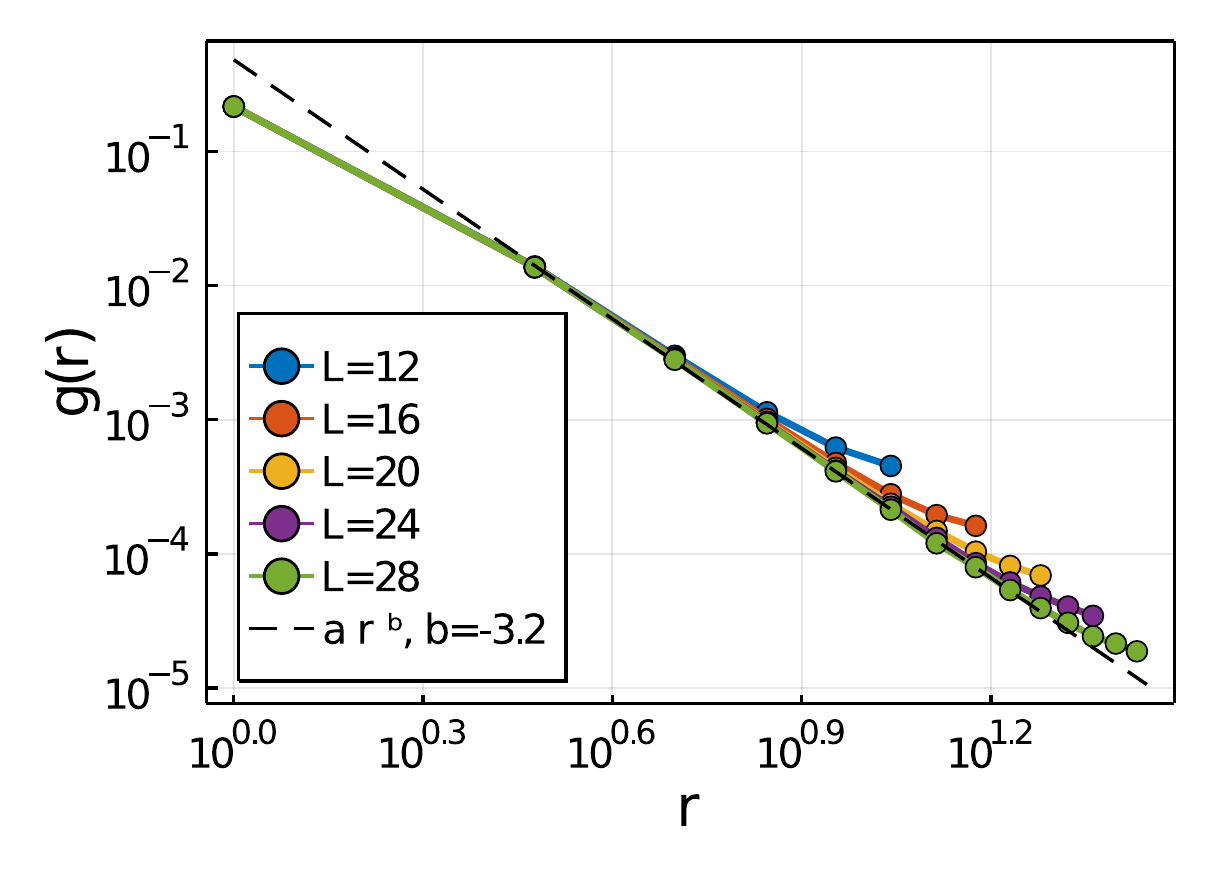}
     \caption{ The Wilson line correlation $g(r)$ in the steady state of the isotropic point $p_x=p_y=p_z$ perturbed with $p_s=0.01$ and $p_\text{plq}=0.1$.   }
     \label{fig_perturb_plq1_iso_gr}
\end{figure}

Fig. \ref{fig_perturb_iso_plq1} on the other hand corresponds to the purification dynamics when $p_x=p_y=p_z=1/3$,  with $p_s=0.01$ and $p_\text{plq}=0.1$, again showing a power-law decay of the entropy density as a function of time.  The steady-state entanglement entropy also shows  $S_{A} \sim L_{A} \log L_{A}$ scaling, as seen in Fig.  \ref{fig_perturb_plq1_Sx} and the Wilson line correlation $g(r)$ defined in Section \ref{sec_wilson} also decays as a power law $1/r^\Delta$ with $\Delta \sim 3$ as shown in Fig. \ref{fig_perturb_plq1_iso_gr}.  On the other hand, we find that TEE computed via the Kitaev-Preskill construction is no longer equal to $1$ but it rather grows slowly with the system size as shown in Fig. \ref{fig_perturb_plq1_iso_TEE}; we note, however, that the interpretation of the TEE for a super-area-law-entangled phase is not clear, and need not be a universal constant as it is in an  area-law-entangled phase. Similarly, as can be seen in Fig. \ref{fig_perturb_plq1_iso_I3} the tripartite mutual information $I_3(A:B:C)$ is also no longer fixed at $-1$, but rather scales linearly with the system size. Note that the argument we provide in Section \ref{sec_mutual_info} for $I_3(A:B:C)=-1$ fails in the presence of single-qubit measurements.  For every single-qubit measurement performed near the boundary of $AB$, the product of plaquette operators spanning the boundary becomes part of the stabilizer group.  As a result, the product of a typical pair of string operators spanning $ABC$ cannot necessarily be reduced in its support, by the action of elements of the stabilizer group, to an operator supported exclusively on $A$ and $C$. From the distribution $Q(r)\sim r^{-2}$ in the critical phases, there should be $O(L)$ such long stabilizers spanning the $ABC$ region, which gives rise to the behavior in Fig. \ref{fig_perturb_plq1_iso_I3}.

Finally, let consider perturbing the original circuit model. So now, at each step, a random qubit would be measured with probability $p_s$, or a bond operator would be measured with probability $(1-p_s)$ such that $p_s+(1-p_s)(p_x+p_y+p_z)=1$. As we have discussed before, plaquette operators are still being measured with constant rate, say $q$, due to the bond operator measurements, though we no longer measure the plaquette operators directly. Therefore, based on the discussion so far, we expect both phases survive as long as $p_s \ll q$. Because $q$ is very small in our model(see Fig. \ref{fig_plq_rate} in Appendix  \ref{apx_sup_figs}), one has to go to very large system sizes to numerically verify this statement. Instead, we provide a more detailed argument in Appendix \ref{apx_link_vs_plq_measurement} to support this claim.
On the other hand, when $p_s \gg q$, we observe that the monitored trajectories of the system settle into a volume law phase(see Fig.  \ref{fig_perturb_S} in Appendix   \ref{apx_sup_figs}).


Another way of perturbing the original dynamics is to add three-qubit measurements, given by the product of adjacent bond operators; this kind of perturbation, however, preserves both the free-fermion nature of the parton dynamics, while also keeping the extensive number of conservation laws.  In the Majorana parton picture it translates into next nearest neighbour coupling of $c$ Majoranas, which in turn translates into next nearest neighbour moves in the classical loop model. From the classical model, we know such a system would flow into another critical phase with long range correlations. This is indeed what we observe by numerically simulating the quantum circuit, where at a each step either a product of two adjacent bond operators is measured randomly with probability $p_3$ or a random bond operator is measured with probability $(1-p_3)$. The entanglement entropy scaling in the steady state are shown for $p_3=0.5$ in Fig.\ref{fig_perturb_3_05} and for $p_3=0.1$ in Fig.\ref{fig_perturb_3_1} in Appendix \ref{apx_sup_figs}. The entanglement entropy clearly follows a $L \log L$ scaling with a $p_3$ dependent coefficient.

\section{Discussion and Outlook}\label{sec:discussion}
In this work, we studied how random measurements of non-commutative observables could give rise to non-equilibrium phases of matter which exhibit long-range entanglement. While these dynamical entanglement phases of matter share some properties with their equilibrium counter parts like spin liquids, they can in principle exhibit novel features which can only appear in the non-equilibrium setting.

An interesting aspect of this work that warrants further investigation, is whether the out-of-equilibrium parton construction introduced here can be used to investigate other regimes of monitored evolution which give rise to long-range-entangled steady-states.  Furthermore, it is important to understand the quantum error correction properties of the area law phase, which are closely related to the recent development of Floquet codes \cite{Hastings2021dynamically,vuillot2021planar,haah2022boundaries, paetznick2022performance}. As shown in Section \ref{sec:purification_dyn}, when the measurement probabilities are highly biased, the random dynamics give rise to two dynamically generated logical qubits, which would be absent if one interprets the bond operators as gauge operators of a subsystem quantum error correcting code. However, it is not yet clear to what extent the area law phase could be used as a quantum error correcting code. In particular, it would be interesting to see whether this model has an efficient decoder with a finite threshold. Note that due to the random and indirect measurement of the stabilizer, the syndrome data of the dynamically generated code lacks structures like $\mathbb{Z}_2$ gauge symmetry, which should be trivially present in the standard surface code syndrome data. On the other hand, the randomness might help the decoder to be more resilient against adversarial errors, compared to simpler models like the Floquet honeycomb code \cite{Hastings2021dynamically, haah2022boundaries}. 

\emph{Note Added:} During completion of this work, we were made aware of forthcoming work \cite{Vedika_to_appear} on a related problem.

\acknowledgments We thank Matthew P. A. Fisher, Yuan-Ming Lu, Ashvin Vishwanath, and especially Adam Nahum for useful discussions.  SV gratefully acknowledges support from the Simons Center for Geometry and Physics, Stony Brook University at which some of the research for this paper was performed. ZXL is supported by the Simons Collaborations on Ultra-Quantum
Matter, grant 651457 from the Simons Foundation. AL is supported by Joint Quantum Institute Physics Frontier Center at University of Maryland (JQI-PFC-UMD). This research was supported in part by the Heising-Simons Foundation, the Simons Foundation, and National Science Foundation Grant No. NSF PHY-1748958.  We acknowledge the University of Maryland supercomputing resources (http://hpcc.umd.edu) made available for conducting the research reported in this paper.

\clearpage
\appendix
\begin{figure}
    \centering
    \includegraphics[width=\columnwidth]{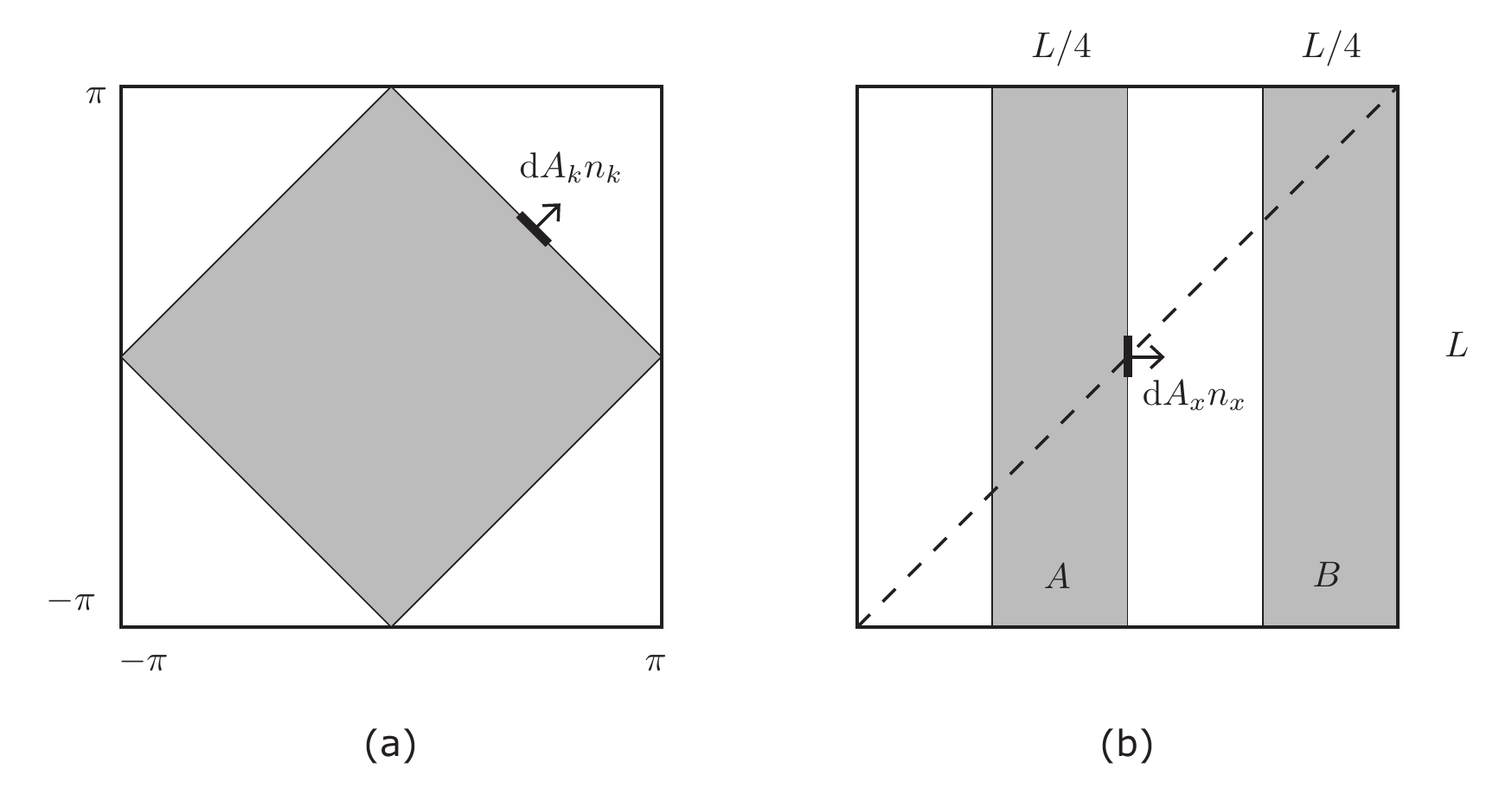}
    \caption{(a) The Brillouin zone and the Fermi surface at half-filling. (b) Two cylinders ($A$ and $B$) of width $L/4$ which are $L/4$ apart.}
    \label{fig_widom}
\end{figure}
\section{Mutual Information in Free Fermion systems}
\subsection{Free Fermions with a Fermi surface}
\subsubsection{1d}
Consider a 1d system of free fermions. Let the subsystem $A$ be composed of disjoint intervals $(u_1,v_1)$, $(u_2,v_2)$, ..., $(u_N,v_N)$. The entanglement entropy of subsystem $A$ is \cite{calabrese2004entanglement}:
\begin{align}
    S_A=1/3\Big[&\sum_{k\le j} \log ((v_k-u_j)/a))\nonumber\\
            &-\sum_{k<j} \log ((v_k-v_j)/a))\nonumber\\
            &-\sum_{k<j} \log ((u_k-u_j)/a))\Big].
            \label{equ_S}
\end{align}
This expression shows that the mutual information is extensive in this system, in the sense that if $A$, $B$ and $C$ are three disjoint intervals, we have,
\begin{align}\label{equ_ext}
    I(A:BC)=I(A:B)+I(A:C).
\end{align}
For two disjoint intervals of widths $w_1$ and $w_2$, separated by distance $d$, the mutual information would be,
\begin{align}\label{equ_IAB1d}
    I(A:B)=\frac{1}{3}\log\qty(\frac{(d+w_1)(d+w_2)}{d(d+w_1+w_2)}),
\end{align}
which falls of as $w_1w_2/d^2$, for $d\gg w_1,w_2$.

In particular if we divide a circle of length $L$ into four regions of equal size $L/4$ and $A$ and $B$ denote two antipodal regions, their mutual information would be,
\begin{align}
    I(A:B)=\frac{1}{3}\log 2.
\end{align}
We can also consider the tripartite mutual information between three adjacent intervals of length $L/4$. Eq.\eqref{equ_ext} already suggests\footnote{Note that Eq.\eqref{equ_ext} is only valid when the intervals are completely disjoint.} that the tripartite mutual information would vanish in free fermions. We can use Eq. \eqref{equ_S} to explicitly compute the tripartite mutual information for three adjacent intervals of width $w$:
\begin{align}
    I(A:B:C)= & I(A:B)+I(A:C)-I(A:BC)\nonumber \\
            = & \frac{1}{3}\qty[\log\qty(\frac{w}{2a })+\log\qty(\frac{4}{3})-\log\qty(\frac{2w}{3a })]\nonumber\\
            = & 0.\label{eq_IABC0}
\end{align}

\begin{figure}
    \centering
    \includegraphics[width=\columnwidth]{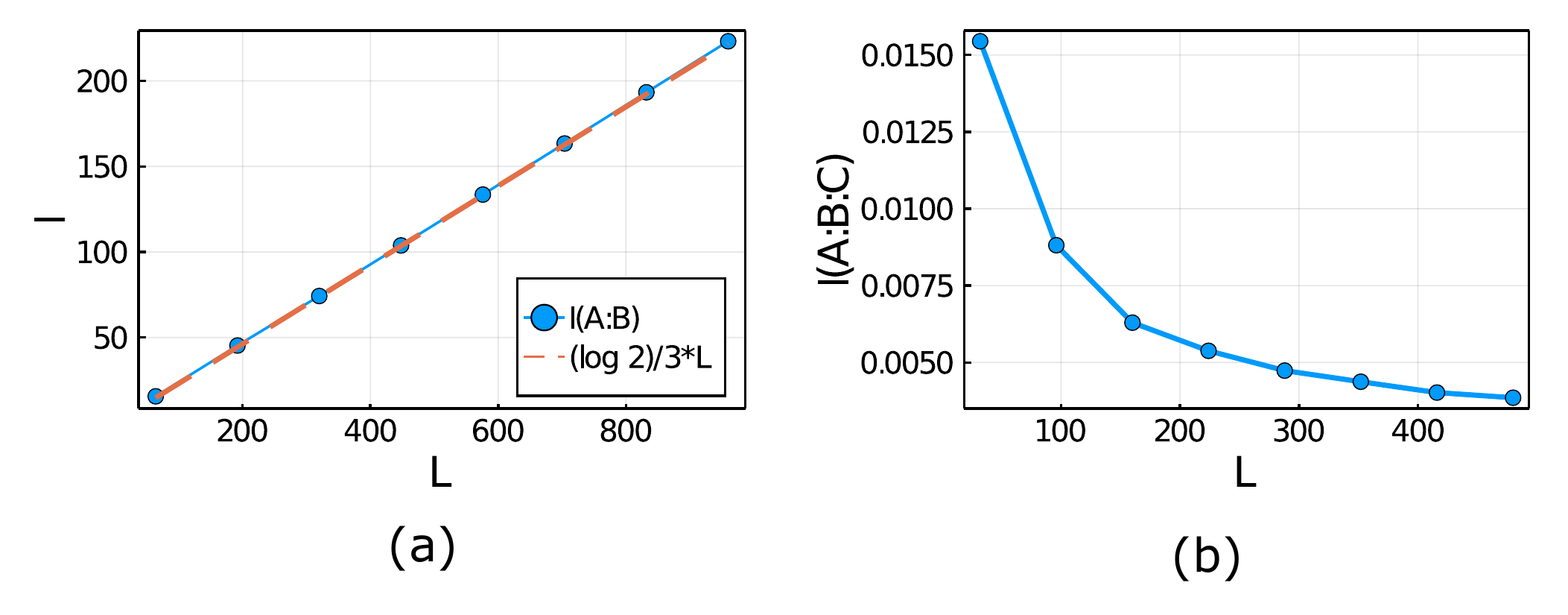}
    \caption{(a) Mutual information of two cylindrical regions of width $L/4$ which are $L/4$ apart in half-filled free fermion on a periodic 2d square lattice. (b) The tripartite mutual information between the aforementioned cylinders and the one in between. Instead of considering half filled system, we set $\mu=-0.3$ to avoid numercial instabilities.}
    \label{fig_I_ff}
\end{figure}

\subsubsection{2d}
One may use the $1$d result alongside the generalized Widom formula \cite{swingle2012renyi} to find the entanglement entropy of non-convex and/or disjoint regions in Fermi liquid systems in higher dimensions where the Fermi surface has co-dimension one:
\begin{align}
    S(R)=\frac{1}{(2\pi)^{d-1}}\int \dd A_x \dd A_k |n_x\cdot n_k| \frac{S_{1+1}(x,k)}{n_\text{int}(x,k)},
\end{align}
where $\dd A_x$ and $\dd A_k$ are area elements on the surface of the region $R$ and the Fermi surface respectively, $n_x$ and $n_k$ are the unit normals at the respective points on those surfaces, $S_{1+1}(x,k)$ is the entanglement entropy of a 1d chiral mode on the subregion $R\cap L$ , where $L$ is the straight line passing through $x$ in the direction of the Fermi velocity $v(k)$ and $n_\text{int}$ is the number of times this line intersects with region $R$.  
\begin{figure}
    \centering
    \includegraphics[width=\columnwidth]{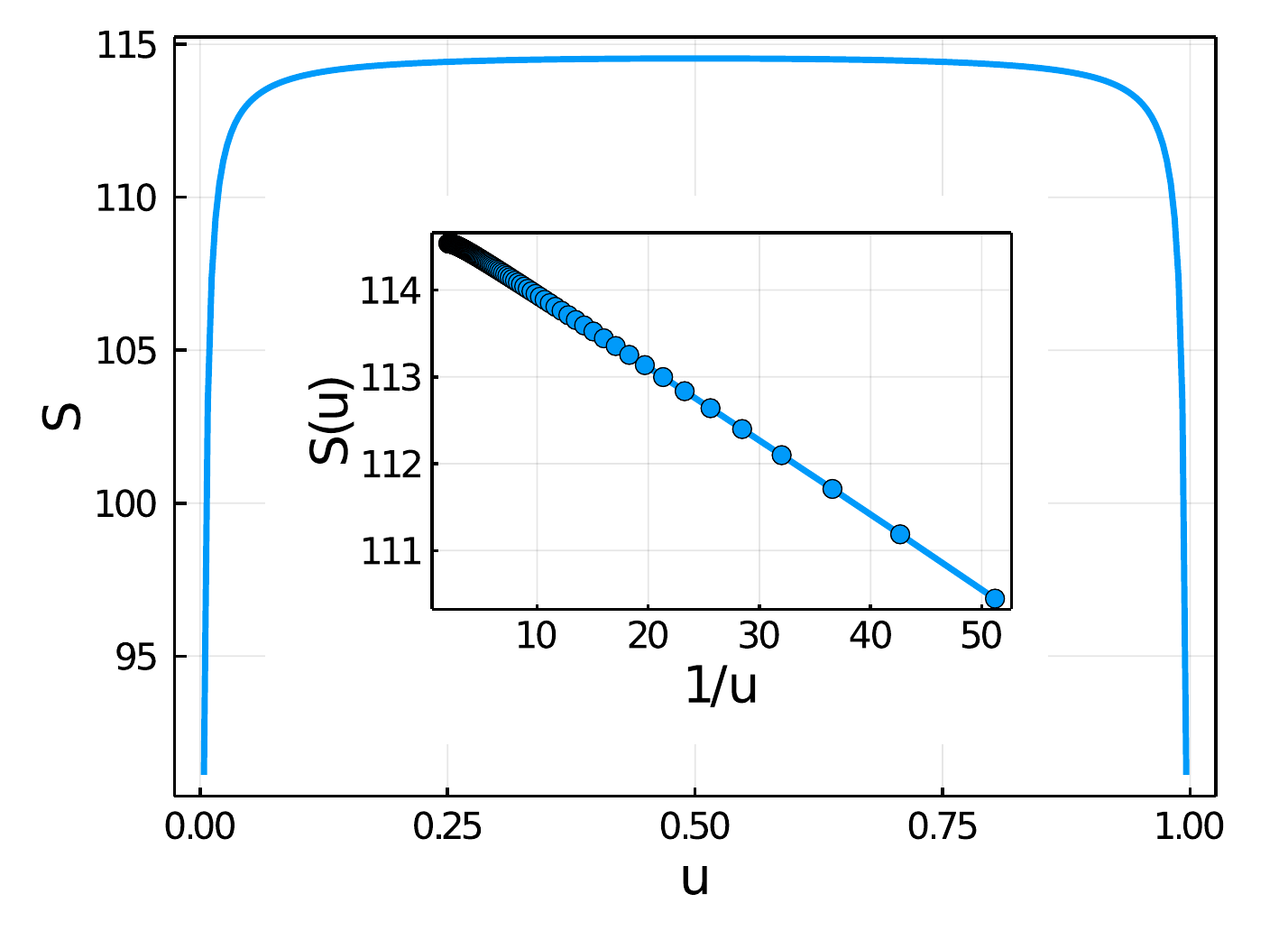}
    \caption{Entanglement entropy of a $x\times L$ cylindrical region of the Kitaev honeycomb model on a $L\times L$ torus as a function of $u=x/L$ at the isotropic point $J_x=J_y=J_z$ for $L=256$.}
    \label{fig_su}
\end{figure}
Consider a tight binding free fermion system on a $L\times L$ periodic lattice at half-filling, where the fermi surface has the simple shape of a rotated square. By using the Widom formula, we find that the mutual information between two cylinders of length $L/4$ which are $L/4$ apart is (Fig. \ref{fig_widom}),
\begin{align}
    I(A:B)&=\frac{1}{2\pi}4\sqrt{2}\pi\times 4L \times \frac{\sqrt{2}}{2}\times \frac{1}{4}\times \frac{1}{2}\times \frac{\log 2}{3}\nonumber \\
    &=\frac{\log 2}{3}L
\end{align}
where the last $\frac{1}{2}$ factor is there since each chiral mode contributes half of the entanglement in Eq. \eqref{equ_S}. As can be seen from the plot in Fig. \ref{fig_I_ff}a, this expression agrees perfectly with the numerical result. Moreover, by using Widom formula and Eq. \eqref{eq_IABC0}, one can see that the tripartite mutual information between the two cylinders and the one in between vanishes (see Fig. \ref{fig_I_ff}b).

\subsection{Kitaev Honeycomb model}
\begin{figure*}
    \centering
    \includegraphics[width=\textwidth]{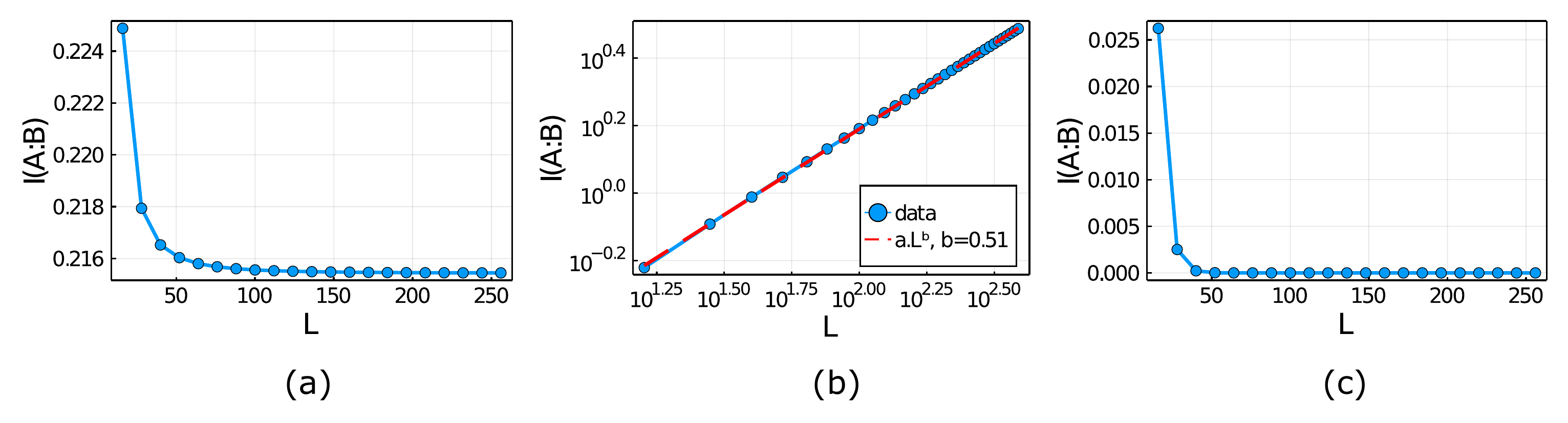}
    \caption{Mutual information in of two antipodal cylinders of length $L/4$ in the ground state of Kitaev at $J_y=J_z=1$ and (a) $J_x=1$, (b) $J_x=2$ and (c) $J_x=3$. }
    \label{fig_I_AB}
\end{figure*}

The Kitaev honeycomb model has two phases, gapped and gapless. Both phases are area law entangled. This means that the entanglement entropy of a $x\times L$ cylindrical region on a $L\times L$ torus is proportional to $L$. In the gapless phase, the subleading correction to the area law is a function of $u=x/L$ and it is proportional to $1/u$ for $u\ll 1/2$ and  $1/(1-u)$ when $1-u \ll 1/2$ similar to the case for free 2d Dirac fermion \cite{chen2015scaling} . The subleading terms in the gapped phase vanish much faster (probably exponentially in the correlation length) as expected (see Fig. \ref{fig_su}). 
The leading term of the mutual information for Dirac fermions is computed in Ref. \cite{chen2017mutual}; for two circular regions of radius $R$ and $R'$ which are distance $r$ apart, the mutual information would be only a function of the cross ratio
\begin{align}
    z=\frac{4R R'}{r^2-(R-R')^2},
\end{align}
and scales as
\begin{align}
    I(z)=\frac{1}{15}z^2+\cdots
\end{align}
for $z\ll 1$. The mutual information being only a function of dimension less $z$ means it remains constant when $R$, $R'$ and $r$ are all scaling proportionally, which is similar to the case we are studying on a toroidal geometry. Fig. \ref{fig_I_AB} shows the mutual information between two cylinders of size $L/4 \times L$ which are distance $L/4$ apart, for the Kitaev model in the gapless phase, on the phase boundary and in the gapped phase.  As we expect from the Dirac fermion, the mutual information saturates to a constant value in the gapless phase. Interestingly, at the phase boudnary the mutual information seems to grow as $\sqrt{L}$. In the gapped phase it goes to zero, as expected from a finite correlation length. 

\begin{figure}
    \centering
    \includegraphics[width=0.7\columnwidth]{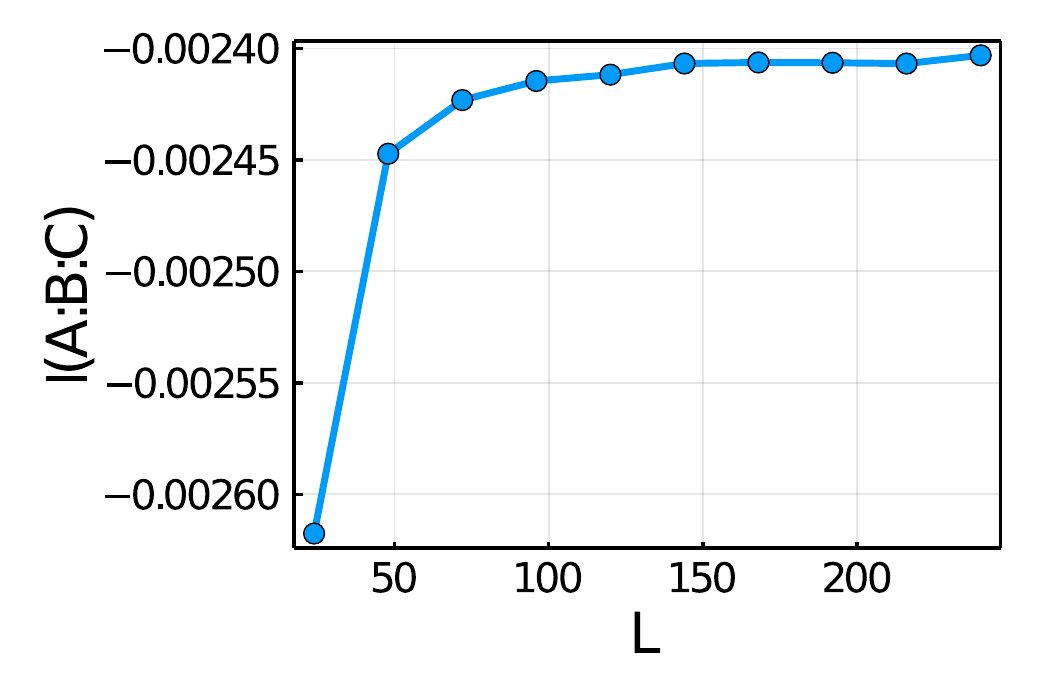}
    \caption{Tripartite mutual Information between three cylinders of length $L/4$ for the Kitaev honeycomb model on a $L\times L$ torus.}
    \label{fig_I_ABC_HC}
\end{figure}

The tripartite mutual information between three adjacent cylinders also vanishes similar to the free fermion case (see Fig. \ref{fig_I_ABC_HC}).

\section{Dimer Number Parity }\label{app:dimer_covering}
\begin{figure}
$\begin{array}{cc}
     \includegraphics[width=0.1\textwidth]{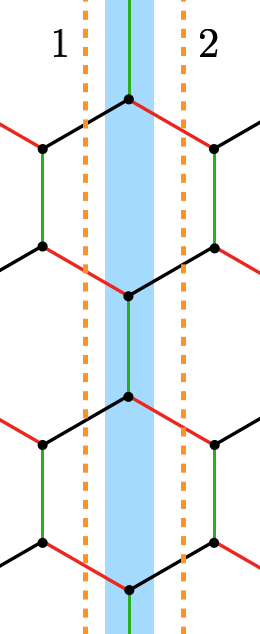} &
     \includegraphics[width=0.25\textwidth]{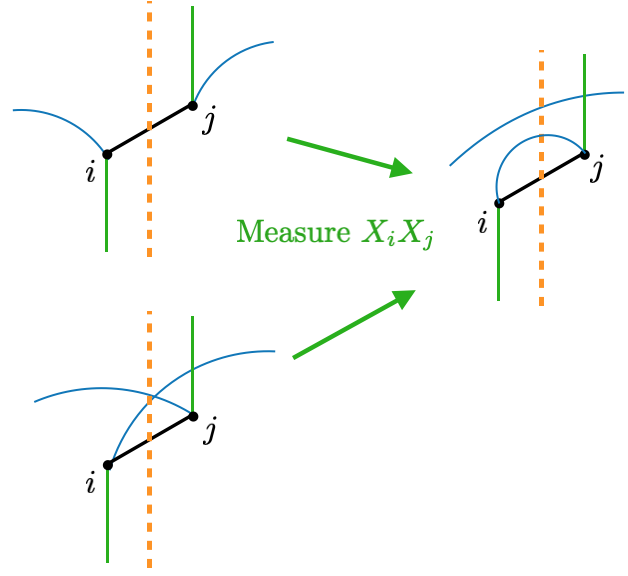}
\end{array}$
     \caption{Under projective measurements of the bond operators on bonds, the dynamical rules for the evolution of dimer configurations leads to the conservation of dimer number parity across vertical cuts, such as the one shown in (a).}
     \label{fig:topo_inv_dimer_covering}
\end{figure}
Consider the honeycomb lattice with periodic boundary conditions and so that each site is paired with another unique site.  We consider a vertical ``cut" through the $x$ and $y$-type bonds of the honeycomb lattice, as shown by the dotted orange line labeled ``$1$'' in Fig. \ref{fig:topo_inv_dimer_covering}.  We now show that the \textit{number parity} of dimers crossing this cut -- given by $(-1)^{n_{1}}$ where $n_{1}$ is the number of dimers which straddle the cut labeled ``$1$" -- is the same for all vertical cuts on the honeycomb lattice.  The number of dimers crossing a cut is a well-defined quantity if the dimers are short-ranged. An example of another such vertical cut, labeled ``$2$", is also shown in Fig. \ref{fig:topo_inv_dimer_covering}a. The proof of this statement is obtained as follows.  First, consider the column of $s$ sites which are shaded in blue in Fig. \ref{fig:topo_inv_dimer_covering}a; $s$ is an even integer since each unit cell on the honeycomb lattice contains two lattice sites.  Let $n_{\ell} (n_{r})$ be the number of dimers which contain one endpoint in the shaded blue column and another endpoint to the left (right) of the column, respectively. The remaining $s - n_{\ell} - n_{r}$ sites in the shaded blue column are dimerized with each other, so that $s - n_{\ell} - n_{r}$ is an even integer.  Since $s$ and $s- n_{\ell} - n_{r}$ are both even, we must have that $n_{\ell} + n_{r}$ is even so that $(-1)^{n_{\ell}} = (-1)^{n_{r}}$.  Let $n_{1}$ and $n_{2}$ be the total number of dimers straddling cut 1 or cut 2, respectively.  These two quantities are related as  
\begin{align}
    n_{2} = n_{1} - n_{\ell} + n_{r}
\end{align}
and as a result
\begin{align}
    (-1)^{n_{1}} = (-1)^{n_{2}}.
\end{align}
This completes the proof. 

Finally, we observe that the dynamical rules for the evolving configuration preserve the parity along the vertical cut.  This is trivially shown by considering the effect of a bond measurement on the dimers crossing a vertical cut passing through that bond.  If the sites were already dimerized with each other, then the measurement has no effect on the state of the system.  Now consider the case where the sites are dimerized with other sites in the system.  The sites could be dimerized with two other sites on ($i$) opposite sides of the vertical cut or ($ii$) on the same side.  The former case, and the dimer configuration after a bond measurement, are shown in Fig. \ref{fig:topo_inv_dimer_covering}b.  The dimer number parity crossing the cut is manifestly preserved by the measurement.  Case ($ii$) can be similarly considered, to show that the bond measurements preserve the dimer number parity across a vertical cut.

\section{Mutual information in the area law phase}\label{apx_mutual_info}

\begin{figure*}
     \centering
    \begin{subfigure}[b]{0.3\textwidth}
         \centering
         \includegraphics[width=\textwidth]{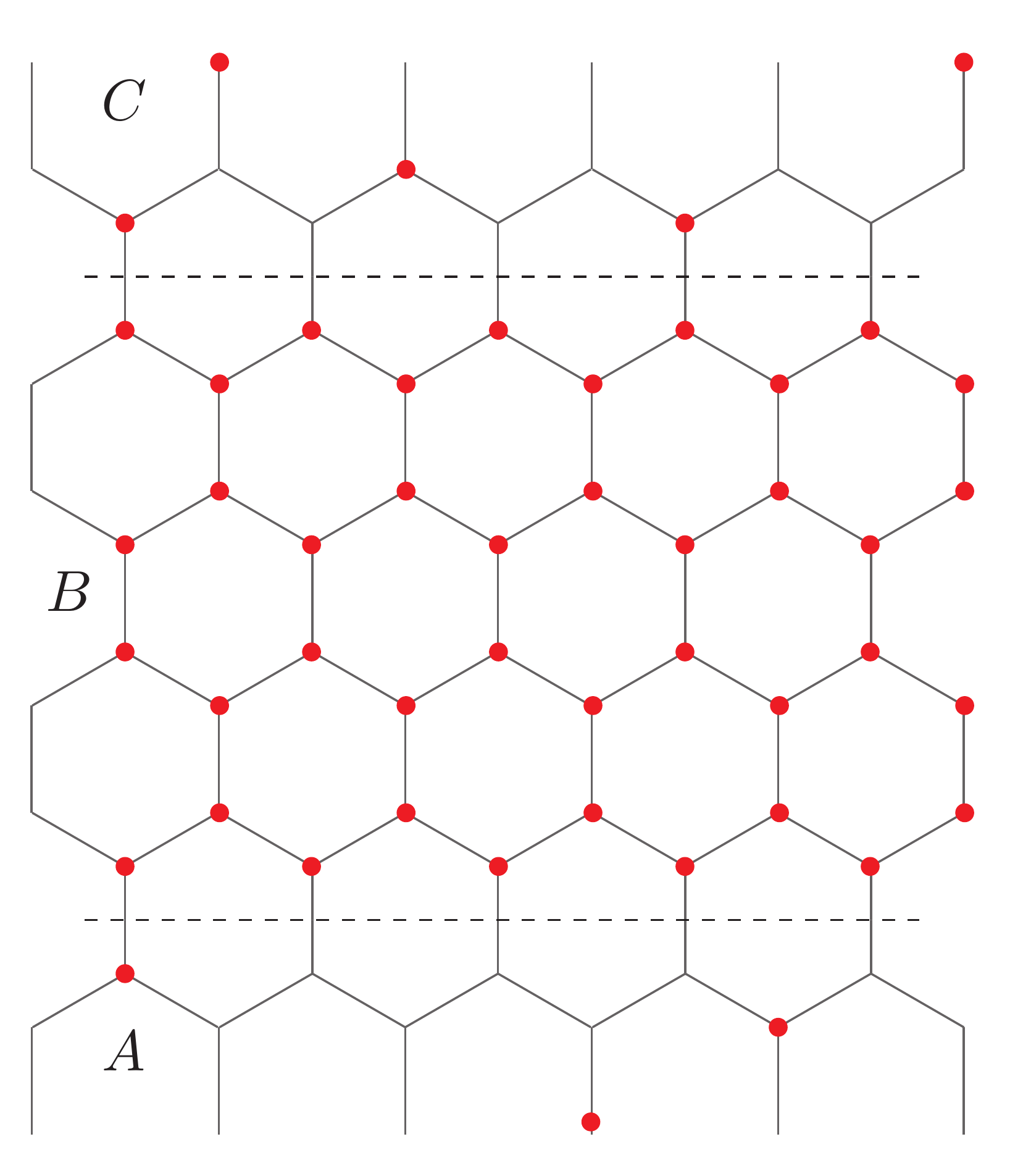}
         \caption{}\label{fig_MIgb}
         \label{fig_SRZ_px02}
     \end{subfigure}
     \hfill
    \begin{subfigure}[b]{0.3\textwidth}
         \centering
         \includegraphics[width=\textwidth]{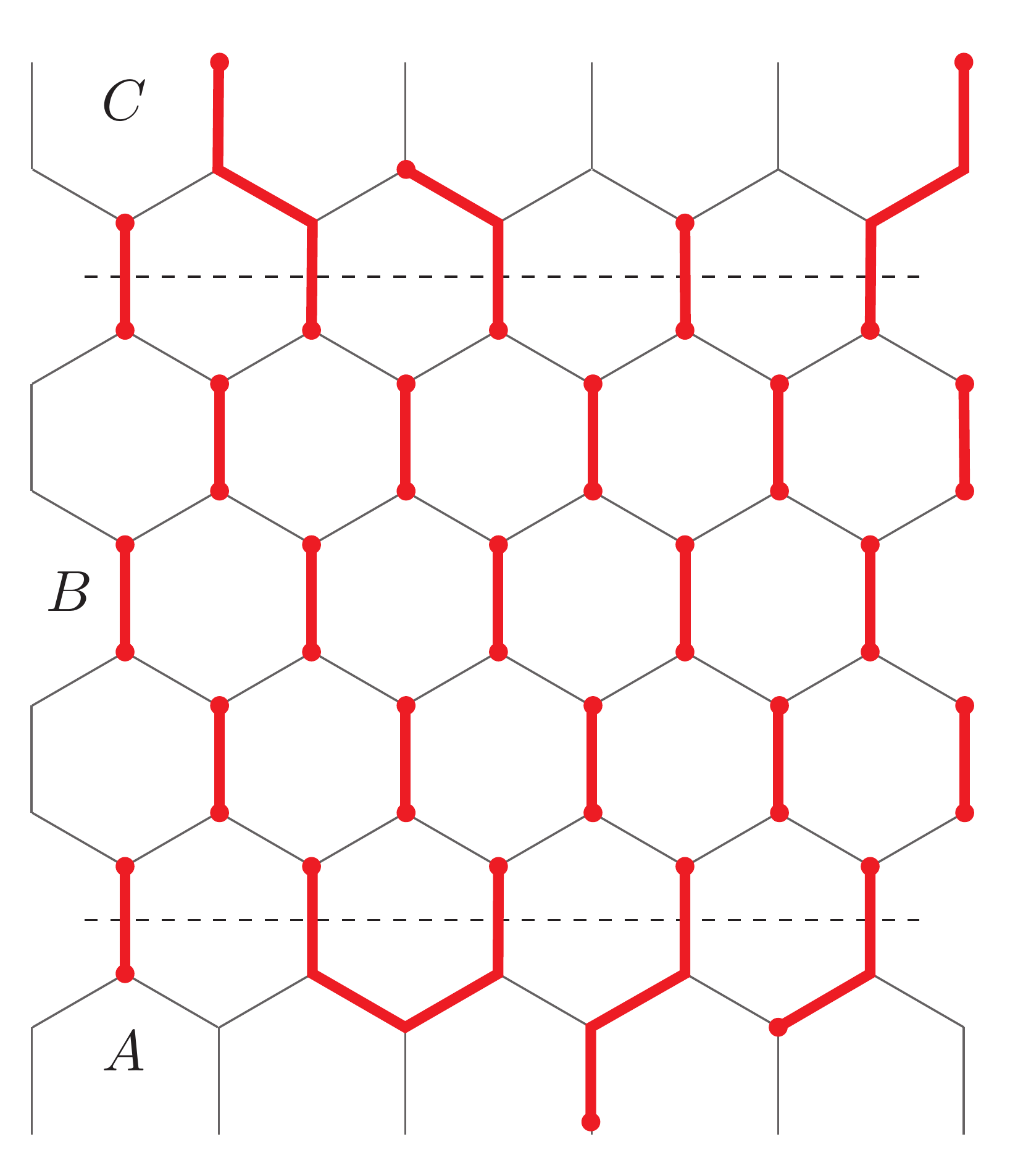}
         \caption{}\label{fig_MIgc}
         \label{fig_SRZ_px02}
     \end{subfigure}
    \hfill
    \begin{subfigure}[b]{0.3\textwidth}
         \centering
         \includegraphics[width=\textwidth]{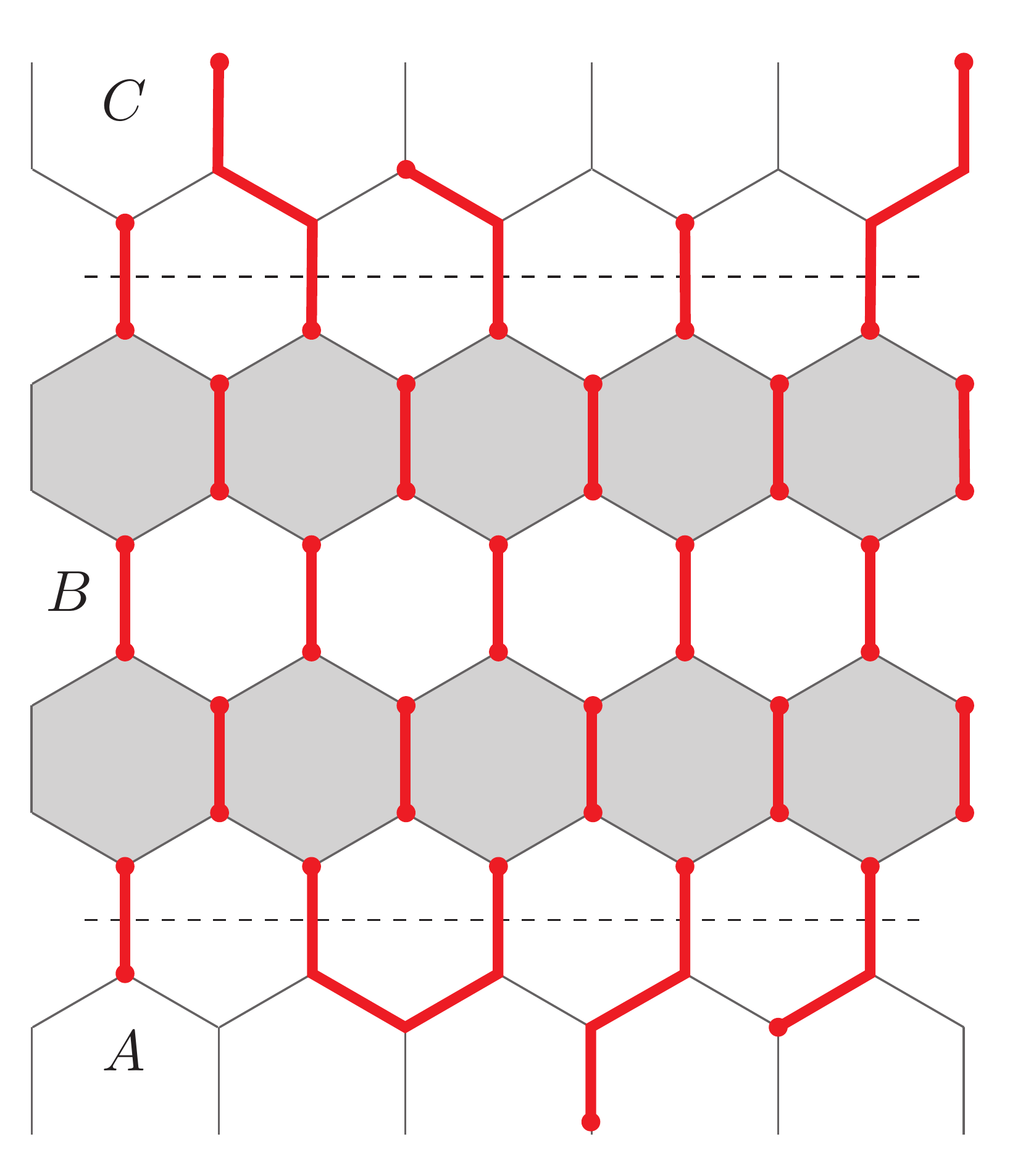}
         \caption{}\label{fig_MIgd}
         \label{fig_SRZ_px02}
     \end{subfigure}
        \caption{Evaluating $I(A:C)$ in the area law phase: Let $P$  be the product of all string operators which has at least one end point in region $B$. A typical set of end points specifying $P$ is shown in panel (a). Up to multiplication by plaquette operators, $P$ can be taken to be the product of the string operators shown in panel (b). The product of the shaded plaquette operators and the red string operators in panel (c) has non-trivial support only in $A$ and $C$ and thus contributes to $I(A:C)$.  }
\end{figure*}

Here we explain why $I_2(A:C)=1$ throughout the area law phase. The $p_y=p_x\to 0$ limit has been explained in Section \ref{sec_mutual_info}. Here we consider the case when $p_x$ and $p_y$ are finite.

Let $g_{a,b}$ denote the string operator which connects $a$ to $b$. It has been already noted that the specific path which connects $a$ to $b$ does not matter in the sense that any two path give rise to the same string operator up to a product of plaquette operators. Now consider the product of two string operators $g_{a,b}~g_{c,d}$. Interestingly, this operators can also be specified only by its end points $a$,$b$,$c$ and $d$. i.e. not only the paths connecting the endpoints do not matter, but all different paring of the endpoints are equivalent up to product of plquette operators. e.g.
\begin{align}
    g_{a,b}~g_{c,d}=g_{a,d}~g_{c,b}~S
\end{align}
where $S$ is a product of plaquette operators (On a torus, $S$ could include the long cycle stabilisers as well). This is due to the simple fact that $g_{a,b}~g_{c,d}\times g_{a,d}~g_{c,b}$ is a string operator without a boundary. More generally, $g_{a,b}~g_{c,d}\cdots g_{e,f}$ is equivalent to $g_{\pi(a),\pi(b)}~g_{\pi(c),\pi(d)}\cdots g_{\pi(e),\pi(f)}$ where $\pi$ is an arbitrary permutation on the set of end points. This means a product of string operators can be specified by its set of endpoints and the specific pairing of the endpoints does not matter.

Now consider a typical state in the area law phase. Let $P$ be the product of all string operators which have at least one end point in the region $B$. As we discussed above $P$ can be specified by a set of end points. This set of end points include all points inside $B$ plus some points in $A$ and $C$ near the boundaries of $B$ (Fig.\ref{fig_MIgb}). Now choose a specific pairing of the end points, where all points in the bulk of $B$ are paired along the $z$ bonds, and the end points next to the boundary are paired with the endpoints which are in $A$ or $C$, and any remaining endpoints on each side (near $AB$ boundary and near $AC$ boundary) will be paired together(Fig.\ref{fig_MIgc}). This pairing will be possible as long as the parity of the number of endpoints in $A$ is the same as the parity of number of $z$ bonds that crosses the $A$ and $B$ boundary (Note that the parity of number of endpoints in $A$ and in $C$ are the same).  For now we assume this is the case and we will provide an argument in its support later. Let $P'$ denote the operator corresponding to this specific pairing. Now consider the operator $C$ which is the product of $P'$ with all plaquette operators on every other row of $B$ (shown as yellow in Fig.\ref{fig_MIgd}). Note that $C$  has non-trivial support only on $A$ and $C$ and it acts trivially in $B$, hence it could contribute to $I_2(A:C)$. 

To see that the parity of end points in $A$ are the same as the parity of the number $z$ bonds between $A$ and $B$, note that we expect the state to be close to the state where all dimers are along the $z$ bonds (as in Fig.\ref{fig_MIga}), in the sense that if one uses the latter state as the initial state of the circuit, it will evolve into the former state with only local rearrangement of the dimers without generating long range strings. In this case, we can use the fact that the parity of dimers crossing a line is an invariant of the circuit dynamics to reach the desired result. 

\section{Entanglement entropy of the steady state}\label{app:entanglement_parton}

In this section, we will explain how to compute the entanglement entropy of a subset of spins using the Majorana fermion picture. In particular, we show that the entanglement entropy of a region $A$ (with smooth boundaries) can be written as 
\begin{equation}
    S(A)=\frac{1}{2}n_p + \frac{1}{2}n_c-1,
\end{equation}
where $n_p$ is the number of plaquettes on the boundary of $A$ and $n_c$ is the number of Majorana dimers with one end in $A$ and the other in $B=\bar{A}$ before projecting back to the spin Hilbert space. 

In general, for a stabilizer state of $n$ qubits with the stabilizer group $\mathcal{G}$, let $\mathcal{G}_A \subseteq \mathcal{G}$ be the subgroup of stabilizers which act trivially on $B$. define $\mathcal{G}_{B}$ analogously. Then let the subgroup $\mathcal{G}_{AB}$ to be the group that is generated by the remaining $n-\rank(\mathcal{G}_A)-\rank(\mathcal{G}_B)$ additional generators needed to generate $\mathcal{G}$, i.e.
\begin{equation}
    \mathcal{G}=\mathcal{G}_A\cdot \mathcal{G}_{B}\cdot \mathcal{G}_{AB}.
\end{equation}
With an abuse of notation, one may write $\mathcal{G}_{AB}$ as the quotient group $\mathcal{G}/(\mathcal{G}_A\cdot \mathcal{G}_B)$. The entanglement entropy of the region $A$ is given as\cite{fattal2004entanglement},
\begin{equation}\label{eq_ee_rank}
    S(A)=\frac{1}{2}\rank(\mathcal{G}_{AB}).
\end{equation}

The stabilizer group of the steady state is generated by the set of all plaquette operators and a set of string operators obtained via projecting Majorana dimers into the spin Hilbert space. Let $\tilde{\mathcal{G}}_{AB}$ be the group that is generated by the plaquette operators on the boundary of $A$ and the string operators with one end point in $A$ and the other in $B$.  Clearly, $\mathcal{G}_{AB}\subseteq \tilde{\mathcal{G}}_{AB}$. However, in general it might be possible to combine some generators of $\tilde{\mathcal{G}}_{AB}$ with other generators $\mathcal{G}$ such that the result is localized in either localized in $A$ or in $B$. In the following we show that there are 2, and only 2 such relations.

Given that the product of a subset of plaquette operators corresponds to a set of closed loops on the lattice, it is clear that the only way such a product with some plaquettes on the boundary can be localized in $A$ or $B$ is to consider the product of all of plaquettes on the boundary. This result in two loops, one inside $A$ and one inside $B$. Then one can shrink away the inside loop via multiplying it with plaquette operators inside $A$ arriving at something with a support only in $B$. 

\begin{figure}
     \centering
     \begin{subfigure}[b]{0.23\textwidth}
         \centering
         \includegraphics[width=\textwidth]{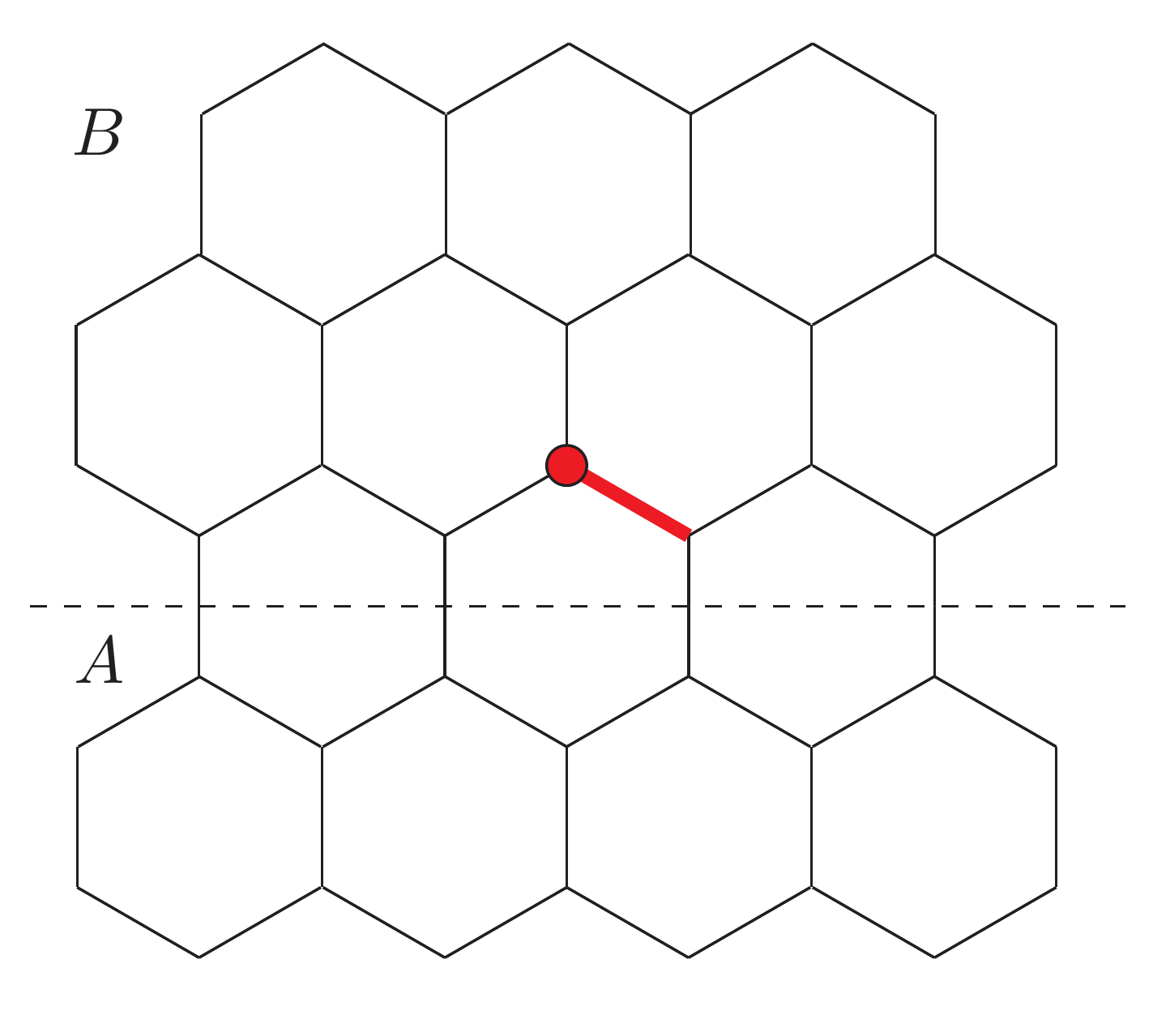}
         \caption{}
         \label{eeproofa}
     \end{subfigure}
     \hfill
     \begin{subfigure}[b]{0.23\textwidth}
         \centering
         \includegraphics[width=\textwidth]{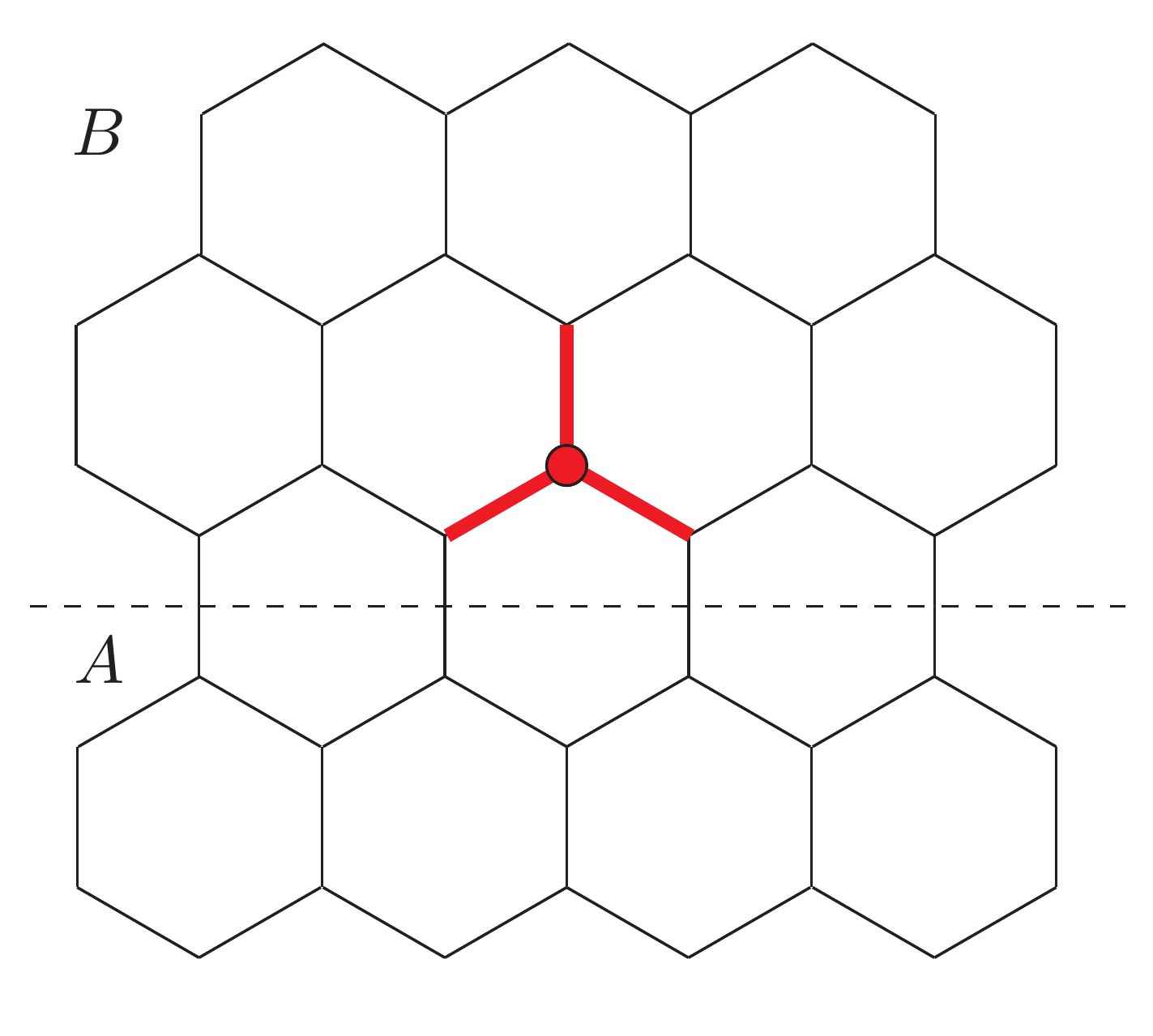}
         \caption{}
         \label{eeproofb}
     \end{subfigure}

     \begin{subfigure}[b]{0.23\textwidth}
         \centering
         \includegraphics[width=\textwidth]{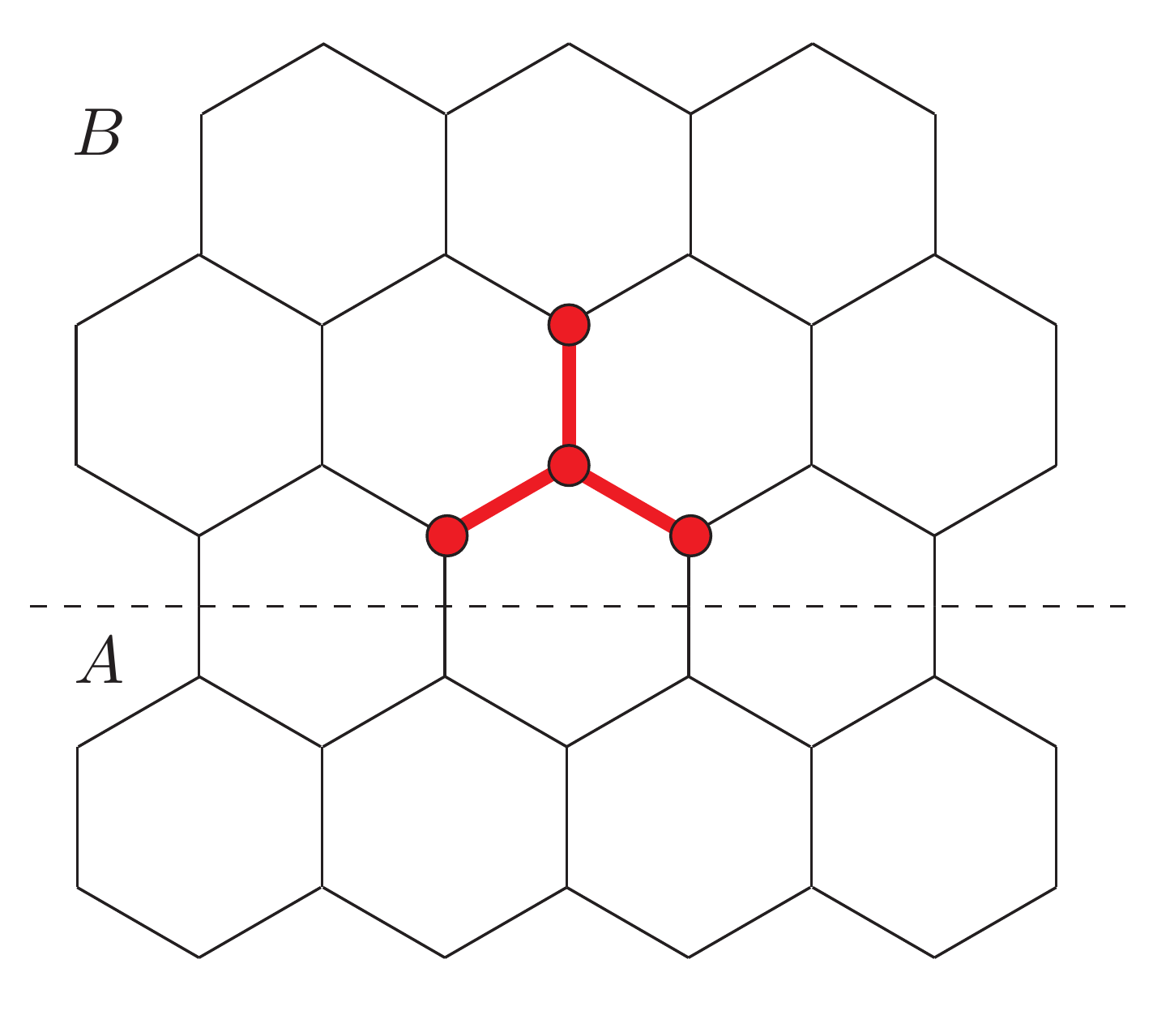}
         \caption{}
         \label{eeproofc}
     \end{subfigure}
     \hfill
     \begin{subfigure}[b]{0.23\textwidth}
         \centering
         \includegraphics[width=\textwidth]{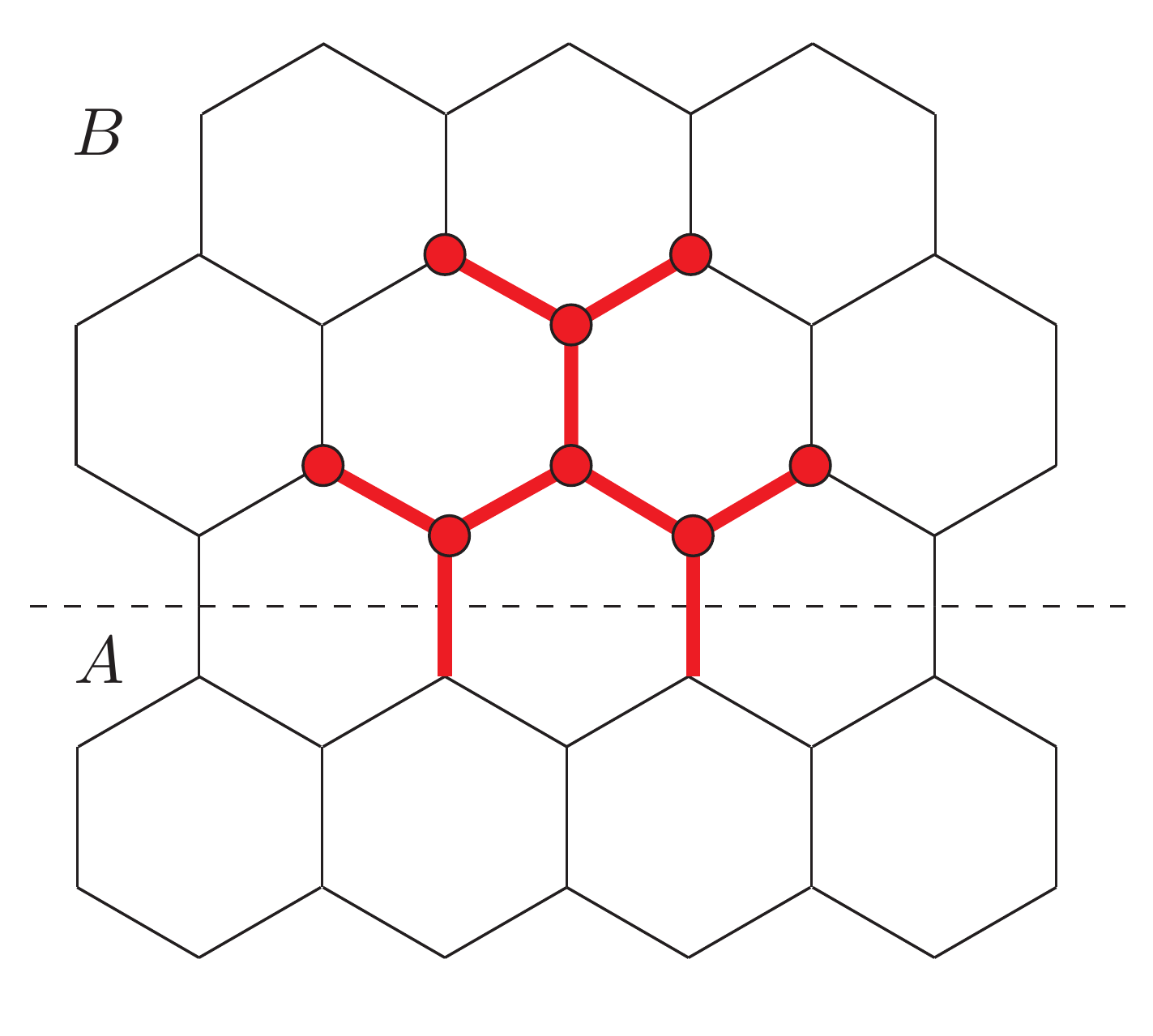}
         \caption{}
         \label{eeproofd}
     \end{subfigure}
        \caption{If $P$ is a non-trivial product of string operators and plaquette operators that acts trivially in region $B$, then it should include all points in region $B$ as end points of string operators. Say $P$ includes an string operator with the end point shown in panel (a). For $P$ to act trivially on the red dot qubit, it should include another string operator which passes through the two remaining bonds as shown in panel (b). But to have trivial support on the neighboring sites as well, the neighboring sites should also be end points of string operators that are included in $P$, as shown in panel (c). By repeating the same argument for these new end points, one can see the neighboring sites shown with red dots in panel (d) should also be end-points of string operators included in $P$. This line of argument then shows that $P$ should included all sites in region $B$ as end point.  }
\end{figure}

Now let us consider a product which also includes a subset of the string operators from $\tilde{\mathcal{G}}$. Such an operator can be represented by a set of end points (which are fixed by the choice of the subset of string operators that appear in the product) and a set of strings that connect them. In what follows we show that to cancel the support of such a product in $B$, the product must include \textit{all} the lattice points in $B$ as end points. Imagine it has a end point denoted by the red dot in Fig.\ref{eeproofa} as an endpoint of a string operator whose last segment is shown by a think red line. Now, the only way for this operator to act trivially on the red dot qubit is if there is another string that passes through the other two bonds connected to the red dot vertex (Fig.\ref{eeproofb}). But since the operator has to act trivially on the neighboring sites too, there has to be endpoints on neighboring sites as well(Fig.\ref{eeproofc}). Now we can repeat the same argument for these new endpoints to show that there should be endpoints on all next neighboring sites in $B$ as well (Fig.\ref{eeproofd}) and so on. Indeed, such a product exists and it can be found with a procedure similar to what was outlined in Appendix \ref{apx_mutual_info}. The important point that the above argument shows is that it is the only non-trivial product with trivial support on $B$. Therefore, we find that,
\begin{align}\label{eq_ranks}
    \rank(\mathcal{G}_{AB})=\rank(\tilde{\mathcal{G}}_{AB})-2=n_p+n_c-2.
\end{align}
Plugging Eq.\eqref{eq_ranks} into Eq.\eqref{eq_ee_rank} then yields the desired result. 

\begin{figure}
     \centering
     \includegraphics[width=0.5\columnwidth]{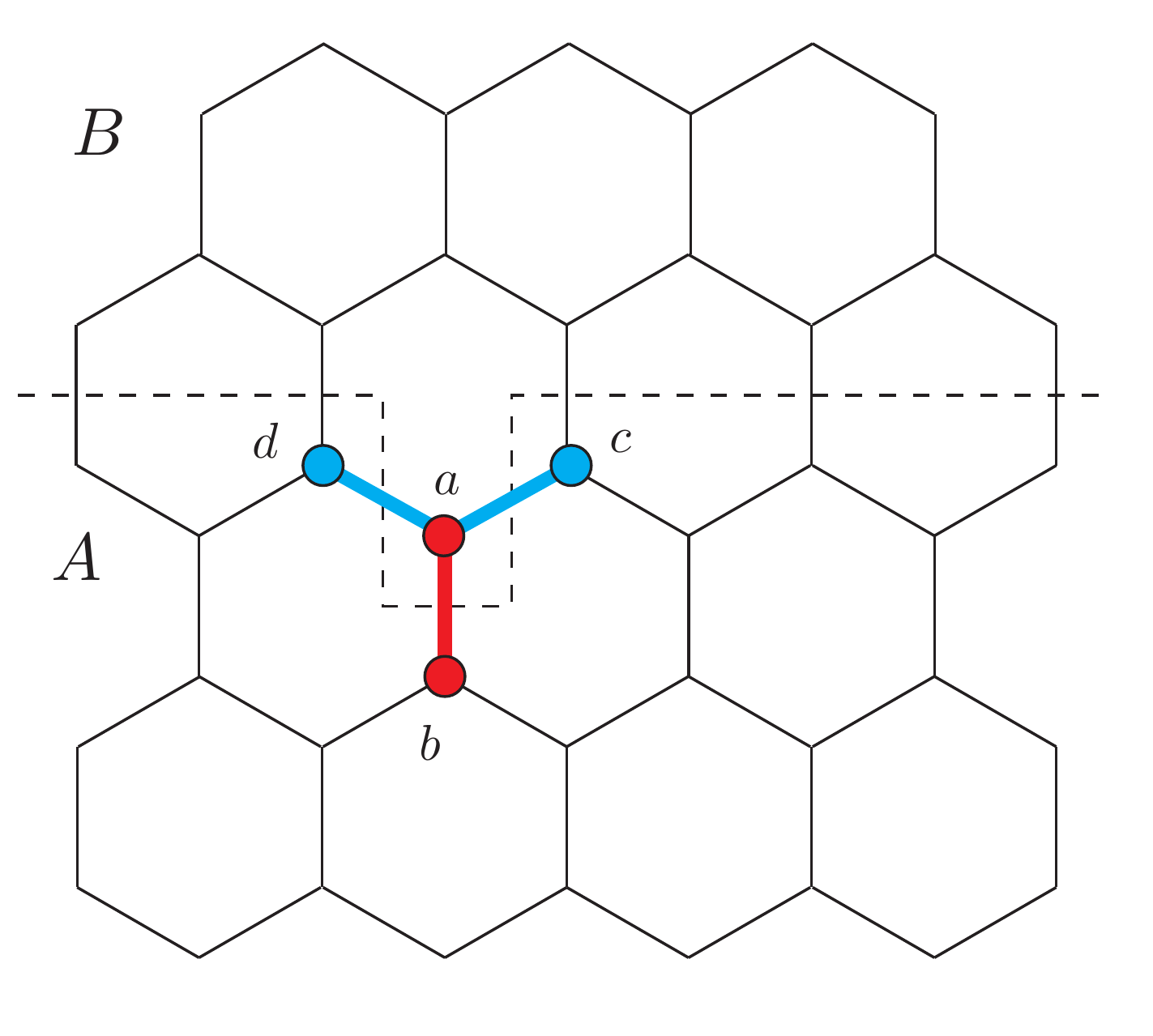}
     \caption{One can cancel the support of $g_{a,b}$ (red) in $B$ by multiplying it with $g_{d,e}$ (blue), resulting in an operator that is entirely supported on $A$.}
     \label{fig_nonsmoothboundary}
 \end{figure}

It is worth mentioning that if the boundary between $A$ and $B$ is not smooth, this argument could fail. A counter example is shown in Fig.\ref{fig_nonsmoothboundary}; $g_{a,b}$ has non-trivial support in both $A$ and $B$. But, one could multiply it with just $g_{c,d}$ (which has both endpoints in $A$) to arrive at an operator with trivial support on $B$.

\section{Percolation phase transitions at the boundary of the phase diagram}\label{apx_perc}
Here, we will show that the entanglement dynamics at the boundary of the phase diagram maps to $L$ decoupled classical 2D bond-percolation problems. 

For concreteness let us focus on the $p_y=0$ boundary. We partition the lattice into $L$ rows, where each row is comprised of $x$ and $z$ bonds. When $p_y=0$, no inter-row operator is going to be measured, hence the entanglement dynamic of each row is completely decoupled from the others. 

In the following we consider the entanglement dynamics of a single row. We label the spins by an index $j=1,\cdots,2L$. The circuit is then comprised of random measurements of the following operators,
\begin{align}
    &A_i=X_{2i-1} X_{2i},\\ &B_i=Z_{2i} Z_{2i+1},
\end{align}
for $i=1,\cdots,L$. On the other hand the following set of operators commute with all $A_i$ and $B_i$ operators
\begin{align}
    S_i=X_{2i}X_{2i+1},\quad \text{for }i=1,\cdots,L, 
\end{align}
and as such we may regard them as symmetries of the circuit. For simplicity, let us assume that the initial state is in the symmetry sector with $S_i=+1$ for all $i$, e.g. $\ket{+}^{\otimes 2L}$. Finally, we map this circuit to the $XX-Z$ measurement-only random circuit which has been studied thoroughly before\cite{PhysRevResearch.3.023200,Ali,PhysRevX.11.011030,PhysRevB.102.094204}. In particular, it has been shown that the entanglement dynamics is described by the 2d classical bond-percolation problem. To this end, we consider the dual circuit under the local unitary
\begin{align}
    U=\bigotimes_i \text{CNOT}_{2i,2i+1}
\end{align}
where $\text{CNOT}_{i,j}$ is the CNOT gate with qubit $i$ as the control and qubit $j$ as the target. Under this unitary, $A_i$, $B_i$ and $S_i$ transform as,
\begin{align}
    A_i&\mapsto X_{2i-1}X_{2i}X_{2i+1},\\
    B_i&\mapsto Z_{2i+1}\\
    S_i&\mapsto X_{2i}.
\end{align}
Since we have assumed the initial state is in the symmetry sector $S_i=+1$, the local unitary disentangles the spins with even index into the $X_{2i}=+1$ state. Therefore, in the dual picture, the circuit is basically consisted of the random measurement of $Z_{2i+1}$ and $X_{2i-1}X_{2i+1}$. 

\section{Stability of phases under single qubit measurements.}\label{apx_link_vs_plq_measurement}
In Section \ref{sec_perturbation} we argued that measuring plaquette operators directly with a rate $p_\text{plq}\gg p_s$ would stabilize both phases against single qubit measurements. Given that in the original circuit model plaquette operators are also measured with constant rate, it follows that both phases are robust against small rates of single qubit measurements. The only caveat is that in the original model, plaquette operators are not measured directly, but rather through a sequence of bond measurements, which in turn might cause proliferation of defects introduced via $Z$ measurements. In this section, we show that this will not happen, i.e. measuring plaquette operators directly or through a series of bond measurements will likely have the same effect. 

In the absence of perturbation, the steady state stabilizer group is generated by two types of stabilizers: 1) the plaquette stabilizers and 2) the string operators. Without perturbations, the plaquette operators do not have any dynamics while the string operators follow a dynamic similar to the parton dynamics described in Section \ref{sec:prop_of_std_state}: when a bond operator is measured, the two string operators  with endpoints on that bond will be replaced by a string operator that is obtained from connecting the two, and the bond operator itself. 

After a single qubit measurement, the two adjacent plaquette operators will be replaced in the generating set by their product and a single qubit $Z$ operator. We denote the latter by $g_z$. The new stabilizer group has three types of generators: 1) plaquette operators, 2) string operators (which have remained unchanged) and 3) one $g_z$ stabilizer. The important observation to make is that, the presence of $g_z$ in the generating set has no effect on the dynamics of the string and plaquette operators under the subsequent bond measurements (although $g_z$ has its own dynamic). As such, after a constant time, the removed stabilizers will be added to the generating set again, at which point $g_z$ will be removed since the set of independent generators can not have more than $N=L^2$ elements. 

\section{Purification in the Critical Phase}\label{app:Levy_Flight}
Starting from a maximally-mixed initial state, we argue that in the critical phase, the system disentangles as a power law $S(t) \sim t^{-1}$ in time as projective measurements are performed.  Within a short time $t_{*}\sim O(\log L)$ after starting to perform measurements of the bond operators, the plaquette stabilizers $W_{p}$ become part of the stabilizer group.  To study the subsequent purification dynamics of the system, it is convenient to consider the following density matrix for the Majorana partons
\begin{align}
    \rho_{f} \sim  \ket{\Psi_{b}}\bra{\Psi_{b}}\otimes\mathds{1}
\end{align}
where $\ket{\Psi_{b}}$ is a pure state of the $b$ Majorana fermions, as described below Eq. (\ref{eq:dynamics_rho}) in which each $b$ Majorana fermion is dimerized with its nearest-neighbor ($ib^{j}_{\rB}b^{j}_{\rB'}\ket{\Psi_{b}} = \ket{\Psi_{b}}$ where $\rB$ and $\rB'$ are sites at the ends of a bond of type $j$).  The $c$
 Majorana partons are in a maximally-mixed initial state.  The density matrix of the spin degrees of freedom
 \begin{align}\label{eq:app:rho_spin}
     \rho \propto \prod_{p}\frac{1 + W_{p}}{2}
 \end{align}
 clearly describes a volume-law-entangled state.  
 
 As measurements of the bond operators are performed, the measured bonds become part of the stabilizer group that describes the evolving, monitored state.  Consider measuring $X_{\rB}X_{\rB'}$ (where $\rB$ and $\rB'$ are sites connected by an $x$-type bond).  Since $X_{\rB}X_{\rB'} = c_{\rB}b^{x}_{\rB}b^{x}_{\rB'}c_{\rB'}$, a measurement of this operator in the state (\ref{eq:app:rho_spin}) is equivalent to adding $\pm ic_{\rB}c_{\rB'}$ as a stabilizer to the evolving state of the Majorana partons.  We refer to $c_{\rB}$ and $c_{\rB'}$ as ``paired" Majorana partons since they are dimerized and belong to the stabilizer group for the density matrix of the fermions after measuring $X_{\rB}X_{\rB'}$.
 
 The ``unpaired" $c$ Majorana partons each provide an $O(1)$ contribution to the entanglement entropy of the entire system.  To understand the purification of the evolving state, we investigate how these unpaired degrees of freedom are ``annihilated" (become paired) as measurements are performed. First, we note that the unpaired Majorana partons can only annihilate when two of them become nearest-neighbors, and the corresponding bond operator connecting the two is measured.  It is easily checked that a measurement of a bond operator connecting two Majorana partons in which at least one is unpaired, will not change the number of independent generators of the stabilizer group for $\rho_{f}$.  Consider, for example, a measurement of $X_{\rB}X_{\rB'} = c_{\rB}b^{x}_{\rB}b^{x}_{\rB'}c_{\rB'}$ in a state $\rho_{f}$ where $ic_{\rB}c_{\sB} = +1$ where $\sB$ is another site in the system, and where $c_{\rB'}$ is unpaired.  After this measurement, $\pm ic_{\rB}c_{\rB'}$ belongs to the stabilizer group, while $c_{\sB}$ is now unpaired.  As a result, the measurement has the effect of moving the unpaired Majorana parton from $\rB$ to $\sB$. 
 
 In the critical phase, the probability distribution of string stabilizer lengths (the Cartesian distance between their endpoints) decays as a power-law, as reviewed in Sec. \ref{sec:prop_of_std_state}; equivalently, in the parton description of the steady-state, the probability that a Majorana $c_{\rB}$ is paired with another Majorana $c_{\rB + \sB}$ at relative separation $\sB$ decays as $P(\sB) \sim |\sB|^{-3}$.  To understand the purification dynamics in the critical phase, we now assume a dilute concentration of the unpaired $c$ Majoranas, and that the probability that a given \emph{paired} Majorana is dimerized with another Majorana fermion at relative separation $\rB$ is again given by $P(\rB)$. As bond measurements are performed, the dynamics of an unpaired Majorana degree of freedom is then described by a random walk, where probability density per unit time for a step in the direction $\rB$ is given by $P(\rB)$.  The variance in the distribution of step lengths $\langle|\rB|^{2}|\rangle = \int |\rB|^{2}P(\rB)d^{2}\rB$ is infinite, so that the wandering of a given unpaired Majorana is dominated by rare ``long" steps; such a random walk is known as a L\'{e}vy flight \footnote{Random walks in $d$ spatial dimensions where a displacement $\rB$ occurs with probability $P(\rB)\sim |\rB|^{-d-\sigma}$ and with $\sigma < 2$ define L\'{e}vy flights, for which $\langle |\rB|^{2}\rangle$ is divergent; see e.g. Ref. \cite{metzler2007evy}}.  
 
The unpaired Majorana degrees of freedom wander and ``annihilate" by pairing with other unpaired Majoranas.  To understand how they wander and annihilate, we define a coarse-grained density $n(\rB,t)$ of the unpaired degrees of freedom and compare two ways in which these Majorana degrees of freedom may annihilate: ($i$) Unpaired Majorana degrees of freedom may annihilate when they are sufficiently close together; in this case, the rate $\Gamma(n)$ at which the Majorana degrees of freedom annihilate is proportional to the local density $\Gamma(n) \propto n$; ($ii$) Unpaired Majorana degreees of freedom annihilate by taking rare ``long" steps.  Let $P(\rB) \sim |\rB|^{-\Delta}$ ($\Delta = 3$ in our case of interest). In a time $t$ an unpaired Majorana degree of freedom will travel a distance $O(t^{1/(\Delta - 2)})$; this can be argued by observing that the evolution of the local density in the absence of annihilation events is given by
\begin{align}\label{app:eq:Levy}
    \frac{\partial n(\rB,t)}{\partial t} = \lambda\int\,d^{2}\rB'\,P(\rB')\left[n(\rB-\rB',t)-n(\rB,t)\right]
\end{align}
The evident re-scaling of space and time that leaves (\ref{app:eq:Levy}) invariant leads to the typical distance travelled by an unpaired degree of freedom in a time $t$. In two spatial dimensions, the typical spacing between unpaired Majoranas in a region with density $n(\rB,t)$ is $n(\rB,t)^{-1/2}$.  As a result, the annihilation rate due to long steps in a L\'{e}vy flight, obtained by estimating the typical time to traverse this distance, is $\Gamma(n) \sim n^{(\Delta - 2)/2}$.  

At long times, these respective processes lead to a decay of the density as $n\sim t^{-1}$ for process ($i$) and $n\sim t^{(2-\Delta)/2}$ for process ($ii$).  As a result, when 
$\Delta < 4$ (for our case of interest, $\Delta = 3$), the the effective annihilation dynamics for the unpaired degrees of freedom leads to the long-time behavior $n \sim t^{-1}$.  We conclude that the entanglement entropy of the system should decay as $S(t)\sim t^{-1}$ in the critical phase.\footnote{A simple generalization of this argument for a distribution $P(\rB) \sim |\rB|^{-d-\sigma}$ in $d$ spatial dimensions can be used to show that process ($i$) dominates over ($ii$) at long times whenever $d>\sigma$ when $\sigma<2$.}


\newpage
\widetext
\clearpage

\section{Supplementary Figures}\label{apx_sup_figs}
\begin{figure}[!h]
     \centering
     \includegraphics[width=0.5\textwidth]{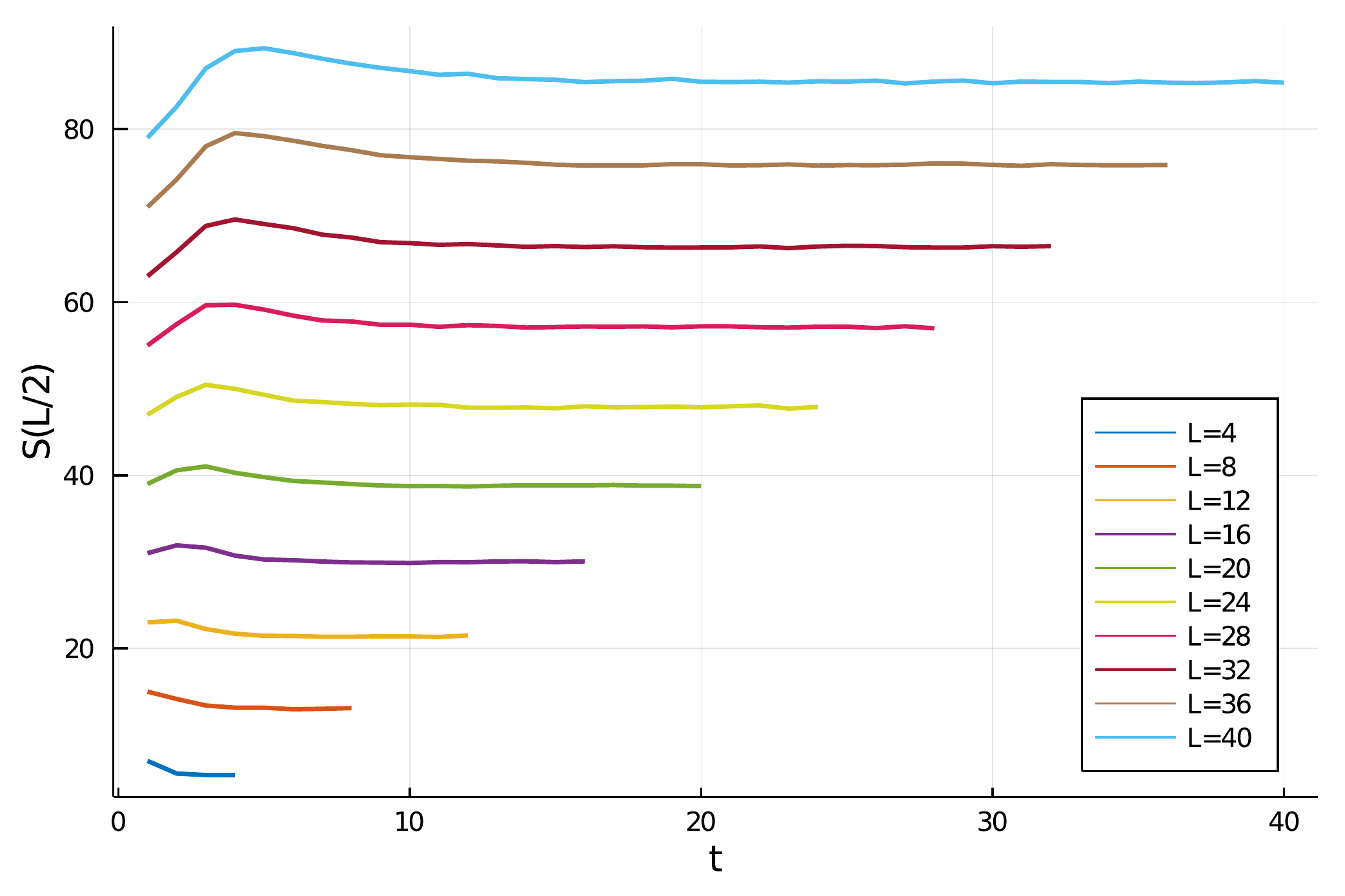}
     \caption{Entanglement entropy of a cylinder of size $L/2\times L$ as a function of time at the isotropic point $p_x=p_y=p_z$, starting from a state which is a projection onto the subspace where all stabilizers (plaquettes + long sycles) and all the $z$-bond operators have some definite value, say $+1$. It is clear from the plot that the entanglement entropy is saturated at the final value after time $t=L$. This should be contrasted to the case when one starts from a totally mixed initial state, where one needs to wait for $t=O(L^2)$ time stpes for the system to reach the steady state. }
     \label{fig_S_half_t}
 \end{figure}
 
 \begin{figure}[!h]
     \centering
     \begin{subfigure}[b]{0.45\textwidth}
         \centering
         \includegraphics[width=\textwidth]{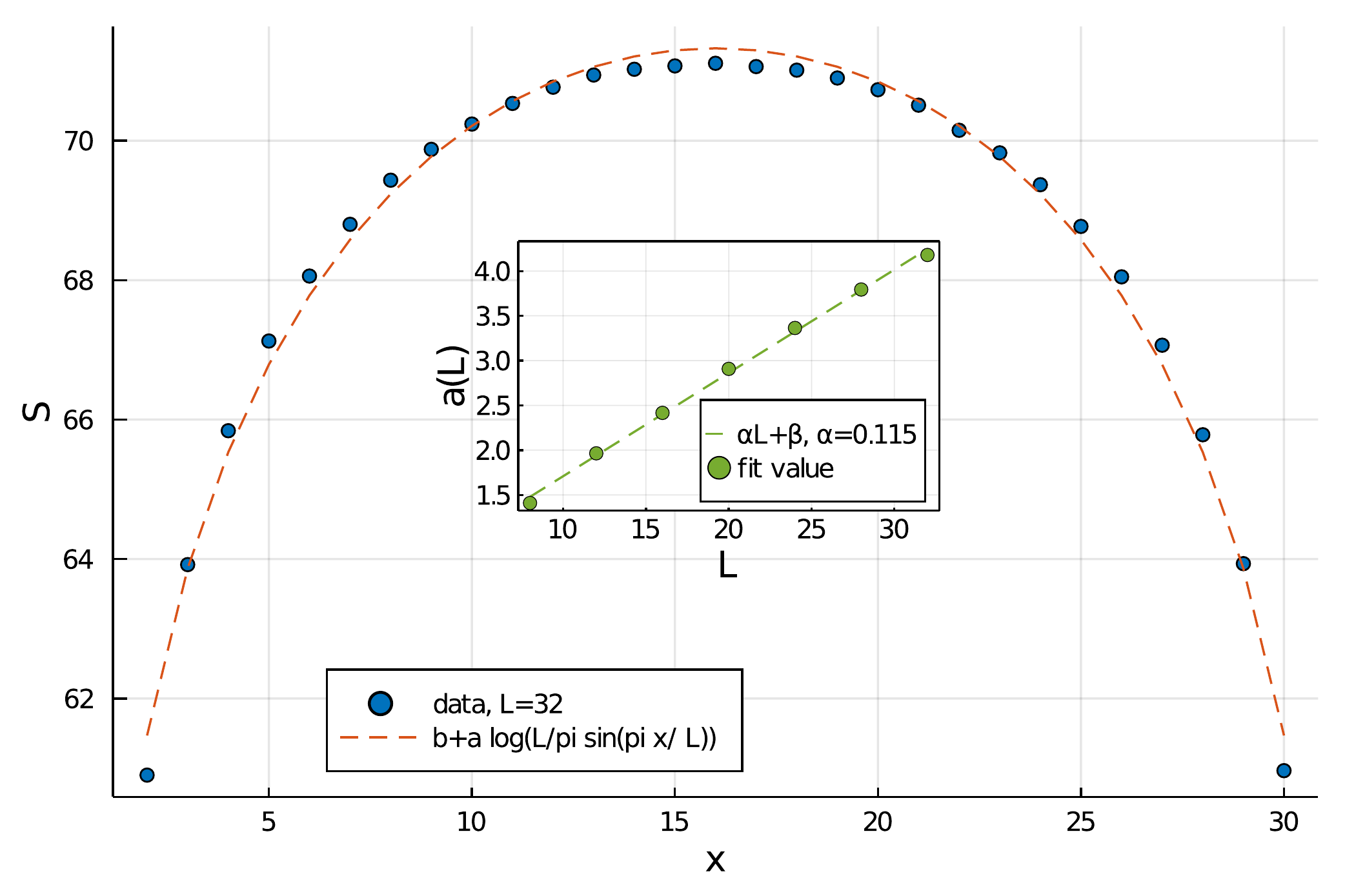}
         \caption{}
         \label{fig_SRZ_px02}
     \end{subfigure}
    \hfill
    \begin{subfigure}[b]{0.45\textwidth}
         \centering
         \includegraphics[width=\textwidth]{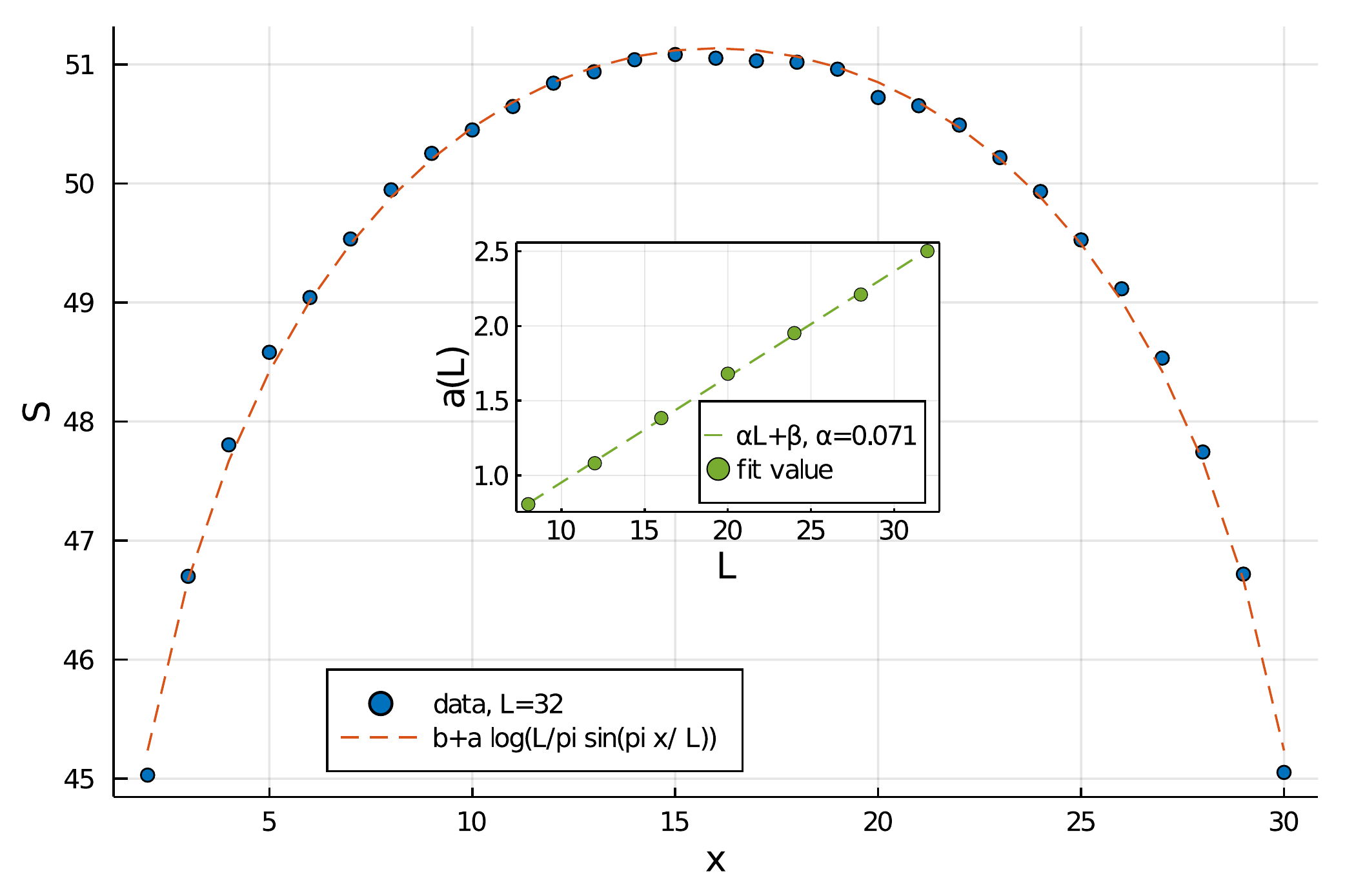}
         \caption{}
         \label{fig_SRX_px02}
     \end{subfigure}
        \caption{Entanglement entropy of cylindrical region of size $x\times L$ for the steady state of the circuit model at $p_x=p_y=0.2$ in the critical phase when  (a) the boundary cuts through  the $Z$-bonds and (b) when the boundary cuts through the $X$-bonds. Note that the best fit value of $a$ is different for different cuts, and they are different from their value at the isotropic point $p_x=p_y=p_z$ (see Fig.\ref{fig_Sr_iso})}
\end{figure}

\begin{figure}[!h]
     \centering
         \centering
         \includegraphics[width=0.5\textwidth]{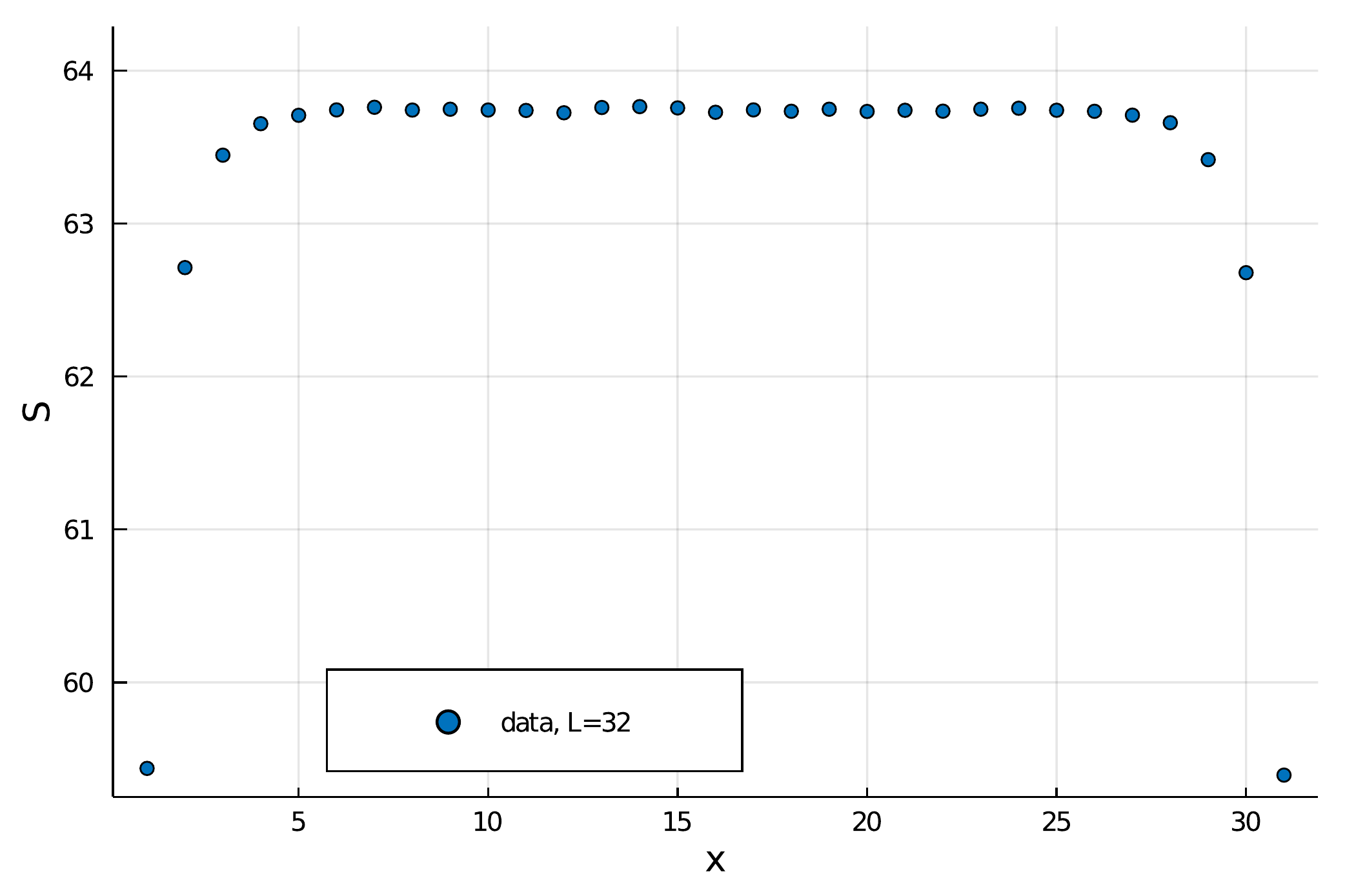}
        \caption{Entanglement entropy of cylindrical region of size $x\times L$ for the steady state of the circuit model at $p_x=p_y=0.1$, which clearly indicates an area law scaling of entanglement.}
        \label{fig_SRZ_px01}
\end{figure}


\begin{figure}
     \centering
     \includegraphics[width=0.4\textwidth]{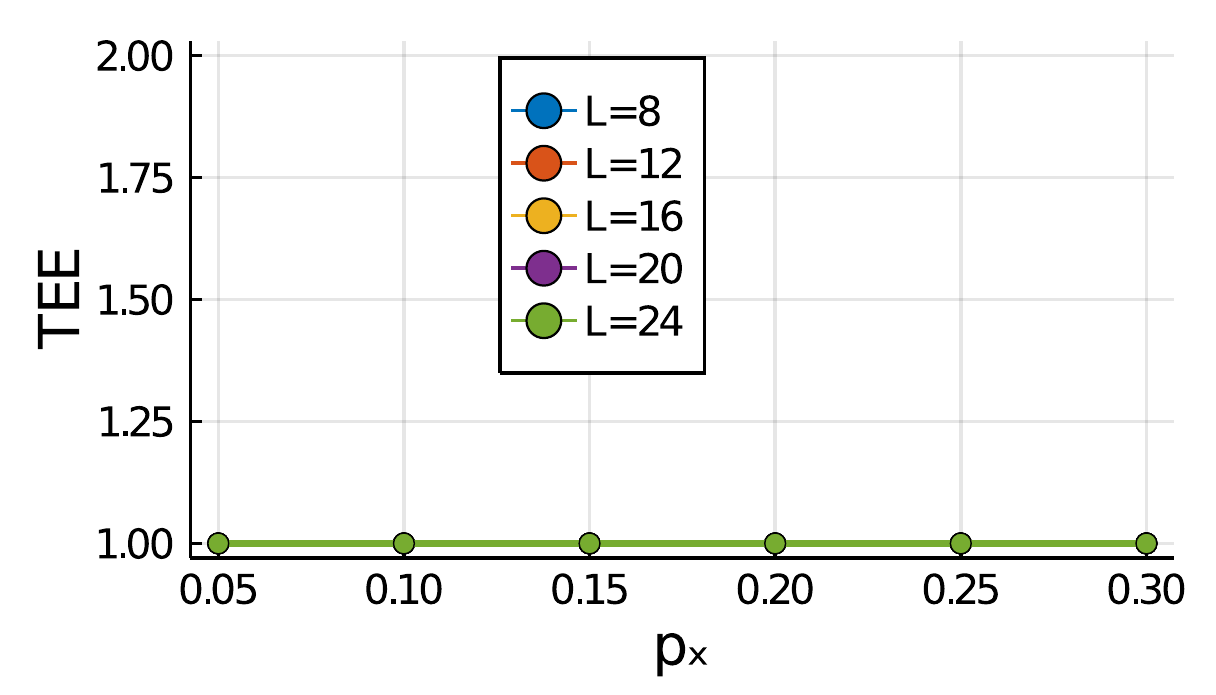}
     \caption{Topological entanglement entropy in the steady state versus $p_x$ on the symmetric line $p_x=p_y$. As is clear from the figure the opological entanglement entropy is always $1$ irrespective of whether the system is in the area law phase or the critical phase. This is due to the fact that the plaquette operators belong to the stabilizer group of the steady state in both phases.}
     \label{fig_topoee_p}
\end{figure}

\begin{figure}[!h]
     \centering
     \begin{subfigure}[b]{0.45\textwidth}
         \centering
         \includegraphics[width=\textwidth]{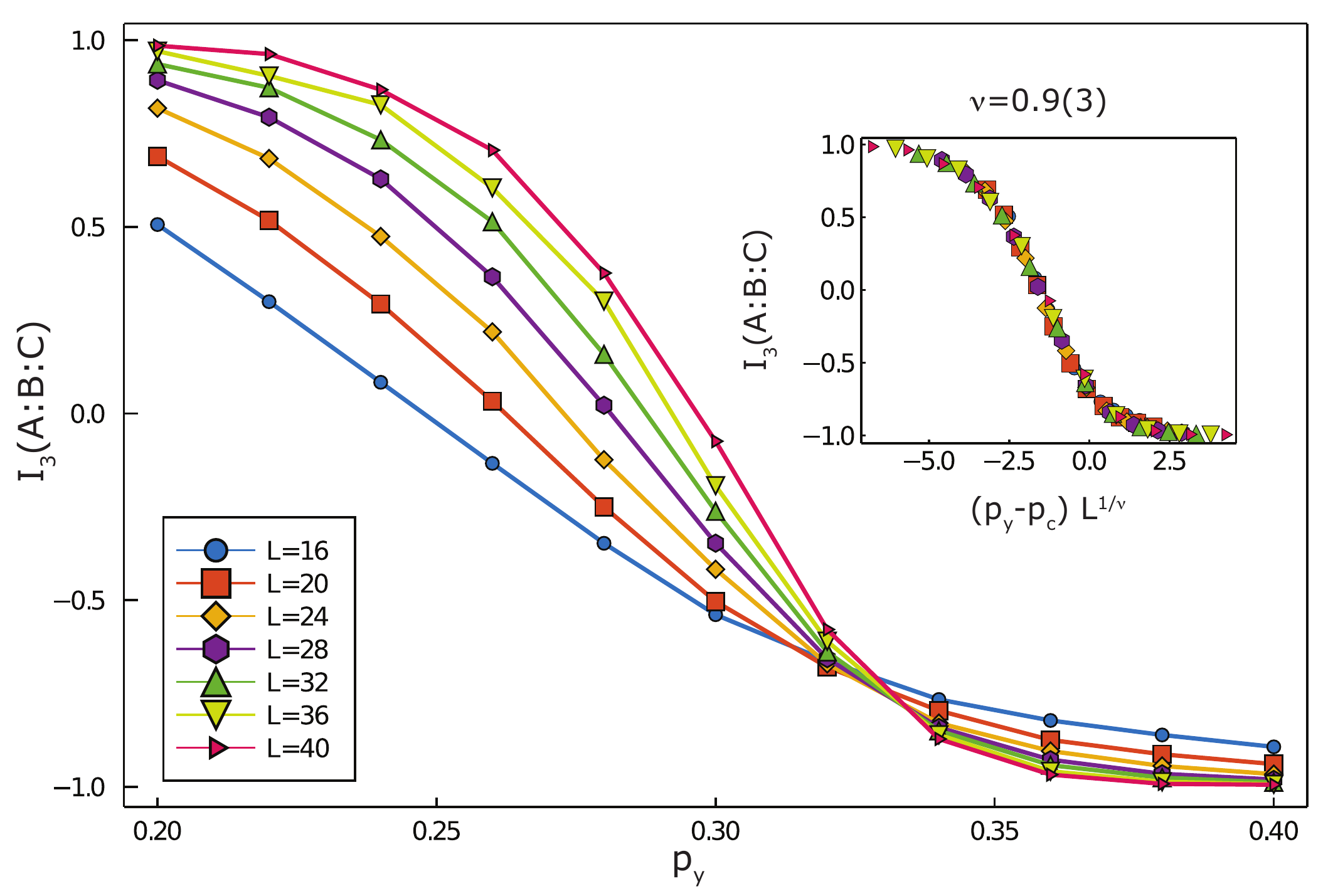}
         \caption{$p_x=0.05$}
         \label{fig_I3_05}
     \end{subfigure}
     \hfill
    \begin{subfigure}[b]{0.45\textwidth}
         \centering
         \includegraphics[width=\textwidth]{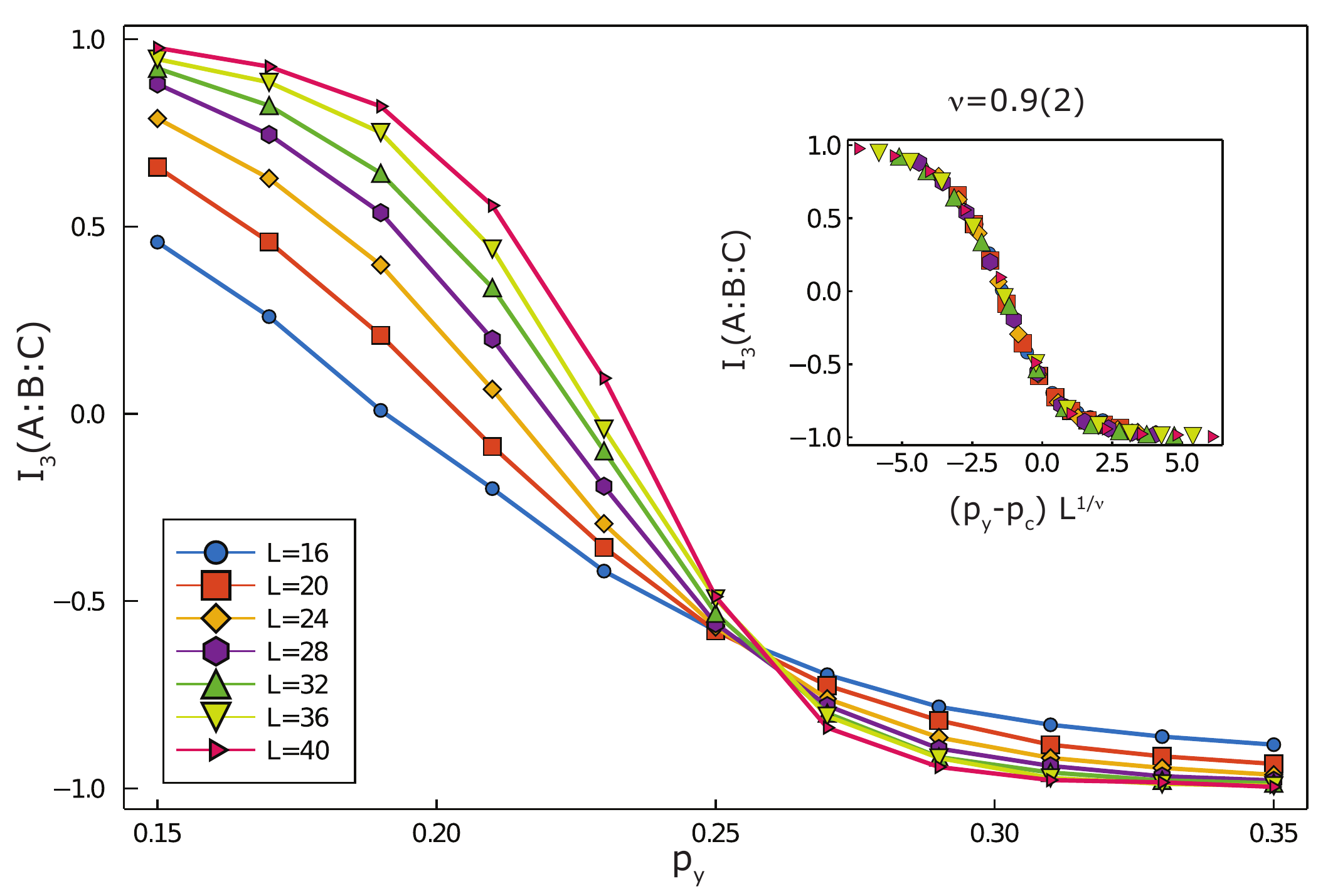}
         \caption{$p_x=0.1$}
         \label{fig_I3_1}
     \end{subfigure}
     
    \begin{subfigure}[b]{0.45\textwidth}
         \centering
         \includegraphics[width=\textwidth]{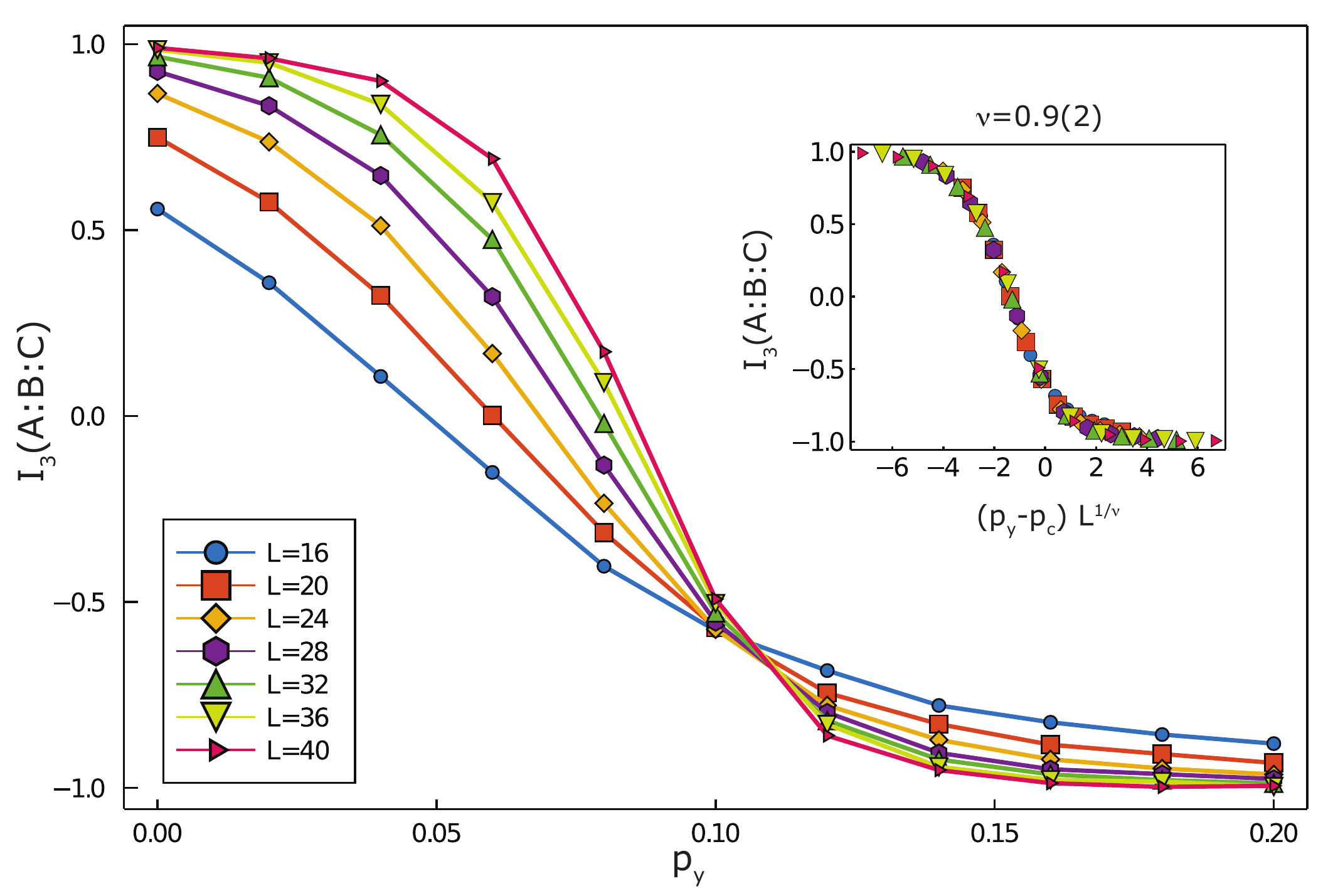}
         \caption{$p_x=0.25$}
         \label{fig_I3_25}
     \end{subfigure}
     \hfill
    \begin{subfigure}[b]{0.45\textwidth}
         \centering
         \includegraphics[width=\textwidth]{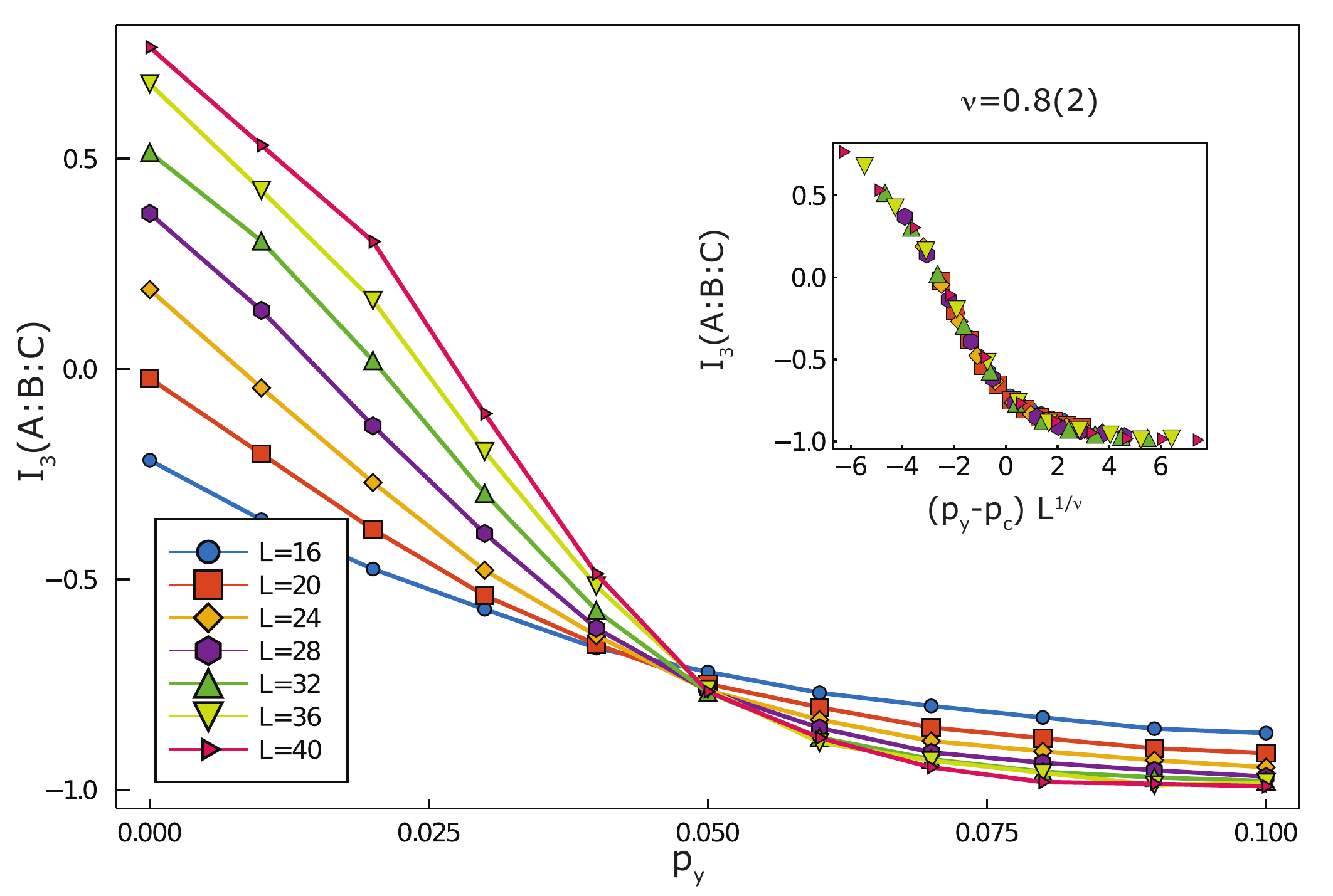}
         \caption{$p_x=0.33$}
         \label{fig_I3_33}
     \end{subfigure}
        \caption{The tripartite mutual information $I_3(A:B:C)$ versus $p_y$ for fixed (a) $p_x=0.05$, (b) $p_x=0.1$, (c) $p_x=0.25$ and (d) $p_x=0.33$. The regions $A$, $B$ and $C$ are chosen according to Fig.\ref{fig_ABC}. The inset in each plot shows the corresponding data collapse. The estimate for the correlation length critical exponent which is found via data collapse is shown above each inset plot. Note that the critical points of subplots (a) and (d) as well as the critical points of subplots (b) and (c) are related by a reflection along $z$ bonds.}
        \label{fig_I3_moreplots}
\end{figure}

\begin{figure}
     \centering
          \begin{subfigure}[b]{0.25 \textwidth}
         \centering
         \includegraphics[width=\textwidth]{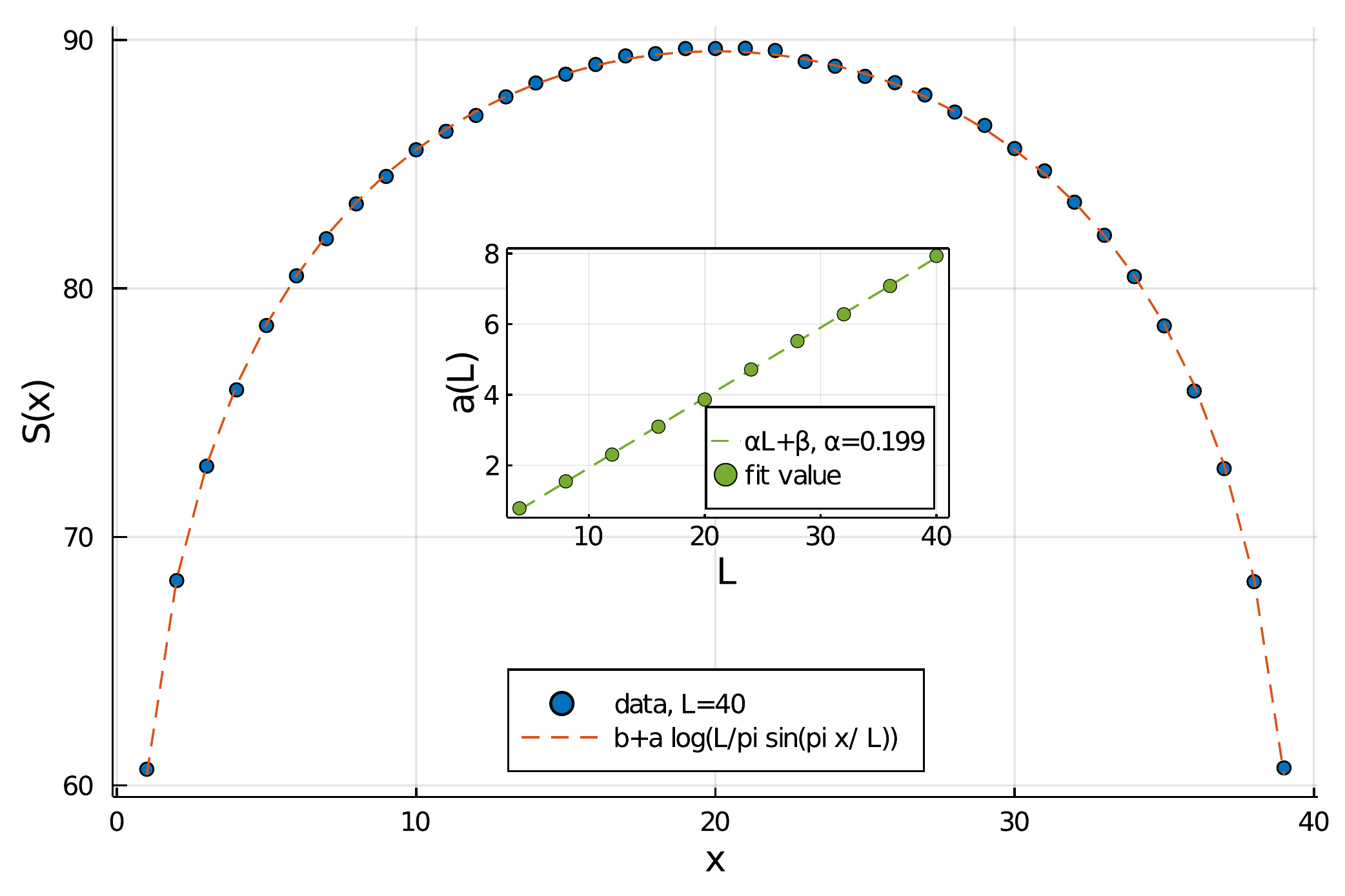}
         \caption{}
         \label{fig_perturb_3_05}
     \end{subfigure}
     \hfill
     \begin{subfigure}[b]{0.25 \textwidth}
         \centering
         \includegraphics[width=\textwidth]{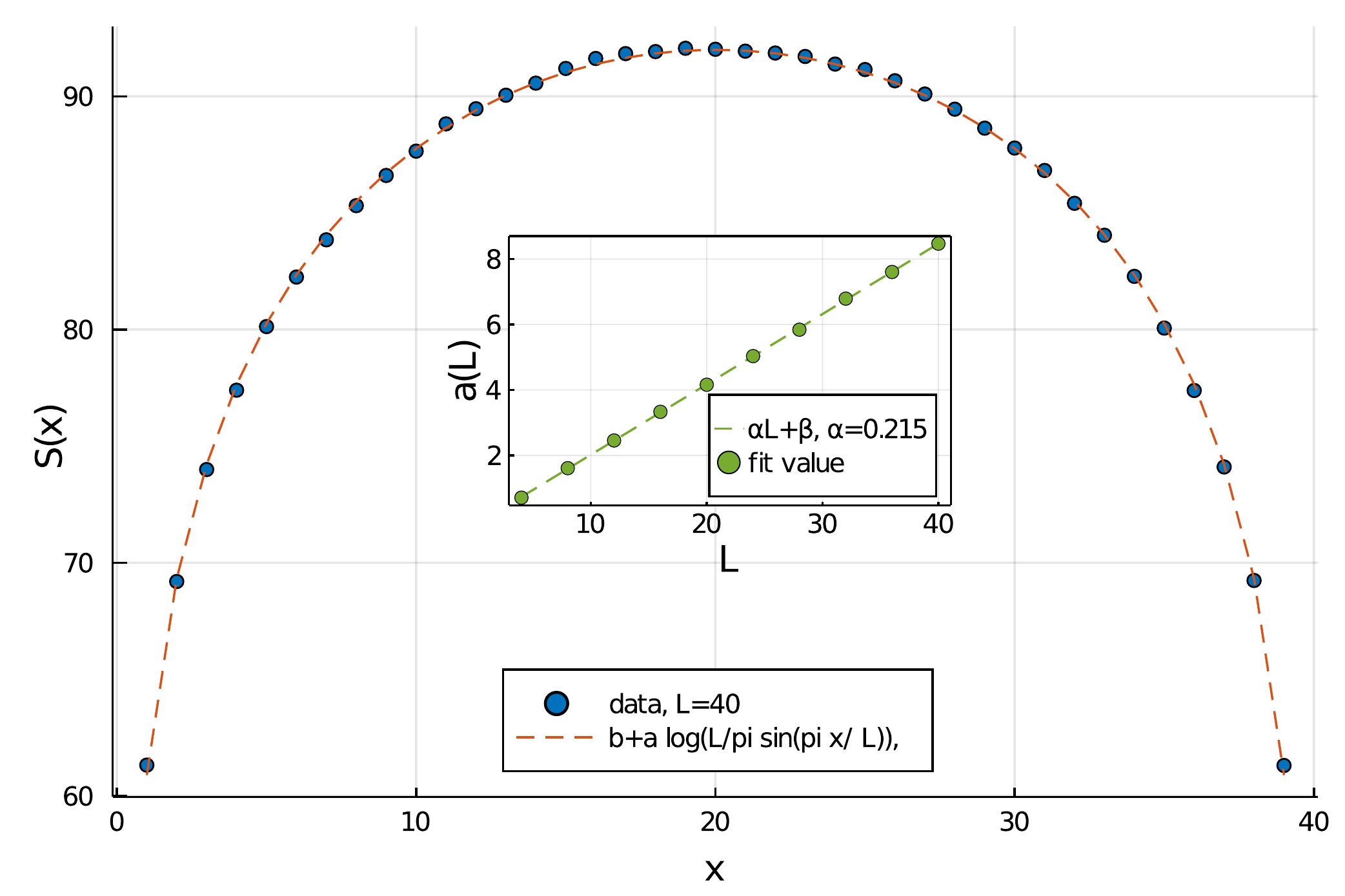}
         \caption{}
         \label{fig_perturb_3_1}
     \end{subfigure}
    \caption{The entanglement entropy of a cylinder of size $x\times L$ in the steady state of the monitored circuit where at each step either a product of two adjacent bond operators is measured randomly with probability $p_3$ or a random bond operator is measured randomly with probability $1-p_3$. (a) corresponds to $p_3=0.05$ and (b) corresponds to $p_3=0.1$. The inset shows the best fit value of $a$ versus system size $L$ which shows a clear linear scaling, indicating a $L \log L$ violation of the area law. }
\end{figure}

\begin{figure}
     \centering
     \includegraphics[width=0.5\textwidth]{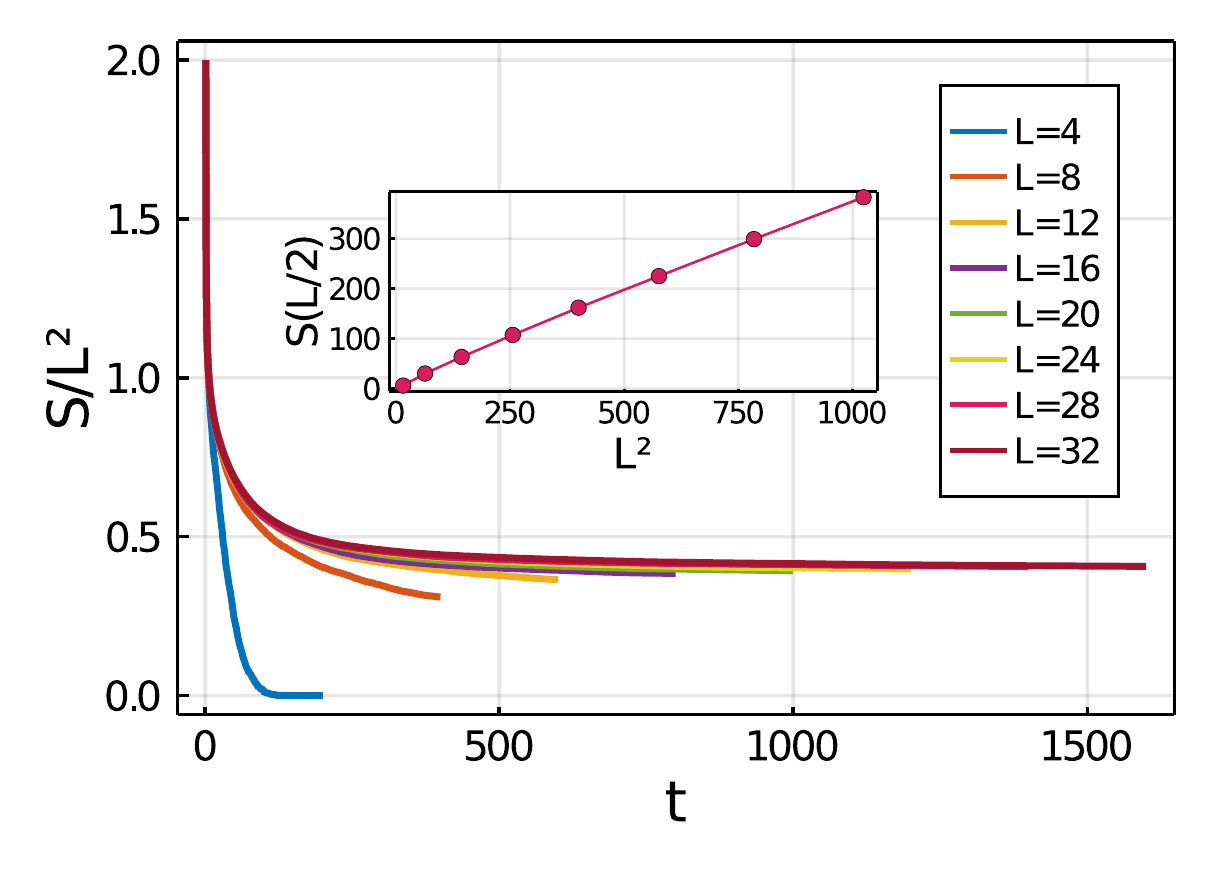}
     \caption{The entanglement entropy of the system at the isotropic point $p_x=p_y=p_z$ subjected to random single qubit $Z$ measurements with probability $p_s=0.01$, starting from a totally mixed initial state. As can be seen from the figure, the entropy plateaus at a value proportional to $L^2$ which means that the steady state has volume law entanglement entropy. The inset shows the entanglement entropy of a half torus cylinder of size $L\times L/2$ in the pure steady state versus $L^2$ which scales linearly.   }
     \label{fig_perturb_S}
\end{figure}



\begin{figure}
     \centering
          \begin{subfigure}[b]{0.49 \textwidth}
         \centering
         \includegraphics[width=\textwidth]{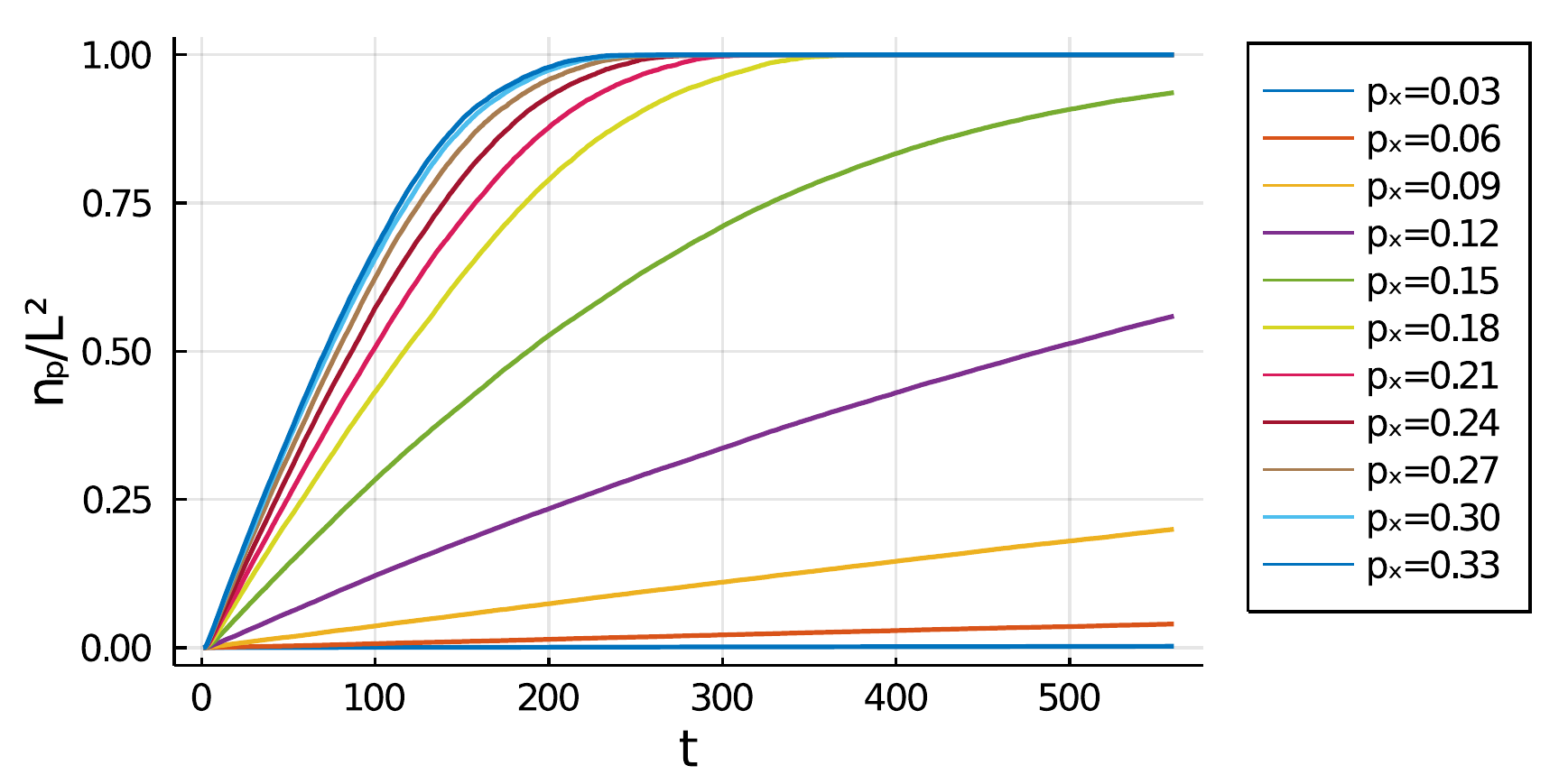}
         \caption{}
         \label{fig_n_plq_t}
     \end{subfigure}
     \hfill
     \begin{subfigure}[b]{0.43 \textwidth}
         \centering
         \includegraphics[width=\textwidth]{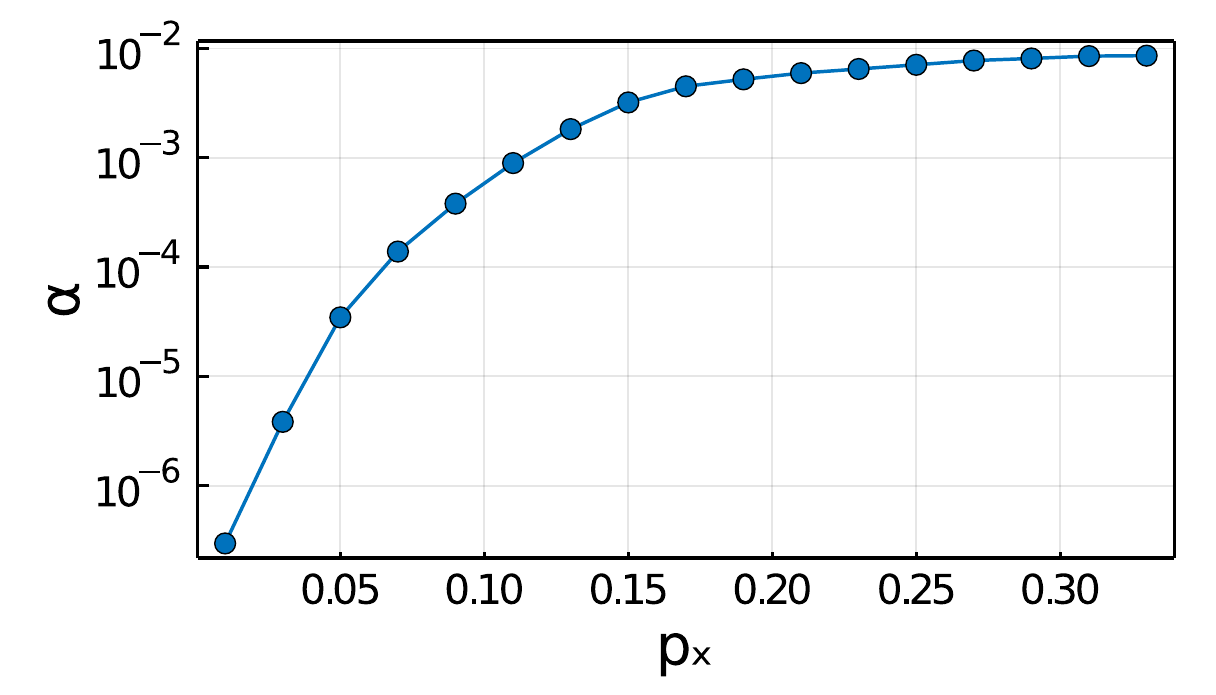}
         \caption{}
         \label{fig_plq_rate}
     \end{subfigure}
    \caption{ (a) The fraction of plaquette operators that are in the stabilizer group as a function of time (b) The rate $\alpha$ of measuring a plaquette operator versus $p_x$ along the symmetric line $p_x=p_y$. For each $p_x$, $\alpha $ is found by fitting the analytic form $n_p/L^2(t)=1-\exp(-\alpha t)$ to the data shown in the left panel, for $t=10-20$. }
\end{figure}

\end{document}